\documentclass{article}
\usepackage{arxiv}
\usepackage[utf8]{inputenc} 
\usepackage[T1]{fontenc}    
\usepackage{hyperref}       
\usepackage{url}            
\usepackage{booktabs}       
\usepackage{amsmath,amsfonts,amsthm}
\usepackage{nicefrac}       
\usepackage{microtype}      
\usepackage{cleveref}       
\usepackage{lipsum}         
\usepackage{graphicx}
\usepackage{natbib}
\usepackage{doi}

\usepackage{float}
\usepackage{graphicx}
\usepackage[english]{babel}
\usepackage{caption}
\usepackage{subcaption}
\usepackage{rotating}
\usepackage{pdflscape}
\usepackage{afterpage}
\usepackage{capt-of}
\usepackage{booktabs}
\usepackage{threeparttable}
\floatstyle{plaintop}
\RequirePackage{natbib}
\usepackage{threeparttable}
\usepackage{rotating}
\usepackage{hyperref}
\hypersetup{
  colorlinks=true,
  linkcolor=blue,
  citecolor=blue,
  urlcolor=magenta,
  hyperfootnotes=true,
}

\usepackage{enotez}
\let\footnote=\endnote
\usepackage{graphicx}
\usepackage[title]{appendix}
\usepackage{color}
\usepackage{graphicx,multicol}

\title{Identifying the post-pandemic determinants of low performing students in Latin America through Interpretable Machine Learning methods}

\date{}

\newif\ifuniqueAffiliation
\ifuniqueAffiliation 
\author{ \href{https://orcid.org/0000-0001-9333-8331}{\includegraphics[scale=0.06]{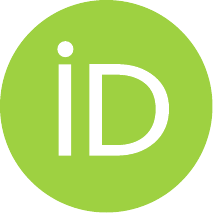}\hspace{1mm}Marcos Delprato}\thanks{Alternative email: \texttt{marcos.a.delprato@gmail.com}} \\
	Department of Education \\
	University of Bath \\
	UK \\
	\texttt{md2645@bath.ac.uk} \\
}
\else
\usepackage{authblk}

\newbox{\orcid}\sbox{\orcid}{\includegraphics[scale=0.06]{orcid.pdf}}
\author[1,2]{%
	\href{https://orcid.org/0000-0001-9333-8331}{\usebox{\orcid}\hspace{1mm}Marcos Delprato\thanks{Emails: \texttt{md2645@bath.ac.uk} and \texttt{marcos.a.delprato@gmail.com}.}}%
}
\affil[1]{Department of Education,  University of Bath, UK}
\affil[2]{Instituto de Investigaciones Educativas, Universidad Nacional de Chilecito, Argentina}
\fi

\renewcommand{\shorttitle}{Determinants of low performing students in Latin America. Machine Learning}

\hypersetup{
pdftitle={Identifying the post-pandemic determinants of low performing students in Latin America through interpretable Machine Learning SHAP Values—Insights from PISA 2022},
pdfsubject={econ.GN},
pdfauthor={Marcos Delprato},
pdfkeywords={learning poverty, interpretable machine learning, SHAP values, SDG4, Latin America, PISA 2022, COVID-19 pandemic},
}

\begin{document}
\maketitle

\begin{abstract}
\textbf{Introduction}. The high prevalence of students not achieving basic learning competencies in Latin America (LAC) is concerning, even more so considering the region's deep structural inequalities and the larger post-pandemic learning losses. Within this scenario, the paper aims to contribute to the identification of the determinants of bottom and low performers (below level 2). \\
\textbf{Methodology}. Based on 2022 data from the Programme for International Student Assessment (PISA) for 10 LAC countries, and using a stacking model integrating binary classification models as well as by applying  Shapley Additive Explanations (SHAP) analysis for interpretability, we identify critical factors impacting on the student performance across low performers groups. \\
\textbf{Results}. We find that a student with the highest probability of being a not achiever speaks a minority language and had repeated, has no digital devices at home, comes from a poor family and works for payment half of the week, and the school the student attends has wide disadvantages such as bad school climate, weak Information and Communication Technology (ICT) infrastructure and poor teaching quality (only a third of teachers being certified). For countries' estimates, we find quite homogeneous patterns regarding the contribution of top ranked factors, with repetition at primary, household wealth, and educational ICT inputs being top ten ranked covariates in at least 8 out of the 10 total countries. \\
\textbf{Discussions}. The paper findings contribute to the broad literature on strategies to identify and to target those most left behind in Latin American education systems.
\end{abstract}



\newpage
\section{Introduction}
\label{section1}

Latin American students' poor achievement at secondary level (\citealp{arias24b}) and weak learning growth (\citealp{angrist21}) are of great policy concern, even more so because the region's low learning trap is rooted in systemic and profound inequalities (\citealp{breton18,fernandez24}). Based on the OECD Programme for International Student Assessment (PISA) 2022 data, \cite{arias23} estimate that 75\% of the students in the region are low math performers\textendash not reaching the basic competencies given by level 2\textendash, and more than half of students (around 55\%) lacks basic reading skills (fail to understand a simple text). Other recent regional learning survey results such the Regional Comparative and Explanatory Study (ERCE 2019) also points in the same direction on the extent of the region's learning crisis for primary school students (\citealp{unicef21}).

This educational crisis has been aggravated post-pandemic (\citealp{azevedo21,betthauser23,bracco25,neidhofer21}) because the Latin America (henceforth: LAC) region experienced the longest period of school closures with poor connectivity (\citealp{gomez22,munguia23}) where, on average between 2020 and 2021, schools were closed for 269 days, which represents a disruption of 1.42 years (\citealp{bracco25}). Hence, the pandemic pushed a large cohort of students down in the learning distribution affecting the more disadvantaged, with estimates suggesting the LAC region lost around 0.9-1.1 years of schooling (\citealp{azevedo21}). On top of  already suffering from preexistent long term and deep educational inequalities (\citealp{brunori24,delprato15,kruger19,murillo23,neidhofer18}), the LAC region experienced an increase on the share of disadvantaged low performers (achieving below level 2) by 5\% in math, a slight decline for reading (see: \citealp{delprato25} using PISA 2022), and with this share being stable for science (\citealp{arias23}). Yet, good cognitive skills are fundamental for quality jobs and to enhance future labour prospects of secondary students which are vital for a steady growth of economies in the region (\citealp{angrist21,hanushek23}). It is within this post-pandemic scenario that the paper addresses the drivers of learning poverty in the LAC region.

In countries where education systems where resources-constrained the negative effects of the COVID-19 pandemic have been more profound because the wider barriers these countries had to tackle given by the triple burden of schooling deprivation, learning inequality and learning poverty (\citealp{asadullah23}). So, in the case of the LAC countries, the pandemic worsened educational inequalities. Prolonged school closures led to loss of learning, increased inequalities, also having negative impacts on psychosocial health and well-being of students and teachers, with younger children and students from more vulnerable groups experiencing the greatest losses (\citealp{asadullah23,unesco24}).

In turn, the COVID19 pandemic's driven households income shocks yielded increasing poverty (which in the region are estimated between 23\%-39\%; \citealp{bracco25}) and their consequences for educational outputs compounded established drivers of learning poverty in LAC region. For instance, \cite{bracco25}'s simulations show how pandemic educational loss in the LAC region is tightly related to households' socioeconomic levels. The authors estimate a mean loss years of education for a student in the bottom decile of the income distribution of 81\%, nearly four times as large than the one experienced by a student in the top decile (losing 22\%). In addition, established drivers of low performance in the region are repetition (\citealp{randall16,vinas21}), parental education and Indigeneity (\citealp{cortina17,delprato21,delprato19}), educational inputs (e.g., households' ICT and books stocks, family educational support: \citealp{lara19,torrecilla20}), as well as school characteristics such as school contextual poverty, school type and poor school climate (\citealp{berkowitz17,delprato17}), schools' ICT investments (number of personal computers$/$PC per student, rate of connectivity; \citealp{agasisti23,cepal20}), quality of teaching \textendash proxied by teachers uptake of professional development (PD) and qualifications (\citealp{mcginn23}).

These drivers, however, are not likely to be homogenous across sub-groups at the bottom of students' learning distributions. Considering the Programme for International Student Assessment (PISA) learning brackets (\citealp{oecd23a,oecd23b}), when targeting students' proximal determinants (i.e., elements that need to be in place in order for a child to have a successful learning experience in school; \citealp{pritchett25}), it is not the same if a student does not reach either level 1 (having insufficient foundational skills) or level 2 (reaching a baseline proficiency level), with the furthest disadvantaged students who perform below level 1 (at level 0) not acquiring foundational skills for most basic problems, struggling with fundamental tasks even following direct instructions. Research on learning determinants for the LAC region (e.g., \citealp{alves20,borchers25}) do not place an emphasis on dissecting these students groups between bottom and low performers, and by following an aggregated approach focusing on the drivers of performance for an average student, they are not able to differentiate and identify which modification of malleable students' factors may allow, for instance, to a student to move up from an extremely low performance (level 0)  to a bordering proficiency level (level 1).

As a result, there is a need to carefully distinguish among low performing groups for targeting, even more so considering they make up for the majority of students population (between 55\%-75\%; \citealp{arias23}). In this paper, we therefore attempt to fill this gap by investigating how these determinants may vary along low performing groups for the whole region and specific countries. Importantly, the paper's empirical contribution is carried out using the latest PISA available data (year 2022) for the LAC region after the COVID-19 pandemic, offering the latest results on students' low learning levels determinants.

Moreover, studies using interpretable machine learning (IML) methods based on learning surveys addressing determinants of students' achievement are rather scarce (e.g., \citealp{bu23,gomez24,huang24,liu22}), with their focus being on richer countries education systems whose realities are entirely contrasting to global south education systems. Methodologically, the current study contributes to this small body of literature, expanding the use of IML to assess what drives students from global south education systems (and at a regional scale) to be more prone to move down towards lower achievement brackets in the learning distribution. The study also contributes to the ML (machine learning) literature of the identification of students at risk of failure or dropout within online environments (see, for instance: \citealp{shi25,shi24}) from a regional LAC perspective.

In particular, relying on post-pandemic PISA data for 10 LAC countries, and using recent developments in the field of explainable machine learning: Shapley Additive Explanations (SHAP) analysis (\citealp{chen23,lundberg17}), this paper answers the following research questions for students at the bottom of the learning distribution (from level 0 to level 2 achievement groups):

\begin{enumerate}
  \item Globally, which are the principal determinants of low performance for the whole LAC region?
  \item What are the learning profiles (namely: leading determinants) of students with the highest and lowest chances of being poor achievers across the region? This question's approach has a local dimension as it deals with specific students (located at extremes).
  \item What are the main drivers at the country level, both globally and locally, of learning poverty?
\end{enumerate}

The rest of the paper is organised as follows. Section \ref{section2} contains the literature review and in Section \ref{section3} we describe the data used, the definition of the working samples for the two educational outcomes and leading covariates employed. Section \ref{section4} describes the methodological approach \textendash i.e., the array of learners (models) and Shap values methodology for interpretability of results and the analytical steps followed. Findings are presented in Section \ref{section5} and in Section \ref{section6} we discuss implications of results for future targeting of low performing students in the region. Section \ref{section7} contains concluding remarks.


\section{Background}
\label{section2}


\subsection{Machine learning in educational research}
\label{section21}

Underpinned by their complexity, ML models are now widely applied to solve supervised learning problems in many different fields and settings (\citealp{chen23,dhal22,olsen24,sarker21}) because supervised ML methods are capable of generating more precise predictions than standard statistical models (\citealp{chen22,lisboa23}). Yet, this may come at the expense of model interpretability (\citealp{rudin19}). Therefore, to account for this absence of interpretation (\citealp{olsen24}), the field of IML has become an area of recent substantial research (\citealp{covert21,linardatos20,holzinger20,molnar20}).

In tandem with this methodological progress, there has also been a recent growth in the application of ML, broadly in the social sciences (\citealp{brand23,grimmer21}) and within the educational field which, although is still limited, it has been continuously growing (e.g., \citealp{elbasi25,ersozlu24,hilbert21}). \cite{grimmer21} advocate for using an agnostic approach, using ML to make social science inferences where new research questions are generated and concepts and hypothesis can be tested with new data, rather than assuming a theory before investigating what is emerging from available data. \cite{brand23} describe how to use ML methods related to causality, mostly in relationship to heterogeneous treatment effects and how to minimize biases when estimating population-level quantities of interest.\footnote{For a ML review  for the economics and econometrics fields, see \cite{athey19}.}

\cite{hilbert21} show how the robustness of educational research can benefit from ML methods and the actual prediction of new data and validation of empirical models, specially when the size of data is large like in international learning surveys (e.g., TIMSS, PIRLS, PISA). So far in the educational research field, most of the focus has been on prediction, such as predicting academic performance in regression tasks (\citealp{hellas18,oz24}) and, to a lesser extent, for the identification of students at risk (e.g., \citealp{alam22,rodriguez23}). \cite{dol23}, in a review of nearly 150 articles over 2010-2020 period, find that classification techniques are mostly used for analysing students performance within the educational data mining (EDM) field. Similarly, \cite{ersozlu24}, in a systematic review of ML methods used for educational data, find that nearly 90\% of publications use ML approaches for predicting students’ performance and learning patterns. In spite of the prediction-centred approach of ML education studies, researchers have produced various approaches for a better understanding of models by quantifying the impact and interactions of features (\citealp{hilbert21}).

The SHAP method (\citealp{lundberg17}), which is employed in the current paper, is a leading example of this, falling into the group of IML techniques. Yet, the majority of educational empirical work using the SHAP method (so far, a small body of studies) are either for specific richer countries or regions, without a scaffolding inductive analytical approach following layers of education systems (\citealp{codiroli19}) and how educational inequalities are produced and entrenched when monitoring inequalities (\citealp{delprato24}).


\subsection{Review of machine learning studies\textendash learning surveys and students at risk}
\label{section22}

Studies using IML methods based on learning surveys addressing determinants of students' achievement are rather scanty and subject specific (\citealp{bu23,gomez24,huang24,liu22,zhu25}). These studies are focused in better-off and more resources-intensive education systems from developed countries (e.g., Spain: \citealp{gomez24}, or the US: \citealp{zhu25}) or groups of countries$/$regions (East Asia: \citealp{bu23,huang24}) whose realities are entirely contrasting to global south education systems. An exception is the study by \cite{ali25}, which addresses factors influencing academic performance among primary school students in Somaliland; though, this study, is a combination of ML and regression methods. The current study contributes to this emerging group of studies, and expanding the empirical evidence on drivers of learning poverty to global south education systems (i.e., the LAC region).

Our paper, through its assessment of determinants of students who are bottom and low performers with pandemic educational data, is related to ML studies dealing with the identification of at-risk students (i.e., failure, dropouts) and, specifically, considering that PISA 2022 data collection addressed issues related to the COVID-19 pandemic and remote learning, it is linked to the prediction of students at risks within online learning environments. A recent review on this topic is given by \cite{shi25}, which reviews 161 studies published in the 2014-2024 period using deep learning (DL), classical machine learning (cML) and cML ensemble approaches under alternative definitions of students at risk. In this review, the authors highlight: (i) the growth on the use of ensemble of cML models over the last 5 years, having superior performance over DL when following an ensemble approach (with Gradient Boosting Tree and Random Forest emerging as leading cases) and using smaller data sizes; and (ii) the majority of studies tend to use as drivers for students at risk engagement variables and achievement covariates (more widely employed than demographic factors), with demographics factors such as age, ethnicity, disability, socio-economic status and parental education being present in different studies. Moreover, in a meta-regression study looking into the success of ML techniques in identifying online at-risk students by data type and DL and cML approaches, \cite{shi24} also find that cML ensemble models exhibit consistently higher prediction accuracy across all evaluated prediction stages.

Specifically, there are various studies about the identification of students at risk (particularity at Higher Education$/$HE) during the pandemic, with \cite{shi25} indicating that around 32\% ML studies are based on online HE courses. For example, \cite{khan24}, is a close study to ours given its focus on the impact of the COVID-19 pandemic on students achievement (using a small dataset from students enrolled during 2020-21 at a Pakistani university), which proposes a regularised ensemble model made up of random forest models to pin down key factors behind their achievement. The authors find that students’ attendance and interaction, internet connectivity, and studying a mandatory entry course during the restrictions are all leading drivers of students’ university achievement. Likewise, \cite{marin25}, employ an array of nine ML models (with XGBoost showing a high degree of accuracy) to predict grade performance of students in a Peruvian university (for the first semester in 2025), finding that prior academic variables stands out as key predictors, though also non-academic factors such as family support, stress, and time management having a moderate impact on performance. The study emphasises the need of ML for educational research (to account and address high-dimensionality, complex and non-linear associations within educational processes coupled with importance of psychosocial and well-being factors) and the relevance of SHAP values, thereby arguing that they provide a deeper understanding of the individual impact of each variable on the prediction and actionable insights leading to specific observations.

Other similar papers dealing with online learning through ML models at HE are, for instance, \cite{chytas23} (for students attending a Greek university) and \cite{asad23} (using data from $\approx$ 12,000 Pakistani HE students), with the latter study relying\textendash as the current paper\textendash on a ensemble model constructed from five ML models. Equally, \cite{Karalar21}'s study relies on an ensemble model (including quadratic discriminant analysis, decision trees, random forest, extra trees, logistic regression, and artificial neural network) to predict students at risk of academic failure in a Turkish public university (n = 2000) during the COVID-19 pandemic, with the model yielding a prediction accuracy of around 90\% and a higher relevance of asynchronous learning activities. In another HE context (i.e., Saudi university), \cite{albreiki22} propose an ensemble modelling approach with explainable ML and rule-based model risk flags to identify students who are at risk of low performance and their factors. All in all, these empirical studies show the capability of ML model to capture non-linear associations behind the chances of students falling at risk of failure, with a performing superiority of ensemble models.

Our paper innovates the current state-of-the-art of the existing literature in three main ways. Firstly, it is the first empirical investigation which assesses the determinants of subject-combined proficiency by disaggregating weak students’ performance (or learning poverty) into bottom (below level 1) and low (below level 2) students groups for the LAC region, with previous studies limited to richer education systems (in either Europe or Asia) and for higher achievement brackets at given subjects, mostly looking into determinants of proficiency (above level 2). We achieve this by looking into these determinants from a global and local perspective. Secondly, the empirical evidence refers to the most recent wave of internationally comparative data on students’ achievement for the region (i.e., PISA 2022) intertwined by the global COVID-19 pandemic. Thirdly, the empirical analysis includes students from 10 LAC countries, whereas earlier literature concentrates on single countries. Thus, the motivation of this study is to provide a complete overview of determinants of learning poverty in the region as well as on dissimilar patterns driven by country heterogeneity. Finally, the analysis relies on recent progress on IML offering a granular snapshot of leading determinants through the SHAP method, using an ensemble model for better accuracy.


\section{Data}
\label{section3}

The analysis is based on the latest wave of PISA for the year 2022 which measures secondary students' learning achievement (at 15 years-old), i.e., their ability to use their reading, mathematics and science knowledge and skills to meet real-life challenges.\footnote{For details on OECD PISA, see \href{https://www.oecd.org/pisa/}{PISA-link}.} The paper is focused on the LAC region, which includes a sample of 10 countries, namely: Argentina, Brazil, Chile, Colombia, Dominican Rep., Mexico, Panama, Peru, Paraguay and Uruguay.\footnote{We exclude an additional country within this sample, i.e., Costa Rica, because its dataset does not contain the key covariate household SES.} Since the paper's aim is to identify the determinants of learning poverty (i.e., those students located at the bottom of the learning distribution in three subjects: math, reading and science) in the region, we employ two comparisons: no competency achieved (level 0 which includes level1b and level1c) versus level 1 (level1a), and level 1 versus level 2 (where those below level 2 are referred to as `low performers'; \citealp[p.~89]{oecd23b}). These two working samples contain, respectively, between 16,236 and 9,484 observations.

It should be mentioned that PISA 2022 data gathering was mediated by the pandemic. Originally, it was planned to take place in 2021 but the disruption caused by COVID-19 global pandemic forced the assessment to be postponed by a year (to 2022) and the content was updated with further questions linking the ``resilience of education systems'' from the point of view of fairness and inclusion within the remote learning shift brought about by the pandemic (\citealp{oecd23a}). The resilience of education systems in PISA 2022 was evaluated against a new framework focusing on students' capacity for continuous learning and adaptability. In addition to the test measuring students' academic performance, there is a comprehensive questionnaire that collects relevant information about the students' family and socioeconomic backgrounds, their attitudes toward learning and study habits and schools' background features. Data submission in the LAC region happened in December 2023.


\subsection{Data coding\textendash rationale}
\label{section31}

The paper's focus on the three bottom levels of achievement is aligned to the adverse long-term learning trajectory of the region where, for example, the rate of low-performers in math among the poorest students either increased or remained unchanged across all countries since 2012 (\citealp{arias24a,arias24b}). As earlier mentioned, we use two binary educational outcomes (or dependent variables) among the low performance groups: level 0 (=1) versus level 1 (=0), and level 1 (=1) versus level 2 (=0). As shown in Table \ref{table1}, these two samples capture most of the LAC students. For instance, levels 0 to 2 (first three rows of Table \ref{table1}) account for 89.5\% (math), 77.5\% (reading) and 80.9\% (science) of total students.

\medskip
\begin{center}
  [Table \ref{table1} here]
\end{center}
\medskip

Regarding data processing, rather than analysing low achievement drivers for specific subjects as in other PISA empirical ML studies (see: \citealp{gomez24,huang24}, and the systematic review: \citealp{wang23}), the paper follows a subject-aggregated definition of dependent variables\textendash for instance, $y_{\text{level 0,level1}}$ = 1 if a student falls into level 0 for math, reading and science, and equals to 0 if a student falls into level 1 for the three subjects. This leads to rates for the first educational outcome (level 0 versus level 1) of 65.3\% and for the second educational outcome (level 1 versus level 2) of 59.3\% (Panel B of Table \ref{table1}). We construct two datasets for each educational outcome to be analysed separately: a dataset with student$/$family variables and a dataset with school variables. Besides, after selecting leading covariates with lower missing rates, we eliminate records with missing values (e.g., $\approx$ 3\% in the student$/$family and school datasets).\footnote{As it can be seen in the Appendix B (Tables \ref{tableB1} and \ref{tableB2}) which contains a detailed account of source variables, most of them are either of a binary or discrete types, with continuous variables constructed as indices grouping similar questions (e.g., family support for education, teacher support on wellbeing, school autonomy, etc.).} Full details on covariates' construction and definitions are shown in the Appendix (Tables \ref{tableC1} and \ref{tableC2}).


\subsection{Summary statistics}
\label{section32}

\subsubsection{Students and family covariates}
\label{section321}

The empirical relevance of students and family covariates as determinants of probabilities of students to fall into the sub-groups of bottom performers (level 0 versus level 1) and low performers (level 1 versus level 2) can be seen by the summary statistics of Table \ref{table2}. On the one hand, in comparison to students whose achievement fall into level 1, level 0 students (Table \ref{table2}, columns 1 to 3), are likely to do less homework, have a lower number of digital devices (= -1.72) and fewer books (= -0.35), being twice as likely to be from an immigrant family and three time as likely to speak another language at home (Panel A, columns 1 to 3). Repetition for the level 0 group of students \textendash either at primary or lower secondary\textendash is much larger (33\% against 9\% and 22\% against 9\%, respectively) and likewise the chances of missing or skipping school, and their intensity to be engaged in paid work (1.66 versus 1.45). Further, students from the level 0 group compared to students from level 1 group are poorer (69\% comes from families in the bottom half of the family SES index distribution in comparison to 59\% in the case of level 1 students) and come from larger families (Panel B, column 3, Table \ref{table2}). On the other hand, these gaps on covariates are similar and hold when comparing level 1 versus level 2 students (Table \ref{table2}, columns 4 to 6), although with some differences in magnitude with the level 0-level 1 comparison. For instance, the gap on possession of books at home and chances of paid work are larger, but repetition gaps are smaller. Notably, gaps on parental education become relevant (and statistically significant) along with the importance of family educational support.

\medskip
\begin{center}
  [Table \ref{table2} here]
\end{center}
\medskip

\subsubsection{School covariates}
\label{section322}

As shown by Table \ref{table3}, school's context\textendash be its composition, inputs and infrastructure, teachers profile or policies implemented\textendash matters when it comes down to a student's chance of falling into either the lower achieving group level 0 (among bottom performers, columns 1 to 3) or level 1 (among low performers, columns 4 to 6). The influence of school's context is slightly more prominent when comparing the bottom two levels of performance. Here, we mention some key differences. Compared to students whose achievement falls into the level 1 performance bracket, level 0's students are more likely to come from rural schools (by 13\%) and public schools (by 12\%), and also from schools where there is less competition with other schools, and from schools which are located in smaller towns and receive less funding from fees, and schools which are smaller in terms of enrolment (Table \ref{table3}, column 3). What's more, students at the bottom of the learning distribution attend schools where there are less PC per teacher connected to the web, and with less involvement of parents and with fewer methods for students' testing.

\medskip
\begin{center}
  [Table \ref{table3} here]
\end{center}
\medskip

Besides, the school rate of disadvantage students now becomes an important factor between level 1 and level 2 groups, being 9\% higher for the latter group (column 6, Table \ref{table3}). The remaining covariates raw differences are, to a large extent, similar to the previous students' comparison. There are some distinct pattern in covariates differences, however. For example, no statistical differences are found for the student-teacher ratio (STR), the proportion of certified teachers, infrastructure, and for some school policies (e.g., transfer, improvement policies, and ability grouping). On the whole, summary statistics from Tables \ref{table2} and \ref{table3} underline to what degree contextual disadvantages stemming from student, family and school levels are critical for the likelihood of students falling into the lowest performing groups.


\section{Methods}
\label{section4}


\subsection{Binary classification models}
\label{section41}

Given the dichotomous nature of dependent variables (level 0-level 1 and level 1-level 2), we employ five binary classification ML models (for details see: \citealp{hastie09,hastie15} and for their implementation: \citealp{pedregosa11,raschka19}). Each model was trained using the stratified $K$-fold (5 folds) method and  train-test split of 80\%-20\% of the data so that to ensure the data is balanced within each fold. Each model was optimised using different hyperparameters setup and the application of grid search to find the top performing model version among each type of model. Below, we briefly describe each ML model employed.

\subsubsection{Logistic regression}
\label{section411}

Logistic regression (or just logit) is one the principal models for binary outcomes. It predicts the probability that the outcome equals to one conditional on a vector of covariates:

\begin{equation}\label{logit}
  \text{Pr}\big(y_{i} = 1 | \mathbf{x}\big)  = \Lambda(\mathbf{x}'\boldsymbol{\beta}) \equiv \text{exp}\big({\beta_{0} + \mathbf{x}'_{i}\boldsymbol{\beta}}\big) / \big(1 + \text{exp}\big({\beta_{0} + \mathbf{x}'_{i}\boldsymbol{\beta}}\big)
\end{equation}

where $\Lambda(.)$ is c.d.f. of the logistic distribution. Logit is simpler and easier to interpret. Here, the hyperparameters used for the two versions are lack of regularisation (no penalty), and with penalty L2 (Ridge) option. The likelihood function for logit is:

\begin{equation}\label{logitLL}
  \mathcal{L} = \sum_{i=1}^{N} \big[ y_{i} (\beta_{0} + \mathbf{x}'_{i}\boldsymbol{\beta}) -  \text{log} \big(1 + e^{(\beta_{0} + \mathbf{x}'_{i} \boldsymbol{\beta})} \big) \big]
\end{equation}

\subsubsection{Logistic Lasso with CV penalty}
\label{section412}

We also rely on a different version of logistic regression; namely, a Logistic Lasso with cross-validation (CV) (or logitlasso), which imposes a penalty (L1) on the size of coefficients to control overfitting and the size of it is chosen by CV. The lasso penalises the absolute size of coefficients via the penalty parameter $\lambda$, yielding sparse solutions where some coefficient estimates are set to exactly zero. We use four values of $C$ (the inverse of regularization strength) for CV ($C$ = 10, 50, 100 and 200). The penalised objective function for logitlasso combines the standard logistic regression loss function (negative log-likelihood) with a L1 penalty term for the coefficients (the second term):

\begin{equation}\label{logitlasso}
Q_{\mathcal{L}} = - \frac{1}{N} \sum_{i=1}^{N} \mathcal{L} + \lambda \sum_{j=1}^{p} |\beta_{j}|
\end{equation}

where $\mathcal{L}$ is the logit likelihood function, $\lambda$ is a non-negative tuning parameter that controls the strength of the regularisation, and $p$ is the number of predictor variables (excluding the intercept).

\subsubsection{Random forest}
\label{section413}

Random Forest (RF) improves modelling by ensembling a combination of multiple decision trees and then aggregating their predictions. RF trains several decision tree models (learners) separately and then let all the learners decide the output of the test samples (i.e., the final classification) by a voting mechanism. The final binary classification prediction is a majority vote across all the individual decision trees. Formally:

\begin{equation}\label{RF}
\widehat{\mathbf{y}} = \text{mode}\big[f_{b}(\mathbf{x}')\big]_{b=1}^{B}
\end{equation}

where $\widehat{\mathbf{y}}$ is the final predicted binary class (bottom or low performer = 1, and 0 otherwise), $B$ denotes the total number of trees, $f_{b}(\mathbf{x}')$ is the prediction for the input $\mathbf{x}'$ for the $b$-th tree. Moreover, the training samples of each learner are randomly chosen and when building the decision tree a portion of the features (variables) is randomly selected from the features to build the decision tree. The hyperparameters configuration we assess are the number of estimators (or trees): 100, 500 and 1,000 trees, and each tree's depth: two maximum depths  given by 5 and 10 leaves.

\subsubsection{Gradient boosting}
\label{section414}

Another ML classifier method we use is Gradient Boosting (GB) which builds decision trees iteratively to minimize prediction errors. GB constructs weak learners (trees) sequentially, and each tree predicts the residual of earlier predictions. That is, GB is an ensemble of predicted decision trees models run in a sequential fashion correcting prediction of previous models, and the update rule is:

\begin{equation}\label{GB}
f_{m} = f_{m-1}(\mathbf{x}) + \sum_{j=1}^{J_{m}} \gamma_{jm} \mathbf{1}(\mathbf{x} \in R_{jm})
\end{equation}

where $f_{m}$ is the model $m$ interaction and $J$ are terminal nodes, $\gamma_{jm}$ the fitted values and $R_{jm}$ the residuals. We assess the GB under six formats (or combination of hyperparameters): the number of trees ranging from 100, 500 and 1,000 and two learning rates (1\% and 10\%).

\subsubsection{Neural networks}
\label{section415}

The final ML model we consider is Neural nets (NN) or Feed-forward neural networks. It consists of hidden layers that link the predictors (or input layers) to the outcome. Each hidden layer is composed of multiple units (nodes) that pass signals to the next layer using an activation function. Central tuning parameters are the choice of the activation function and the number and size of hidden layers. We consider six specifications: one NN with one hidden layer of 200 nodes each, and another neural net with three hidden layers with 200, 100 and 50 nodes, and both using three activation rules (i.e., logistic, relu, tanh).


\subsection{Assessment of binary classification models}
\label{section42}

We assess the performance of each ML model version by relying on an array of evaluation metrics (\citealp{naidu23}). Specifically, six metrics are employed when comparing models: Accuracy (ACC), Recall (RC), Precision (PR), F1 Score (F1), Specificity (SP) and the Area Under the Curve (AUC). These metrics provide overall measures of a model's performance. Noteworthy is the AUC indicator because, by calculating the area of under the Receiver Operating Characteristic (ROC) curve, it can efficiently balance the true positive rate and the false positive rate. We include a description of the six metrics in Appendix \ref{appendixA}.


\subsection{Stacking approach for classification}
\label{section43}

When faced with a classification task as in the current paper\textendash and after choosing the best model configuration for each of the five ML models based on evaluation metrics\textendash the next decision is to assess which model is best suited. But this decision is not obvious and, combining multiple learners into a final ensemble model, is quite likely to lead to a superior performance compared with each individual learner. Ensemble models represent a paradigm in which multiple algorithms are strategically combined to improve prediction accuracy and robustness beyond what can be achieved by any single model (\citealp{qiao20}). Our choice of the five types of ML classification models to build up the ensemble model is due to their diversity, so that errors or better fittings across models are complemented. For instance, logit and logitlasso are linear models whereas RF and GB are tree-based ensemble methods, and NN can account for model highly complex, non-linear relationships. This choice is somehow validated by systematic reviews on the topic of ML for the identification of students at risk where ensemble models combining RF, GB and SVM are more commonly employed (\citealp{shi25}).

Stacked generalisation (or stacking), a concept introduced by \cite{wolpert92}, allows a form of model averaging using a technique that utilises the predictions of various models to train a meta-model. Theoretically, stacking performs as well as individual learners (\citealp{vanderlaan07}). Here, we follow the stacking approach among the best five binary classification ML models following the work of \cite{ahrens23}.\footnote{For stacked generalisation we use the Stata command: \texttt{pystacked}.} The stacking process for binary classification exercises consists of two steps.\footnote{It should be noted that there are another recent progress on stacking models for binary classification tasks. For instance, \cite{ali24} introduce a new ensemble classifier using several base $k$ nearest neighbour ($k$NN) relying on an extended neighbourhood rule and randomly projected bootstrap sample which outperforms other ensemble classifiers because it includes a bootstrap sampling of all features. See also, \cite{ali23}, for further details.}  First, cross-validated predicted probabilities for each of the five base learners and each observation $\big(\widehat{y}^{-k(i)}_{(j),i}\big)$ are obtained for all folds ($K$ data partitions of similar size), expect from the fold where the current observation is located (i.e., $-k$($i$)). Second, the final learner (stacked model) is fit by using the observed $y_{i}$ outcome and cross-validated predicted probabilities as predictors $\big(\widehat{y}^{-k(i)}_{(1),i}, \widehat{y}^{-k(i)}_{(2),i},...,\widehat{y}^{-k(i)}_{(J),i}\big)$ of the full sample. The final learner is obtained by minimising the least squares with the additional constraint that coefficients sum to one:

\begin{equation}\label{stacking}
\begin{split}
\underset{w_{1},...,w_{J}}{\text{min}} & = \sum_{i=1}^{N} \Big[(y_{i} - \underbrace { \sum_{j=1}^{J} w_{j} \widehat{y}^{-k(i)}_{(j),i} }_{\widehat{y}_{i}^{*}} \Big]^{2} \\
  \text{s.t.   } & w_{j} \ge 0  \text{     and      } \sum_{j=1}^{J} w_{j} = 1
  \end{split}
\end{equation}

where here $J$ = 5 (the number of models$/$learners), stacking predicted probabilities are defined as $\widehat{y}_{i}^{*} = \sum_{j} \widehat{w}_{j} \widehat{y}_{(j),i}$ and  $\widehat{w}_{j}$ is the estimated stacking weight for learner $j$, and $\widehat{y}_{(j),i}$ are the predicted probabilities when refitting learner $j$ on the full sample.


\subsection{SHAP values\textendash interpretable ML estimates}
\label{section44}

In the body of educational research using ML (see: \citealp{ersozlu24,hilbert21}), there has been a latest growth in the application of Shapley values to identify main features behind educational performance in learning surveys (e.g., \citealp{gomez24,huang24}). Importantly, we rely on Shapley values to present leading estimates in Section \ref{section5}. Shapley values have their origin in game theory (\citealp{shapley53}), measuring the average marginal contribution of a player in a cooperative game. In the context of ML, the player is interpreted as a characteristic (or attribute), and the cooperative game becomes the prediction task performed by the model. Through recent developments (\citealp{vstrumbelj14,lundberg17}), Shapley values have been adapted as an explicative framework for ML models. For the estimation of the main determinants of learning poverty in the LAC region, in particular, we follow the SHAP method (\underline{SH}apley \underline{A}dditive ex\underline{P}lanation) (\citealp{lundberg17}). SHAP values offer a local explanation of a model output by calculating the degree to which each feature contributes to a given prediction value when conditioning on that feature.

Formally, the change on prediction or SHAP value $\phi_{i}$(.) for feature (or covariate) $i$ in model $f$ for data point $x$ is given by:

\begin{equation}\label{shap1}
  \phi_{i}(f, x) = \sum_{R \in \mathcal{R}}\frac{1}{M!}\big(f_{x}(P^{R}_{i} \cup i) - f_{x}(P^{R}_{i})\big)
\end{equation}

where $\mathcal{R}$ is the set of all feature permutations, $P^{R}_{i}$ is the set of all features before $i$ in the ordering $R$, $M$ is the number of input
features, and $f_{x}$ is an estimate of the conditional expectation of the model’s prediction. The additive property of SHAP values means that SHAP values add to the output of the model:

\begin{equation}\label{shap2}
  f(x) = \phi_{0}(f) + \sum_{i=1}^{M}\phi_{i}(f, x)
\end{equation}

Attributions based on the Shapley interaction index result in a matrix of feature attributions. SHAP values main effects (on the matrix's diagonal) are obtained as the differences of SHAP values and the off-diagonal elements (i.e., SHAP interactions)  for a given feature. That is:

\begin{equation}\label{shap3}
  \Phi_{i,i} = \phi_{i}(f,x)-\underbrace{\sum_{j \neq i}^{}\Phi_{i,j}(f,x) }_{\text{SHAP interactions}}
\end{equation}

There are various algorithms to define the value function for a set of attributes (e.g., DeepSHAP, LinearSHAP, KernelSHAP, TreeSHAP). In the paper, and due to computational issues and the fact that we use a stacked ML model (which is a combination of classification models), we rely on the LinearSHAP (\citealp{chen20}), the default \texttt{Explainer} of the SHAP Python package.


\subsection{Analytical steps}
\label{section45}

\begin{figure}
\centering
\includegraphics[width=\textwidth]{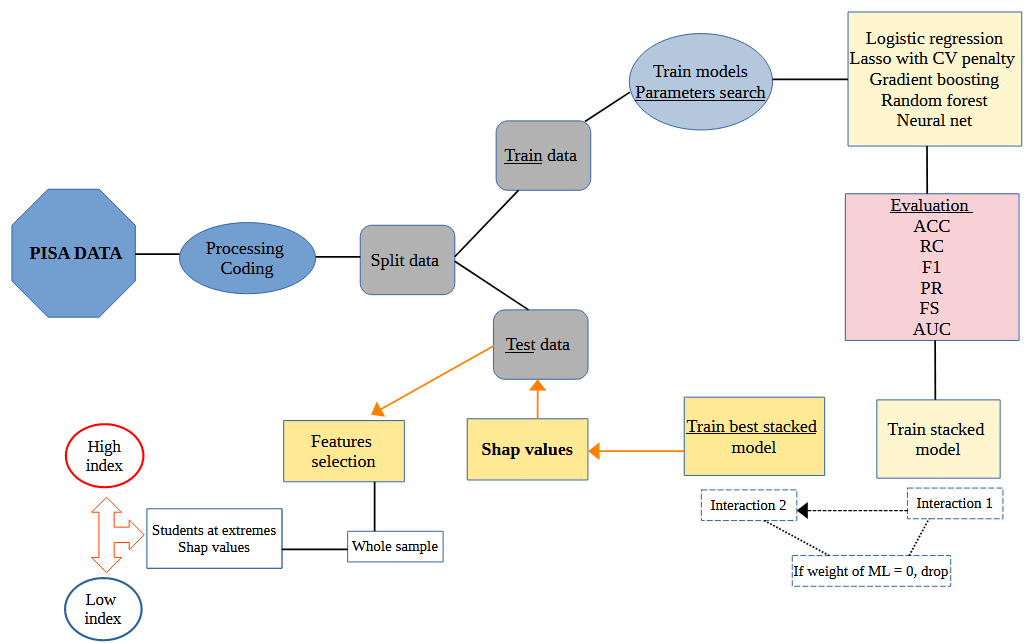}
\caption{Analytical steps}
\label{figure1}
\end{figure}

The steps for the analysis for the PISA 2022 data are displayed in Figure \ref{figure1}. First\textendash and after carefully processing the PISA data and generating a dataset for each dependent variable\textendash the dataset is divided into a trained dataset (80\%) and a test dataset (20\%). This split, coupled with the stratified $K$-fold (number of folds = 5) cross-validation, ensures that each fold retains proportionality values for $y$ educational outcomes and covariates avoiding skewed results. Second, using the train dataset, we run the hyperparameters search fitting different configurations for the five ML classification models, and we obtain their best model specifications based on evaluation metrics. This yields the stacking model for the train sample. Third, we re-estimate the chosen models excluding a model if its weight during stacking is equal to zero (interaction 2 in Figure \ref{figure1}) obtaining SHAP values. Fourth, the final stacked train data model SHAP estimation is then applied to the test data leading to main features covariates selection at a global level. In the final step, we produce a local version of SHAP values by examining the profile of leading explanatory variables (covariates$/$features) for students at the extremes of the distribution of the aggregated SHAP values (i.e., $\sum_{j=1}^{J}\phi^{(i)}_{j}$ equal to the minimum and maximum across all $i$ samples or observations).

\medskip
\begin{center}
  [Figure \ref{figure1} here]
\end{center}
\medskip


\section{Results}
\label{section5}

This section presents the findings via the SHAP analysis for the two samples at the bottom of the learning distribution: level 0 (no achievement) versus level 1, and level 1 versus level 2, both globally (across all samples) and locally (bottom and high performers students within each working sample), dividing the explanatory variables into two sets: (i) students and family covariates, and (ii) school covariates. Before, we evaluate the five binary classification models through evaluation metrics. Countries estimates are also shown as local explanations for specific students based on the SHAP contributions index.


\subsection{Models' comparison}
\label{section51}

We present a comprehensive comparison of the efficacy of the five models using a range of evaluation metrics. Metrics for all models' configurations, the grounds for selection of the best performing version of each of the five ML models, are shown in Appendix C. For the level0-level1 dataset (the bottom two performing groups), less parsimonious models outperform other simpler models in terms of the degree of accuracy of the classification of students into achieving levels. Notably, GB and RF models consistently show being more effective in terms of the SP and AUC (columns 5 and 6, Table \ref{tableC1}), by using more trees, depth and higher learning rates. Results for the level1-level2 comparison dataset (low performing groups comparison; Table \ref{tableC2}), follow a similar pattern, with the GB model in particular showing metrics as a leading model.

\begin{figure}
\centering
\includegraphics[width=0.90\textwidth]{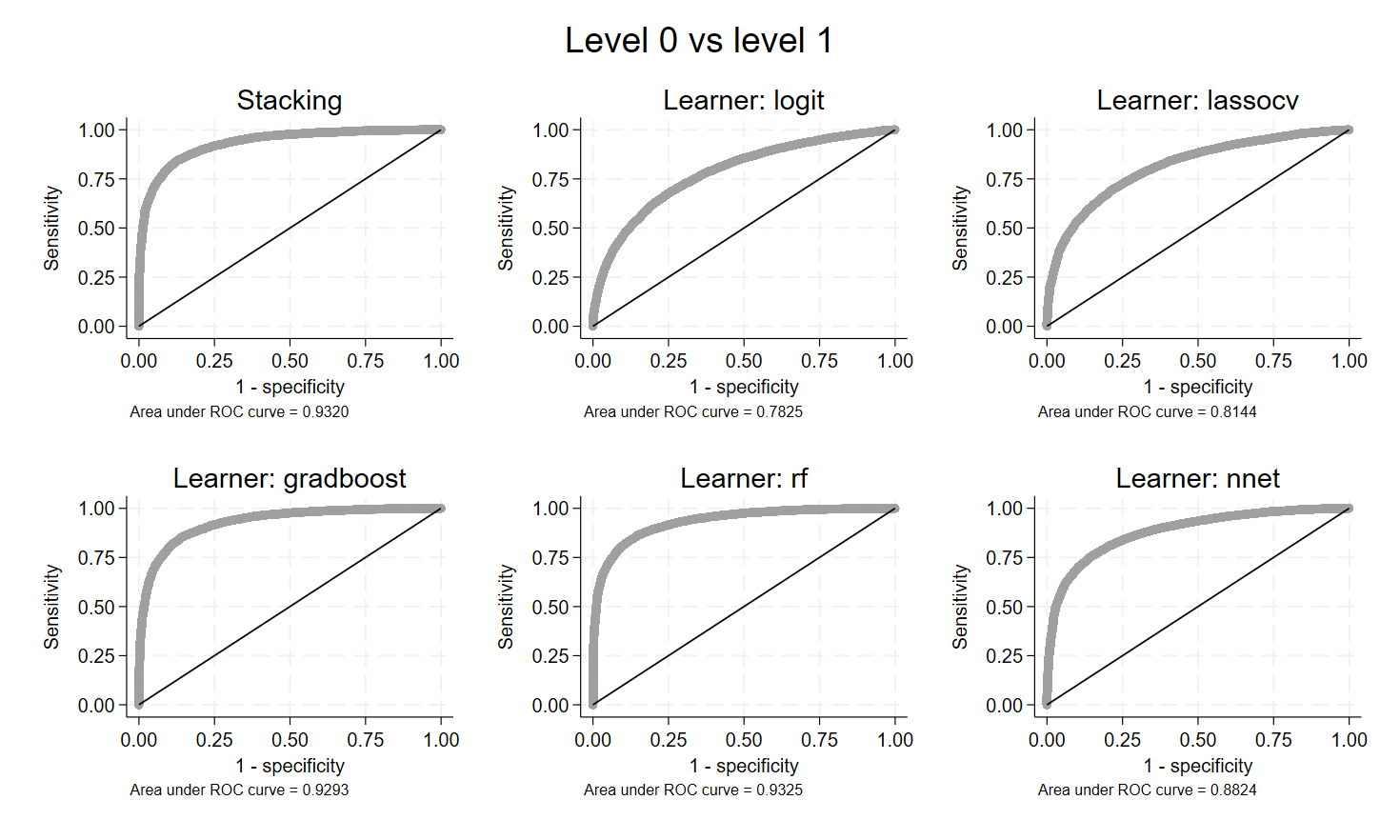}
\includegraphics[width=0.90\textwidth]{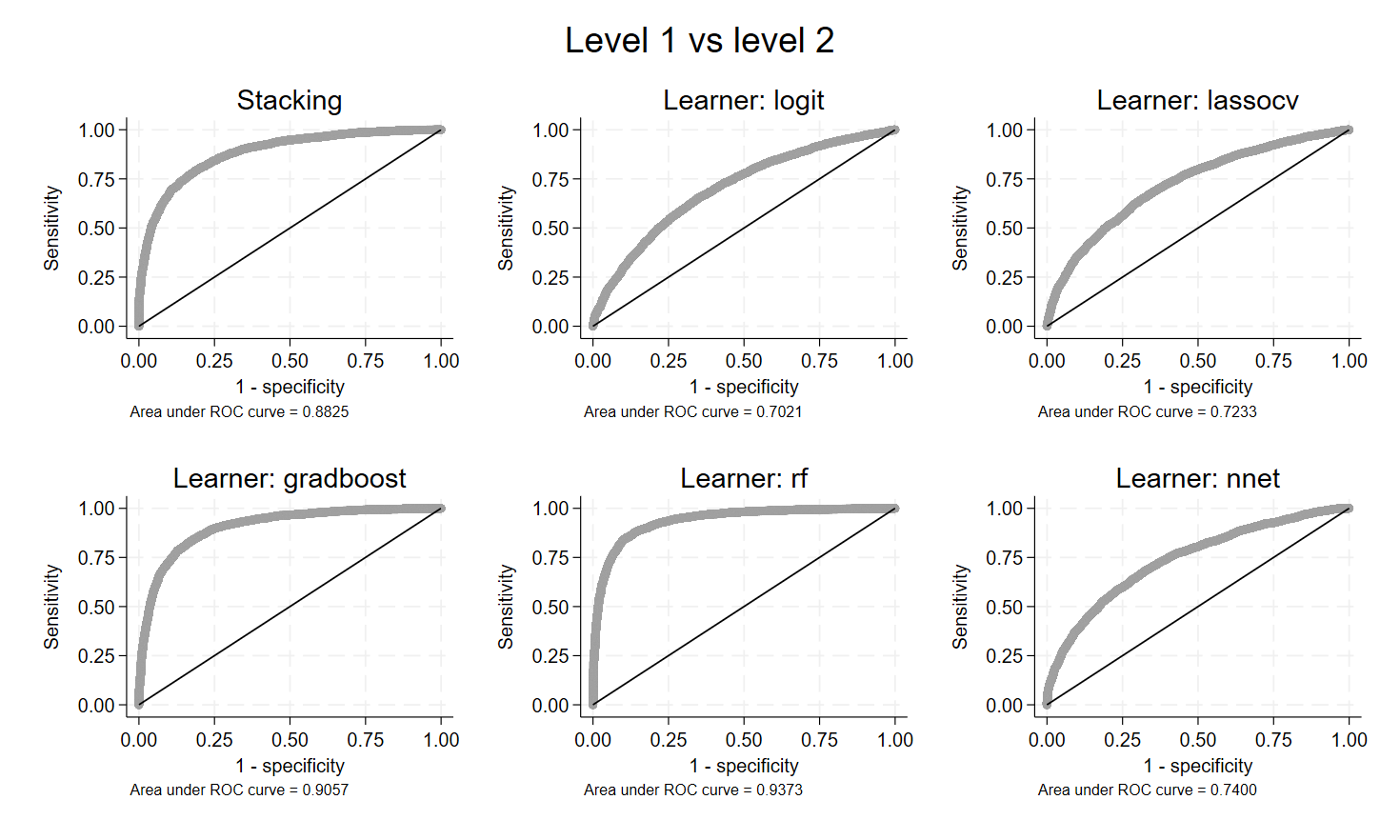}
\caption{Area under the ROC curve, different ML models}
\label{figure2}
      \medskip
\begin{minipage}{0.99\textwidth}
\footnotesize{
Notes: (1) The $y$ axis in a ROC plot corresponds to sensitivity (true-positive rate), and the $x$ axis corresponds to 1 - specificity (false positive rate). (2) Displayed ROC plots are those models with the better evaluation metrics across each model configuration (which are highlighted in bold in Tables \ref{tableC1} and \ref{tableC2}).
}
\end{minipage}
\end{figure}

Table \ref{table4} displays evaluation metrics estimates for the selected best performing models (columns 1 to 6) as well as each model's weight (column 7) for the stacked (ensemble) modelling approach where, in the first interaction, it includes all models. Vitally, the stacking model has the benefit of combining the strengths of the individual models it comprises. Figure \ref{figure2} displays the AUC for all models, including the stacking model where one can see how the classification and identification of the binary categories of educational outcomes improve by the amalgamation of models. In the first interaction, Table \ref{table4} shows that models (or learners) with zero weight\textendash i.e., not contributing to the overall prediction of the stacking model\textendash are logit (for the level0-level1 comparison, Panel A, column 7), and logit and NN (for the level1-level2 comparison; Panel B, column 7). These models are therefore excluded in the second interaction. This leads to the following two stacking models used for predictions and derivation of Shapley values: lassologit, GB, RF and NN (level0-level1 data), and lassologit, GB and RF (level1-level2 data). Also, note that the higher estimated weights within the final ensemble models for GB and lassologit models suggest their superior relevance in overall predictions.\footnote{Alternative stacked models and their performance are shown in Table \ref{tableC3}. We re-run the two stacked models of Table \ref{table4} (from interaction 2) using the range of hyperparameters of the GB (for the bottom performers sample) and lassologit (for the low performers sample) models obtaining evaluation metrics for each specification. This exercise validates our initial choice of the two stacked models. Specifically, the initial choice of the GB's specification (trees = 1000, learning rate = 0.10) has the largest AUC (as, too, for the PR and SP metrics) for the level-level1 comparison, whereas for the level1-level2 comparison lassologit has similar AUC with the selected version ($C$ = 10) having as well the largest PR and SP metrics.}

\medskip
\begin{center}
  [Table \ref{table4} and Figure \ref{figure2} here]
\end{center}
\medskip


\subsection{Whole sample\textendash global explanations. Student, family and school determinants}
\label{section52}

Global explanations on leading determinants of LAC students' learning poverty for the two samples are shown in Table \ref{table5}. The ranking is based on absolute values of average Shapley values across all students; i.e., for covariate (feature) $j$:  $I_{j} = \big( (1/n) \times \sum_{i=1}^{n}|\phi^{(i)}_{j}| \big)$. This provides, regardless of the direction of covariates' associations with model outputs and the specific value of a given covariate, an overall assessment of which are the more powerful determinants for the probability of students' falling into performance levels 0 (against level 1) and level 1 (against level 2). We divide the covariates by student$/$family and school types when running the stacking model.

On the one hand, focusing on the top ten most impactful student and family explanatory variables with the highest SHAP average values (columns 1 and 3 of Table \ref{table5}), there is a similarity on determinants for bottom and low performers (with 7 out of 10 covariates being in the top ten for these two groups). Primary repetition, sense of belonging to school and wellbeing support from teachers are key drivers of a student chances to fall into the lowest achievement group among bottom performers (level 0, no competency, column 1), but not in low performing group comparison (column 3). In the reverse direction, engagement in paid work, the intensity of homework and mother's educational level are among the top ten SHAP values for the level 1 and level 2 comparison, although not for the bottom two levels of achievement comparison. Furthermore, there are some discrepancies on SHAP values' ranking among the two samples. For instance, family SES and family educational support are the top two determinants for low performers; yet, in the case of bottom performers sample, they are ranked fifth and sixth in order of importance. The number of digital devices and school climate (ranked second and third) are also more relevant for bottom performers, and gender has a larger impact when comparing level 1 versus level 2 achievement groups. This suggests distinct factors for those students at risk of falling into either level 0 or level 1 of each sample, and therefore the necessity of differential strategies when identifying and targeting those students.

\medskip
\begin{center}
  [Table \ref{table5} here]
\end{center}
\medskip

On the other hand, as regards to school characteristics SHAP values (Table \ref{table5}, columns 2 and 4), school size is equally important for both samples, although the rate of school disadvantage and the STR for pedagogic support are higher in the ranking for low performers compared to their ranking among bottom performers. Distinctly, school type and government funding are more powerful school features for the low performing sample (column 4) than for the bottom performing sample (column 2), whereas the rate of students with a second language and wider support for inclusivity are more relevant for the latter comparison (level 0 versus level 1). In both groups of samples, policies (ability grouping, transfer, admission) are located at the lower end within the SHAP ranked values.\footnote{In Figure \ref{figureD1} (Appendix D), we include the SHAP values interactions for student$/$family and school variables. For level 0 versus level 1, highest associations between these two group of covariates are: school autonomy with repetition and school climate, PC connected to internet for teaching and students' number of books and STR of pedagogic staff with teacher support for students' wellbeing. For the sample level 1-level 2, leading cross-level intersectionalities of SHAP values are: school prevalence of disadvantages students and ICT infrastructure with family educational support and family SES and school autonomy.}

\begin{figure}
\begin{multicols}{2}
\centering
\includegraphics[width=0.50\textwidth]{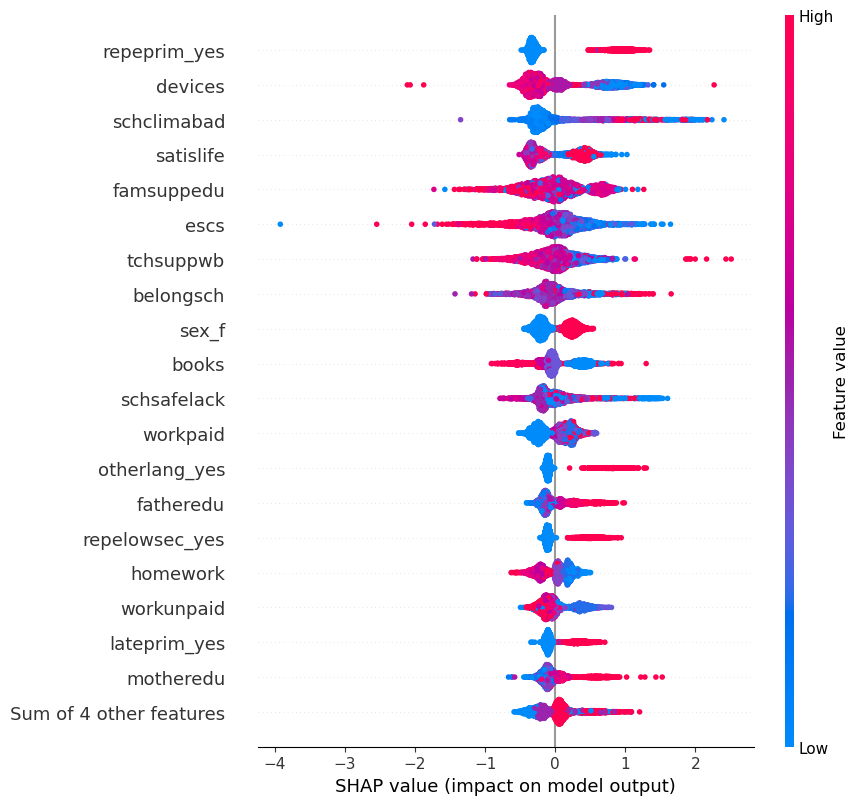}
\includegraphics[width=0.50\textwidth]{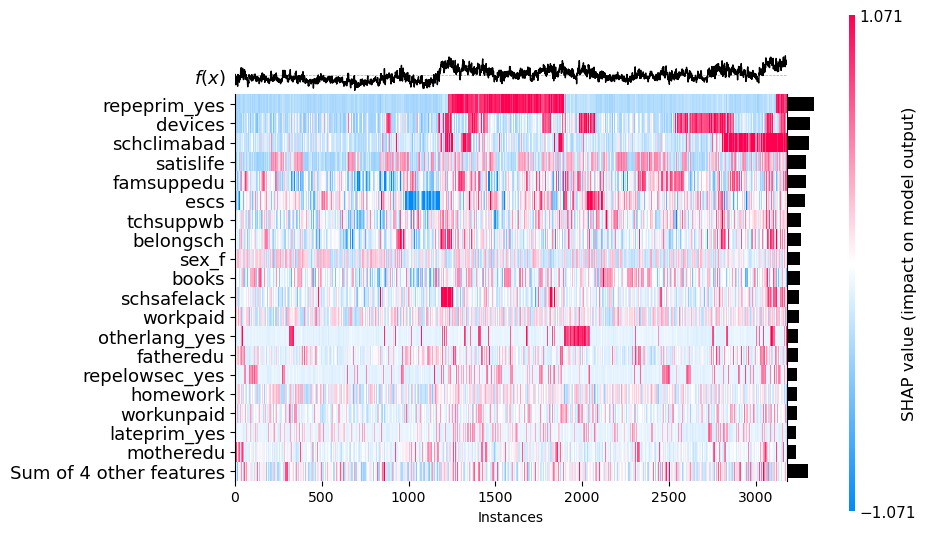}
\includegraphics[width=0.50\textwidth]{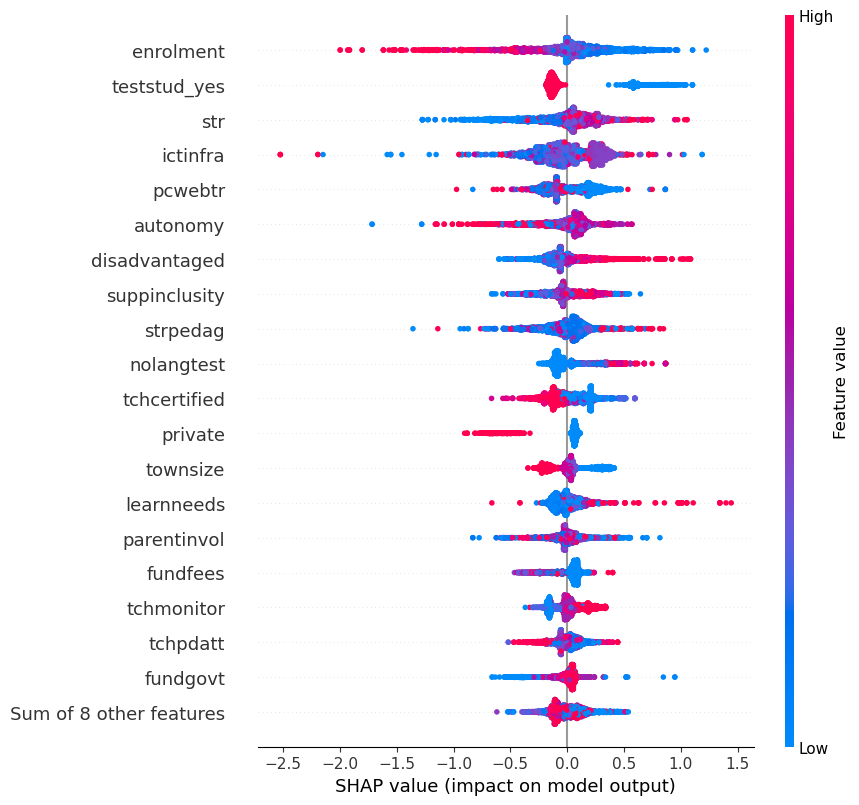}
\includegraphics[width=0.50\textwidth]{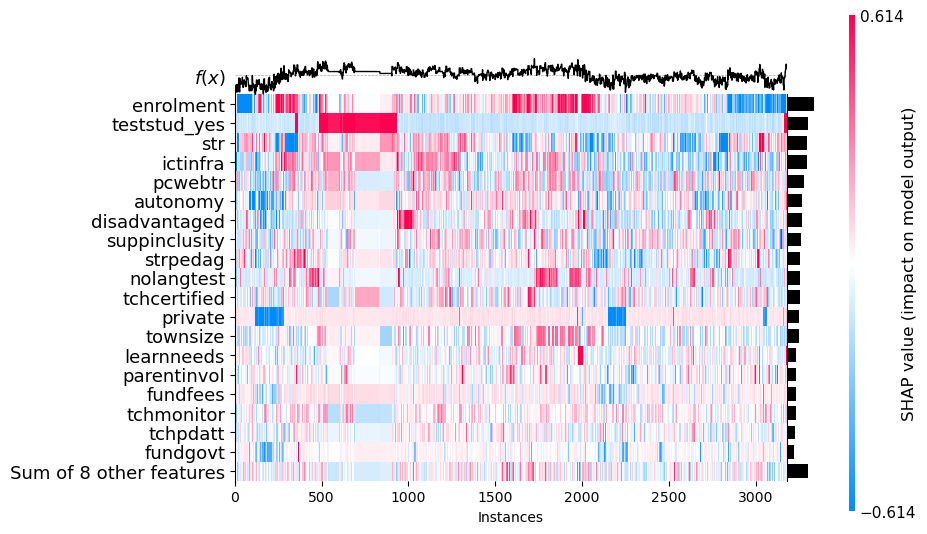}
\end{multicols}
\caption{Model: level 0 versus level 1 (bottom performing group). Beeswarm dot plot and heatmap of individual SHAP values for each student$/$famly and school covariate across observations}
\label{figure3}
\medskip
\begin{minipage}{0.99\textwidth}
\footnotesize{
Notes: (1) Beeswarm plot. Red represents that covariates is = 1 and, blue (= 0) otherwise. (2) Heatmaps. Colour intensity reflects the degree of impact on student performance, ranging from low (blue) to high (red). (2) See Tables \ref{tableB1} and \ref{tableB2} (Appendix B) for variables' acronyms and definitions.
}
\end{minipage}
\end{figure}

Figure \ref{figure3} displays the distribution of SHAP values for each student and covariate for the bottom performing group using beeswarm (top figures) and heatmap (bottom figures) plots\footnote{Beeswarm plots are essentially dot plots where each student's specific SHAP value has one dot on the line of each covariate, and dots are stacked horizontally showing the SHAP value density for each $X$ (explanatory variable) arranged from low to high values. The bottom plots are heatmaps further showing SHAP values distribution for each covariate across the sample, with colour intensity reflecting the degree of impact on student's performance, ranging from low (blue) to high (red).} which offer a more granular approach than the aggregated SHAP values of Table \ref{table5} in terms of impacts on predictions.\footnote{For completeness, the same plots for the low performing group, level 1 versus level 2 working sample, are shown in Appendix D (Figure \ref{figureD2}).} All variables are shown in the order of global feature importance. Below, we outline main findings emerging from these plots.

\medskip
\begin{center}
  [Figure \ref{figure3} here]
\end{center}
\medskip

Figure \ref{figure3}'s beeswarm plot (on the left) shows how high values (displayed in red) of repetition at primary (=1) have a positive contribution on the prediction (that is, the probability of student's achieving level 0 instead of level 1), and an equivalent pattern is found for repetition at lower secondary, speaking other language and for those who started primary late and female students. Most of the number of devices' SHAP values are low and fairly spread (shown in blue), and they have a positive contribution on the chances of students' falling into level 0 achievement as well as family SES (poorer students) and any engagement in paid work (mostly blue dots) while, for teacher's wellbeing support, estimates are equally divided between high (red) and low (blue) values, something similar is found for number of books. This is complemented by the SHAP values heatmap for student$/$family covariates (bottom left) where one can see that covariates with significant homogenous segments (mostly in red) are repetition, devices number, school climate and safety and minority language spoken at home.

For school's variables (plots on the right of Figure \ref{figure3}) the beeswarm plot shows that, school's features specific estimates withholding students moving from level 1 to level 2 (with the majority of SHAP values located in the positive section of the x-axis), are low values of testing methods of students (or no testing), high students' contextual disadvantage, low values of either the ICT index (or weaker ICT infrastructure) or STR (pedagogic staff). And, in the oppositive direction, high values (shown in red) are found, for instance, for autonomy, private school and testing, rate of certified teachers and teachers PD attendance.\footnote{All these covariates can lead to negative predictions of the model output (or higher chances of reaching level 2 achievement across the three subjects).} The heatmap for school covariates, in turn, shows that covariates with darker red (with positive impacts on prediction) or blue (negative impacts on prediction) compact segments are: testing, school size (enrolment) and school type, and ICT infrastructure.


\subsection{Experiment\textendash local explanations. Students at extremes of SHAP index distribution}
\label{section53}

Here, we move away from global explanations by shifting our focus towards local explanations of SHAP values which are given by specific observations. This exercise is very relevant policy-wise. For this, we proceed as follows. First, we obtain the overall average contribution of SHAP values across the sample for each student, namely:  $\Phi_{i} = \frac{1}{J} \sum_{j=1}^{J} \phi_{j}^{(i)}$ (for $i$ = 1,...,$N$) and, secondly, we sort out these values and, finally, we select students with the minimum ($\Phi^{\text{min}}_{i}$) and maximum ($\Phi^{\text{max}}_{i}$) contributions so that to profile the impact of covariates at extremes. Once student pairs are identified we plot leading covariates impacts for each student, with graphs showing specific values that each covariate takes. This exercise is important because, by choosing students at extremes of the SHAP contributions, it is possible to identify students who are the most likely to be performing at level 0 or level 1 (or at level 1 and level 2 for the low performing sample) and their related covariates pushing students' performance towards either of them.\footnote{Appendix B includes, as well as the definitions and how variables are constructed, the meaning associated to each label of variables (see: column 5 of Tables \ref{tableB1} and \ref{tableB2}).}

\subsubsection{Level 0 versus level 1\textendash estimates}
\label{section531}

\begin{figure}[ht!]
        \centering
        \subfloat[\footnotesize{Student$/$family variables (high)}]{%
            \includegraphics[width=0.45\textwidth]{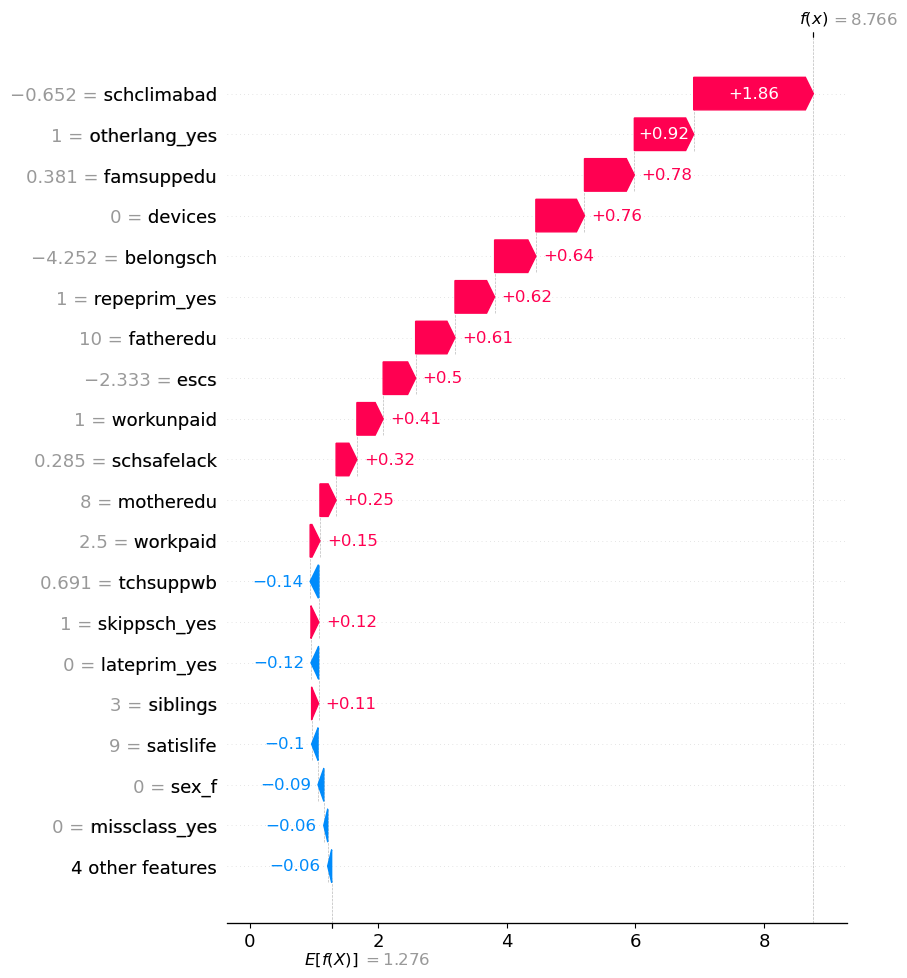}\label{figure4a}
        }
        \subfloat[\footnotesize{Student$/$family variables (low)}]{%
            \includegraphics[width=0.45\textwidth]{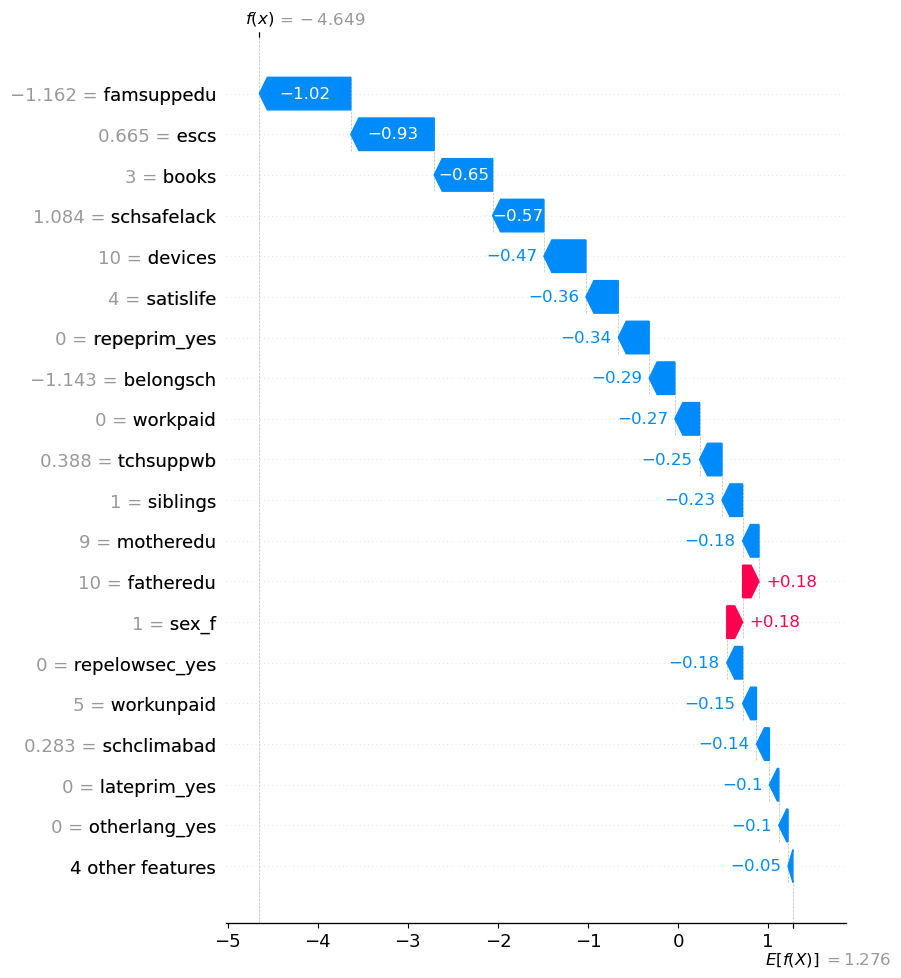}\label{figure4b}
        }\\
        \subfloat[\footnotesize{School variables (high)}]{%
            \includegraphics[width=0.45\textwidth]{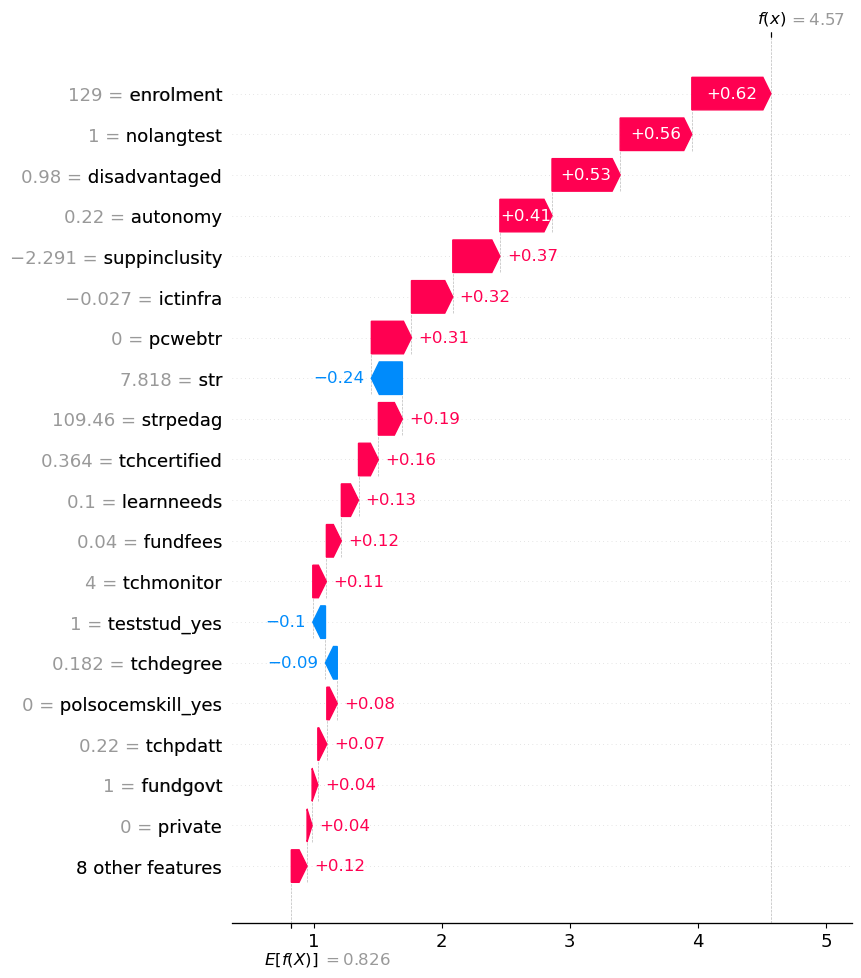}\label{figure4c}
        }
        \subfloat[\footnotesize{School variables (low)}]{%
            \includegraphics[width=0.45\textwidth]{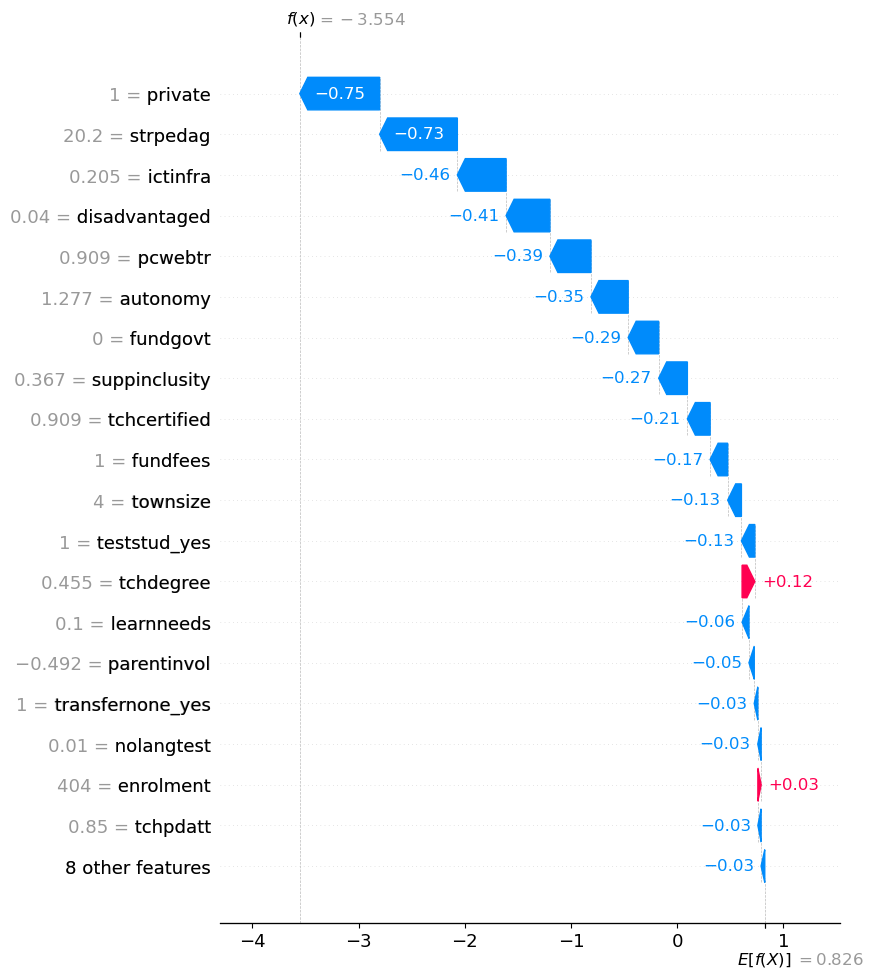}\label{figure4d}
        }\\
\caption{Model: level 0 versus level 1 (bottom performing group). Student$/$famliy and school covariates individual SHAP values for students with the lowest and highest average SHAP values contributions}
\label{figure4}
\medskip
\begin{minipage}{0.99\textwidth}
\footnotesize{
Notes: (1) See Tables \ref{tableB1} and \ref{tableB2} (Appendix B) for variables' acronyms, definitions and associated labels to values (in grey, left to the name of covariates).
}
\end{minipage}
\end{figure}

We compare the profile of students ($\Phi^{\text{max}}_{i}$ = +7.49 and $\Phi^{\text{min}}_{i}$ = -5.92) for the bottom performing sample. In the top two plots of Figure \ref{figure4} are SHAP values for the student$/$family group of covariates. A comparison of Figure \ref{figure4a} and Figure \ref{figure4b} demonstrates how inequality operates at the bottom of the learning distribution. A student with the highest probability of being a level 0 achiever attends a school with bad school climate (index of -0.65 falling into high interval); speak a minority language; has no digital devices; has a very weak sense of belonging to school; repeated in primary; and comes from a very poor family (SES index of -2.33 falls into the bottom half of SES distribution); and is engaged in unpaid work (1 day) and paid work (=2.5; between 2 and 3 days per week) (see: Figure \ref{figure4a}). In the oppositive direction, a student whose performance falls into level 1, comes from a better off family (SES index falling into top half); has books (=3) and high number of devices (=10); is satisfied with life and supported by teachers; has not repeated a grade and does not work; and comes from smaller families (Figure \ref{figure4b}).

\medskip
\begin{center}
  [Figure \ref{figure4} here]
\end{center}
\medskip

School covariates are also dissimilar for these two students' profiles (Figures \ref{figure4c} and \ref{figure4d}). Consider the first ten ranked covariates in terms of impacts. Smaller schools (enrolment = 129), with nolangtest variable equals to 1 (hence an Indigenous school) and a huge degree of disadvantage (98\%) and with weaker support for inclusivity and low ICT investments (no PCs) and only a third of teachers certified, are all features of schools linked to a level 0 achievement student. On the contrary, a student attending a private school with low STR for pedologic staff (20.2 students per staff versus 109.5 for the comparison level 0 student), with low contextual disadvantage and higher autonomy (index = +1.27), nearly all PCs connected to the web (90.9\%) and around 55\% more qualified teachers, are all school features pushing up the chances of a bottom performer student towards level 1 achievement.

\subsubsection{Level 1 versus level 2\textendash estimates}
\label{section532}

\begin{figure}[ht!]
        \centering
        \subfloat[\footnotesize{Student$/$family variables (high)}]{%
            \includegraphics[width=0.45\textwidth]{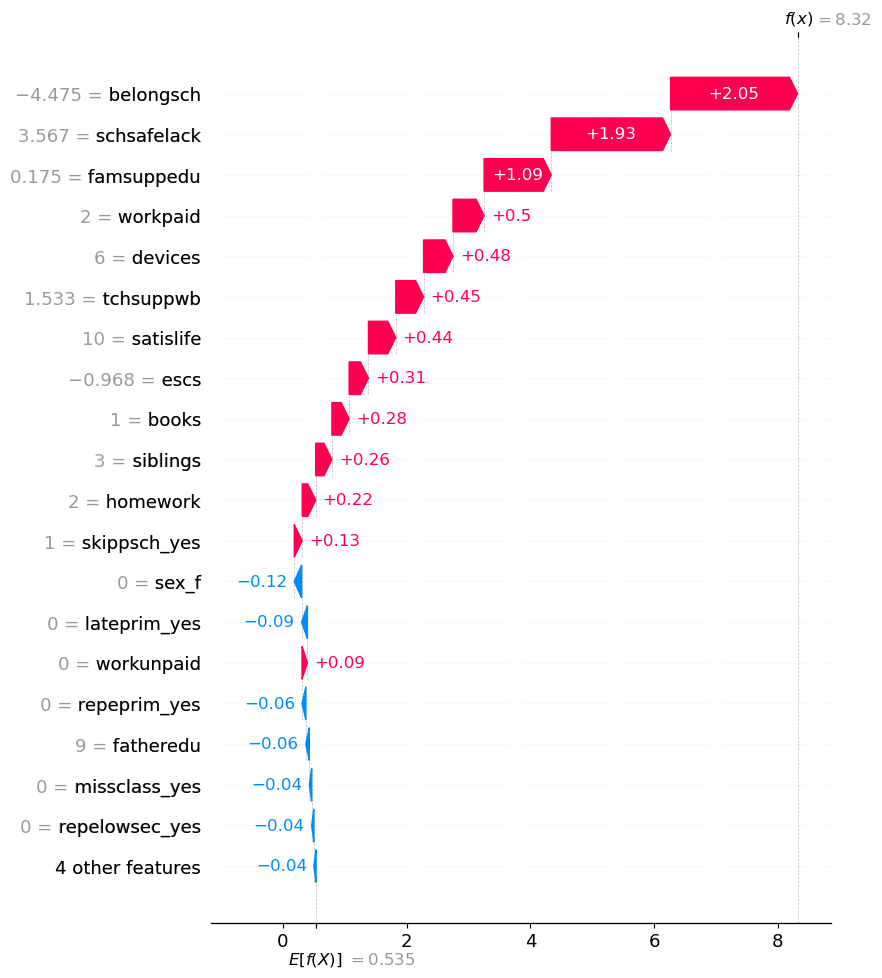}\label{figure5a}
        }
        \subfloat[\footnotesize{Student$/$family variables (low)}]{%
            \includegraphics[width=0.45\textwidth]{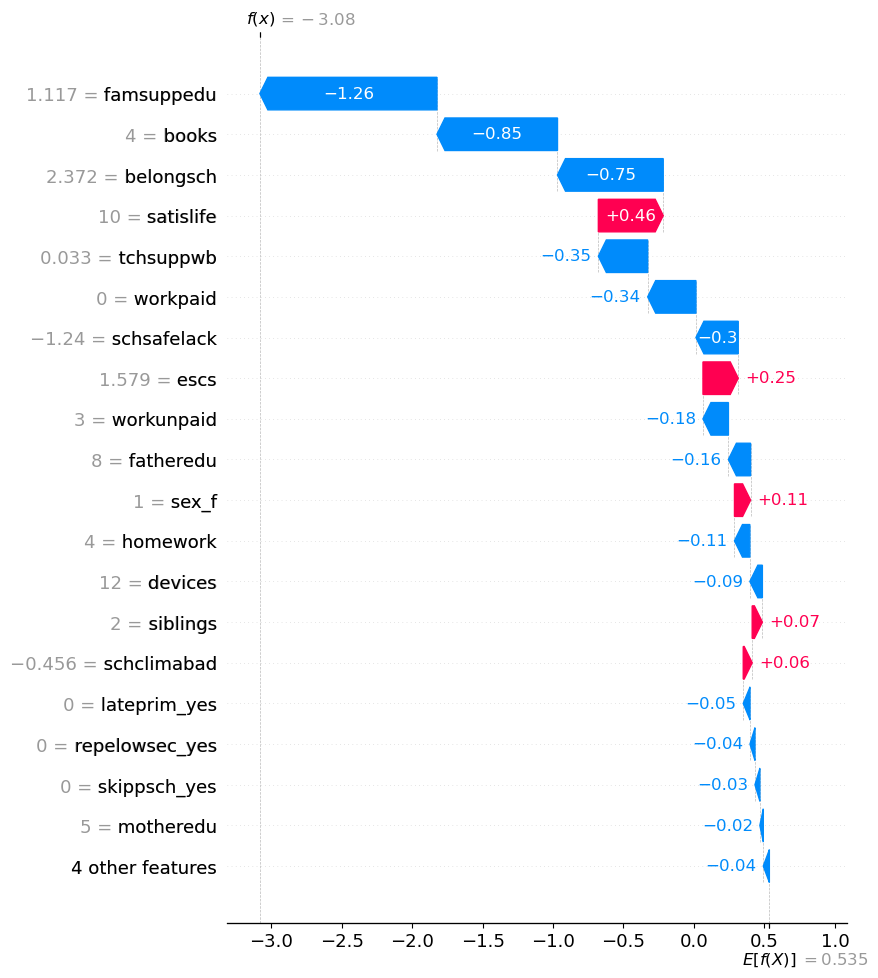}\label{figure5b}
        }\\
        \subfloat[\footnotesize{School variables (high)}]{%
            \includegraphics[width=0.45\textwidth]{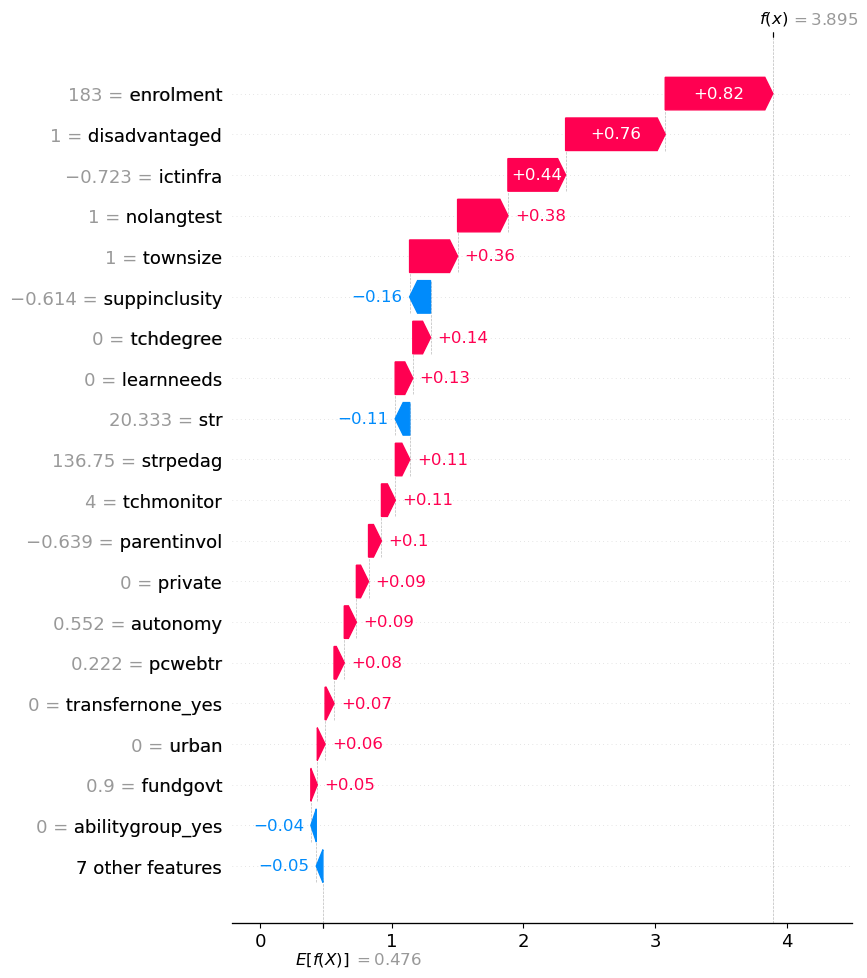}\label{figure5c}
        }
        \subfloat[\footnotesize{School variables (low)}]{%
            \includegraphics[width=0.45\textwidth]{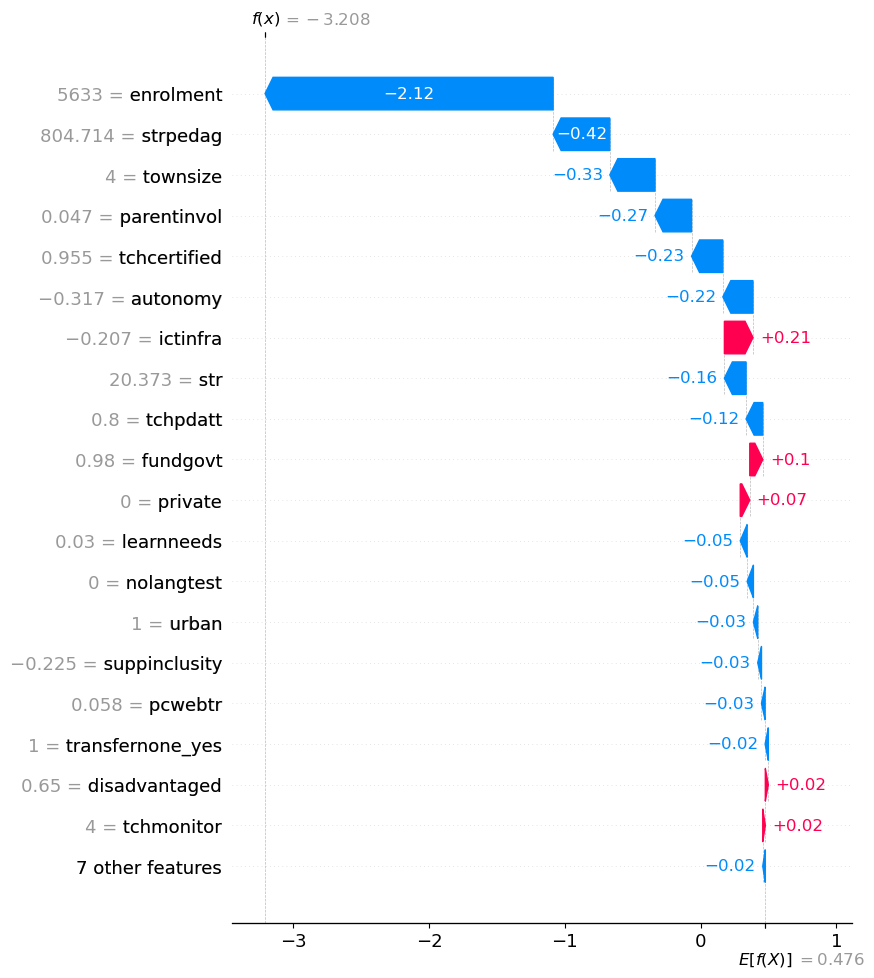}\label{figure5d}
        }\\
\caption{Model: level 1 versus level 2 (low performing group). Student$/$famliy and school covariates individual SHAP values for students with the lowest and highest average SHAP values contributions}
\label{figure5}
\medskip
\begin{minipage}{0.99\textwidth}
\footnotesize{
Notes: (1) See Tables \ref{tableB1} and \ref{tableB2} (Appendix B) for variables' acronyms, definitions and associated labels to values (in grey, left to the name of covariates).
}
\end{minipage}
\end{figure}

We repeat the same exercise for the low performer sample (level 1 and level 2). Results are displayed in Figure \ref{figure5}. First, compared to a student with the highest chance of reaching no more than level 1 (Figures \ref{figure5a} and \ref{figure5b}), a student reaching level 2 with the lowest index ($\Phi^{\text{min}}_{i}$) has larger family educational support (index = +1.11); 4 books (instead of 1) and 12 digital devices (instead of 6); a high index regarding school belonging (+2.37 versus -4.47); is not engage in paid work and comes from a richer family (SES index = +1.57 versus -0.96); and do twice as much homework (2-3 hours versus less than 1 hour).

\medskip
\begin{center}
  [Figure \ref{figure5} here]
\end{center}
\medskip

Second, a level 1 student with $\Phi^{\text{max}}_{i}$ comes from a small Indigenous rural school (as rate for the nolangtest is equal to 1 and urban = 0), while the student with the highest chance of reaching level 2 comes from a large public school (without language minority students) which is located in an urban (large town) area and with parents involvement fostered by school (index = +0.047), where the latter is less encouraged at the Indigenous school where the level 1 student is attending (index = -0.63) (see Figures \ref{figure5c} and \ref{figure5d}). The school rates for certification and PD attendance of teachers are high (between 80-95 percent) and they are leading SHAP values for the school of the level 2 student, but this is not the case for the comparison level 1 student's school.


\subsection{Countries' results}
\label{section54}

\begin{figure}
\centering
\includegraphics[width=0.86\textwidth]{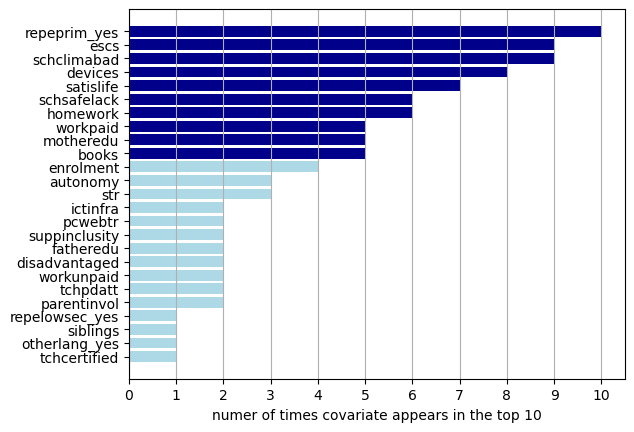}
\caption{Frequency of student$/$family and school covariates with countries' SHAP values appearing in the top 10 ranked estimates}
\label{figure6}
        \medskip
\begin{minipage}{0.99\textwidth}
\footnotesize{
Notes: (1) See Appendix B (Tables \ref{tableB1} and \ref{tableB2}) for variables' acronyms and definitions.
}
\end{minipage}
\end{figure}

Here we examine SHAP values at the country level by running a similar analysis per country for the level 0 and level 1 sample\footnote{Results for the level 1 and level 2 countries' comparison can be obtained from the authors upon request.} yielding local explanations of SHAP values. But, first, we present the global contributions, i.e., the average SHAP (absolute) values for all covariates.\footnote{Plots for each country, with the top 20 Shapley values and their corresponding covariates, are shown in Figure \ref{figureD3}.} As a summary, we plot the frequency of the highest SHAP values contributions on predictions across countries. In Figure \ref{figure6} we select the top 10 ranked covariates  for each country and then obtain the number of times a given covariate appears in the top 10 list. The figure shows that primary repetition appears in the top ranked list for all 10 countries, while family SES and school climate (in 9 countries) and digital devices (in 8 countries). This is followed by lack of school safety and homework intensity (in 6 countries) whereas the covariates paid work, mother education and number of books appears in the top 10 ranked covariates for 5 countries.

\medskip
\begin{center}
  [Figure \ref{figure6} here]
\end{center}
\medskip

As regards to local explanations, the main takeaways concerning leading drivers for students more likely to either reach level 0 or level 1 achievement per country are outlined below.\footnote{Due to space constraints, we discuss the first five covariates with highest SHAP values of each group.} For the first set of five countries (Table \ref{table6}):

\begin{itemize}
  \item \textit{Argentina}. (a) The student with the largest probability of reaching just level 0 (high-index $\mapsto$ level 0) is a repeater, has no devices, comes from a small size school with low parental involvement in the school running and bad school climate. (b) The student with the highest probability of reaching level 1 (low-index $\mapsto$ level 1) comes from a better-off family (with positive SES index), has 4 books (2.5 books above mean), attends a private school with just 12\% rate of disadvantaged students, and has 11 devices (3 above the country's mean).
  \item \textit{Brazil}. (a) Profile of student with achievement $\mapsto$ level 0: attends a school with full disadvantage, comes from low SES family, has zero books. (b) Profile of student with achievement $\mapsto$ level 1: does more regularly homework (above average), attends a school with below average index of bad school climate, has a satisfaction with life in the middle of the scale, has a low educated mother and is likely to be engaged in paid work everyday.
  \item \textit{Chile}. (a) Profile of student with achievement $\mapsto$ level 0: is a repeater who also skipped school  (1 whole day last 2 weeks), whose mother only completed lower secondary, attends a school of a size in the bottom quartile with good safety in its surroundings areas. (b) Profile of student with achievement $\mapsto$ level 1: attends a school located in a even safer area and better school climate, has one sibling, and did not repeat primary.
  \item \textit{Colombia}. (a) Profile of student with achievement $\mapsto$ level 0: comes from a poor family (SES into the bottom half of distribution), has none digital devices and repeated primary, attends a school with bad school climate, and has a mother who is educated. (b) Profile of student with achievement $\mapsto$ level 1: comes from a better-off family and does not engage in paid-work, has 3 books (top 90\%), and attends a school without disadvantaged students and with a high degree of autonomy (index large and positive).
  \item \textit{Dominican Rep}. (a) Profile of student with achievement $\mapsto$ level 0: repeated primary, attends a school with poor school climate, comes from low-SES family, is satisfied with life, and in the school there is a support of inclusivity. (b) Profile of student with achievement $\mapsto$ level 1: comes also from a school with poor school climate but with the school located in safer area, the extent of support of inclusivity in the school is larger, does not work and is satisfied with life.
\end{itemize}

\medskip
\begin{center}
  [Tables \ref{table6} here]
\end{center}
\medskip

For the second set of five countries (Table \ref{table7}), main findings are as follows:

\begin{itemize}
  \item \textit{Mexico}. (a) Profile of student with achievement $\mapsto$ level 0: is engaged fully in paid work, does not have any books, missed school (3 months in a row), and repeated primary. (b) Profile of student with achievement $\mapsto$ level 1: comes from relatively poor family with low SES index, but is not engaged in paid work and does regularly homework, attends a school located in an unsafe area, and in a school where parents have low participation.
  \item \textit{Panama}. (a) Profile of student with achievement $\mapsto$ level 0: attends a highly disadvantaged school with a lack of PCs connected to the internet, does hardly any homework, comes from really poor household whose mother did not complete primary. (b) Profile of student with achievement $\mapsto$ level 1: comes from a family whose SES is in the top quartile and has 9 digital devices (large than the median by 2), though the school he$/$she attends has very low autonomy and no qualified teachers.
  \item \textit{Peru}. (a) Profile of student with achievement $\mapsto$ level 0: comes from an Indigenous background who repeated primary and does just less than 1 hour homework per week, and attends a school with severe low school climate and very weak ICT infrastructure. (b) Profile of student with achievement $\mapsto$ level 1: does not work for payment, has 8 digital devices at home (above median), the family wealth falls into the top half of the SES distribution, an attend school with a STR of 32 and high autonomy.
  \item \textit{Paraguay}. (a) Profile of student with achievement $\mapsto$ level 0: comes from a family with 3 devices and high SES, only engages 1 day in unpaid work, but attend a school with low autonomy which has just a rate of teacher attending PD of 16\%. (b) Profile of student with achievement $\mapsto$ level 1: is richer than a level 0 student (higher value of SES index) and has more digital devices at home, attends a private school with larger autonomy, and a school's enrolment of $\approx$ 900 students.
  \item \textit{Uruguay}. (a) Profile of student with achievement $\mapsto$ level 0: is a primary repeater, has low life satisfaction and no devices at home. (b) Profile of student with achievement $\mapsto$ level 1: did not repeat primary, family SES is larger (in the 3rd quartile), has 3 books at home (median for this covariate is just 1), attends a school with more than 70\% of either certified teachers or teachers attending PD.
\end{itemize}

\medskip
\begin{center}
  [Tables \ref{table7} here]
\end{center}
\medskip


\subsection{Heterogeneity's results}
\label{section55}

\begin{figure}[ht!]
        \centering
        \subfloat[\scriptsize{Family SES-gender}]{%
            \includegraphics[width=0.33\textwidth]{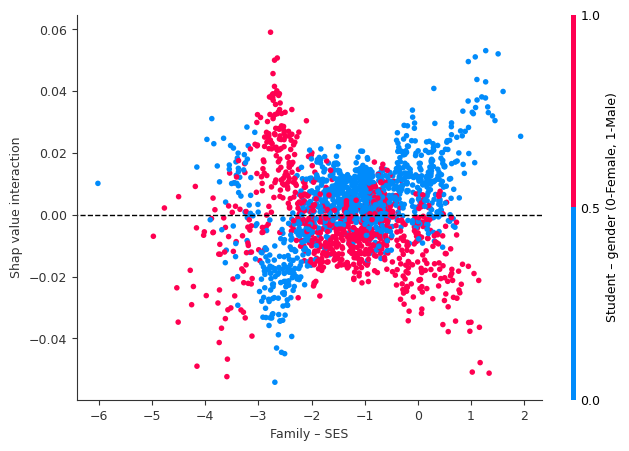}\label{figure7a}
        }
        \subfloat[\scriptsize{Family SES-primary repetition}]{%
            \includegraphics[width=0.33\textwidth]{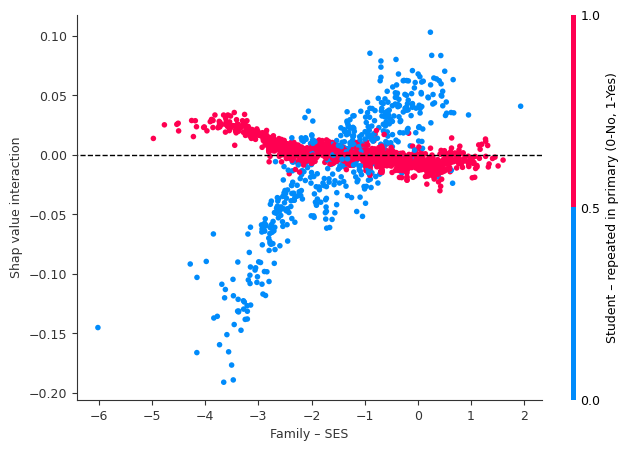}\label{figure7b}
        }
        \subfloat[\scriptsize{Family SES-ICT}]{%
            \includegraphics[width=0.33\textwidth]{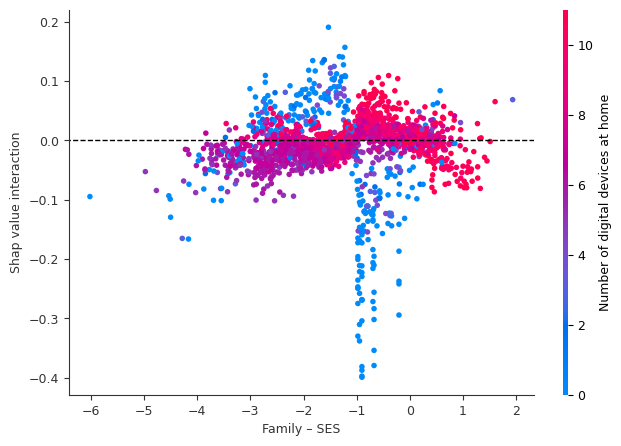}\label{figure7c}
        }\\
        \subfloat[\scriptsize{Family support-gender}]{%
            \includegraphics[width=0.33\textwidth]{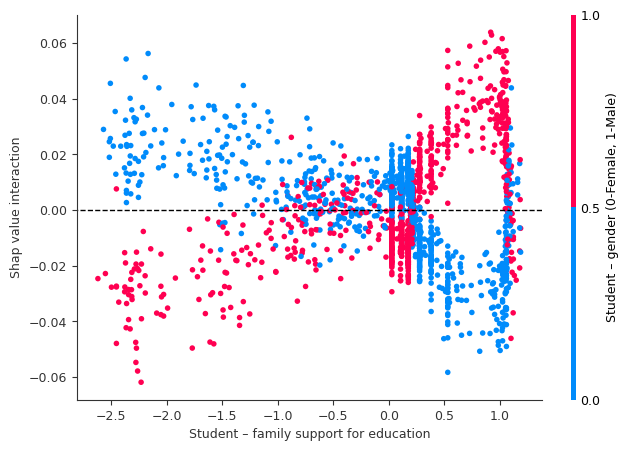}\label{figure7d}
        }
        \subfloat[\scriptsize{Family support-ICT}]{%
            \includegraphics[width=0.33\textwidth]{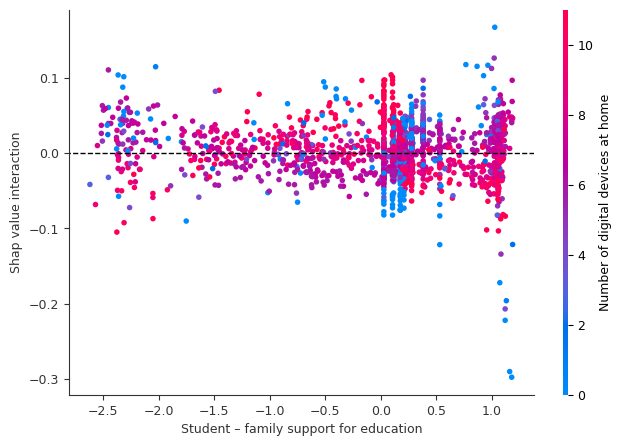}\label{figure7e}
        }\\
        \caption{Model: level 0 versus level 1 (bottom performing group). SHAP values interactions of drivers of student's disadvantages (i.e., household wealth and parental educational input)}
        \label{figure7}
\end{figure}

\begin{figure}[ht!]
        \centering
        \subfloat[\scriptsize{Family SES-gender}]{%
            \includegraphics[width=0.33\textwidth]{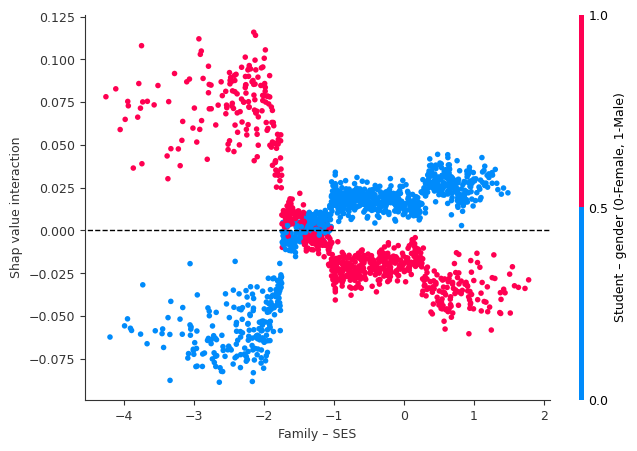}\label{figure8a}
        }
        \subfloat[\scriptsize{Family SES-primary repetition}]{%
            \includegraphics[width=0.33\textwidth]{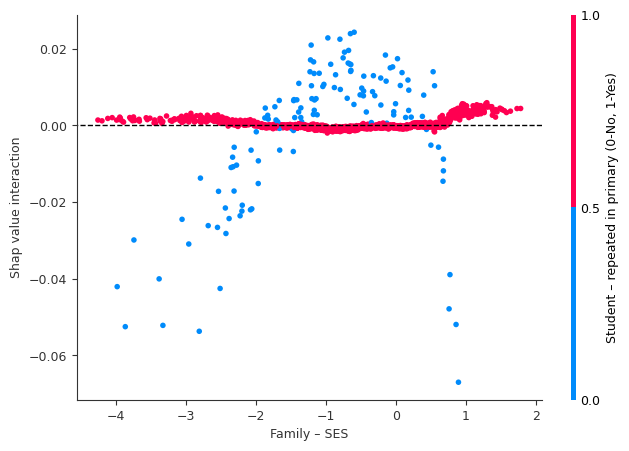}\label{figure8b}
        }
        \subfloat[\scriptsize{Family SES-ICT}]{%
            \includegraphics[width=0.33\textwidth]{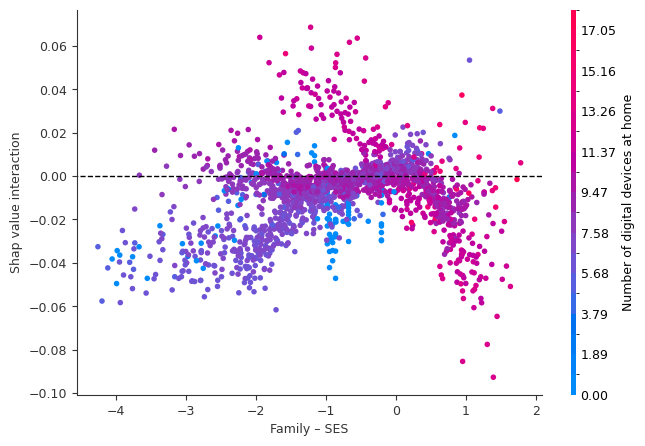}\label{figure8c}
        }\\
        \subfloat[\scriptsize{Family support-gender}]{%
            \includegraphics[width=0.33\textwidth]{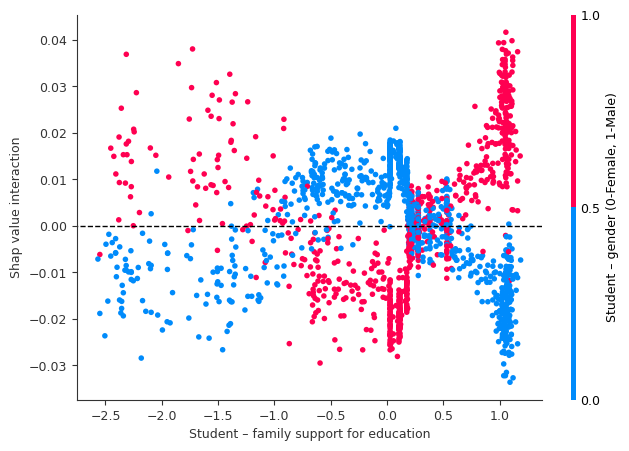}\label{figure8d}
        }
        \subfloat[\scriptsize{Family support-ICT}]{%
            \includegraphics[width=0.33\textwidth]{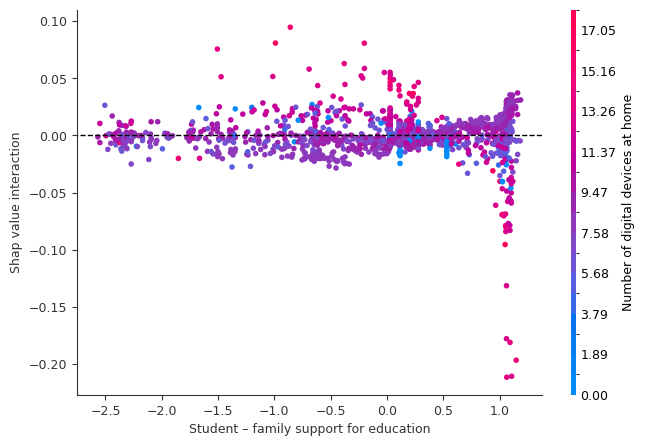}\label{figure8e}
        }\\
        \caption{Model: level 1 versus level 2 (low performing group). SHAP values interactions of drivers of student's disadvantages (i.e., household wealth and parental educational input)}
        \label{figure8}
\end{figure}

In this section we briefly address some aspects of intersectionality of the whole sample estimates. Namely, we show how estimations for the relationship of two key drivers behind low and bottom achievement of students (that is, family SES and parental educational input) vary by students' gender, repetition and ICT stock.\footnote{We are grateful to a referee for this suggestion.} Interaction effects are derived using the term $\sum_{j \neq i}^{}\Phi_{i,j}(f,x)$ from Eq. (\ref{shap3}). Results are displayed in Figure \ref{figure7} (for the bottom performers comparison) and in Figure \ref{figure8} (for the low performers comparison).

On the one hand, Figures \ref{figure7a}, \ref{figure7b} and \ref{figure7c}  describe the interaction of family SES with a given student's feature: gender, primary repetition and with ICT stock at home (i.e., number of digital devices). When the effect of family SES is compounded with gender (with blue dots denoting female students and red dots male students) it can be seen decreasing SHAP values along the wealth index for male students and increasing for female students; in other words, coming from a poorer household affects relative more a male student than female student and, conversely if a student stems from a better-off household, and without important gender gaps for students in the middle of the wealth index (Figure \ref{figure7a}). Specifically, the plot shows gaps on SHAP values of around +4 (and so a higher chances to fall into level 0 achievement group) for a female student when family SES is equal to 1 (a richer family) and, conversely, gaps on SHAP values of +6 for quite poor families (wealth index = -3) with male students being much more likely to be at the bottom of the learning distribution. In the case of primary repetition (Figure \ref{figure7b}) there is clear pattern about how wealth compounds the effect of repetition: poorer and repeater students have positive SHAP values (higher chances to be level 0 performer), with gaps on SHAP values against a poor but not repeater of around +15. Figure \ref{figure7c} shows that, when family SES is interacted with a student's ICT stock, a poorer student with low number of digital devices (in blue) has slightly larger SHAP values whilst a richer student with larger number of devices (in red at wealth = + 1) has negative SHAP values of -0.1, with better chances to fall into the better performing group (level 1). On the other hand, estimated SHAP values for the interactions with family educational support have not much of a differential effect in the case of ICT stock (Figure \ref{figure7e}), though they do for gender. Figure \ref{figure7d} shows that lack of family educational support affects negatively female students (as most blue dots are above zero at values of the family support index $<$ 0), but they don't for male students.

\medskip
\begin{center}
  [Figure \ref{figure7} here]
\end{center}
\medskip

Figure \ref{figure8} contains estimates for SHAP values interactions for the level1-level 2 sample comparison. The gender gap is more marked here when the comparison moves up in the learning distribution. Household poverty impacts relatively more male students, particular for very poor families (index below -2) as gaps with female students SHAP values is $\approx$ +0.10, although from quite a long range and family disadvantage (wealth index $\ge$ -2) all female students are more prone to fall into level 1 (rather than the reaching proficiency at level 2) (Figure \ref{figure8a}). Primary repetition interaction with wealth (Figure \ref{figure8b}) does not seem to significant impact for a student falling into level 1 as SHAP interactions (red dots) hovers around zero. Along the same vein, interactions of ICT with both family SES and family educational support (Figures \ref{figure8c} and \ref{figure8e}, respectively) and their estimated SHAP values are not different by level of ICT (or number of digital devices), thereby suggesting a lack of intersectionality among these factors. Yet, for gender and family educational support (Figure \ref{figure8d}), a distinct pattern on SHAP interactions values emerges boosting gender gaps (in favour of female students) at higher degree of family support, and in the reverse direction (detrimental for female students) at lower educational family support levels.

\medskip
\begin{center}
  [Figure \ref{figure8} here]
\end{center}
\medskip


\section{Discussion}
\label{section6}

Underpinned by broad inequalities in the education sector, Latin American students' limited achievement at secondary level is concerning. This inequality-driven learning crisis has been made worse by the COVID-19 pandemic (\citealp{betthauser23,unesco24}), with the region experiencing the larger period of school closures (\citealp{bracco25,gomez22}). The region learning crisis is severe, with only between 5 and 7 out of 10 students reaching adequate competency in reading and math (\citealp{arias24b}). Is within this bleak scenario that the paper's contribution should be understood, namely, through the profiling of low achievers, so that adequate interventions can be put forward to mitigate the extent of the learning crisis in the region.

In particular, this study follows a mechanism to identify poor performing students through explainable ML methods (\citealp{chen22,lundberg17,vstrumbelj14}), i.e., Shapley values. The paper contributes to the emerging body of studies using interpretable ML results in education (\citealp{ali25,gomez24,huang24,liu22,zhu25}) and broadly to the literature about the identification of at-risk students (\citealp{shi25,shi24}); though, expanding this literature to global south education systems at a regional scale.

The paper, equally relying on global and local explanations at a regional level and at a country level, yields several insights into the LAC region learning crisis by offering a detailed account of which factors may sturdily place a students in the lowest bracket of performance (level 0, no competency achieved) or, alternative, which malleable factors may allow a bottom performer to transition to the next achievement bracket (level 1). The same analysis is carried out for the level 1 and level 2 students (where level 2 performance denotes students achieving basic subjects competencies). This granular analysis at the bottom of the learning distribution is fundamental for successful targeted interventions pushing up students to the next level of achievement. In essence, the paper's approach and its policy implications are close to the existing literature of the region for the detection of secondary students at risk of dropout (e.g., \citealp{haimovich21,rodriguez23,vinas21}) via the setup of automatic alert systems (Early Warming Systems).

The utility of local SHAP values cannot be underestimated. They are a powerful policy tool as they offer an identification strategy of students locked into lower performing groups (or at risk of falling into that group) because they, not only rank leading covariates revealing positive or negative impacts of each variable on the model output for each observation (student), but also because they give the specific values each covariates take in that ranking, thereby given precise profiles of students. Results from the SHAP method can also further guide follow-up qualitative studies attempting to gather more in-depth information on the reasons behind poor performance. This sequential combination of ML approaches and interviews has been applied in certain studies (e.g., \citealp{gonzalez21,jayachandran23}). The sequential application of estimated SHAP values in learning surveys with qualitative interviews is an interesting avenue for future research by offering more insights into the learning crisis problem. Below, we flesh out some policy implications from leading regional estimates which can be framed as a gradual two-way process of interventions: (i) first, to identify those at the bottom of achievement; and (ii) to introduce leverage mechanisms to modify covariates of level 0 (level 1) students towards profiles of level 1 (level 2) students.\footnote{Within this two-way process it is relevant to emphasise that is possible to further add the timing of these mechanisms behind the shift on covariates. For example, short term policies are linked to an educational process's proximal covariates which can be more easily modified  (e.g., repetition, gender gaps, supporting Indigenous schools, books investments), whereas schools’ physical and ICT infrastructure and connectivity can be deemed as medium-term targets, and the overall wealth-driven inequality of schooling outcomes, for instance, a longer-term societal objective. We are thankful to an anonymous referee for pointing this out.}

In the local regional analysis we find that the profile for a student who is locked-into the lack of competency group (level 0) is Indigenous, a repeater, is engaged in paid work (half of the week) and comes from a very poor household without digital devices. Thus, an entry point of a policy is to target poor Indigenous students and$/$or poor repeaters with financial support so that to minimise the impact on learning of paid work, likely leading to repetition. Culture responsive programs on bilingual education, raising Indigenous agency (\citealp{lopez21}) and the attraction of good teachers to schools with mostly Indigenous students (\citealp{cortina17,delprato22}), tackling repetition earlier on in primary (see: Learning Spaces program in Dominican Rep; \citealp{chavez21}), and cash transfers (see examples for Brazil and Peru \citealp{draeger21,gaentzsch20}) alleviating disadvantaged students financial constraints, would all be beneficial interventions for level 0 achieving students.

Conversely, we find that students performing at level 1 do not work, are not repeaters, have a higher number of books and digital devices at home. Hence, shifting leading covariates from the profile of level 0 performing student to a level 1 performing student would also require additional public investment on educational inputs. This shift, in turn, is related to programs such as the one laptop and investment in books in the region (e.g in rural schools in Peru: \citealp{cristia17}, or in Uruguay: \citealp{diaz22,melo17}). Also, improving school climate and boosting students' well-being support from teachers, are features of level 1 student profile which can benefit a lowest level 0 performer. The overall importance of school climate for learning is in line with the reviews of \cite{varela24} and \cite{chavez21}.

Besides, a comparison of school features between the two level 0 and level 1 students' profiles\textendash and as found by earlier studies (\citealp{agasisti23,delprato25b,mereles22})\textendash highlights the need of higher investments in schools' ICT stock and connectivity, and promoting the collaboration on the use of digital devices by teachers (\citealp{arias24a}). Poor regional ICT access is a sturdy barrier for those students attempting to move up in the learning distribution towards performance levels where subjects competencies are in fact reached. For instance, \citealp{cepal20}'s report indicates that only 70\% of people in the LAC region poses internet connection and 46\% of children (aged 5-12) live in households without an internet connection. High gaps on ICT access by socioeconomic status are also observed, with students in the top wealth quartile being at least four times as likely (in comparison to the students from the bottom quartile) to have a laptop at home (\citealp{berlanga20}). This comparison also reveals that more investment is required on teachers preparation and teachers' numbers in pedagogic roles (\citealp{ainscow25,miranda25}), ensuring quality and promoting continuous professionalisation.

Policies seeking out to move up students from level 1 to level 2 would need to focus on (as well as on enhancing level 1's students stock of educational inputs and alleviating poverty and so the chances of being involved in paid work) offering better conditions for level 1 students to be able to carry out more homework (e.g., after school classes), and boosting family educational support. At the school level, policies should reshape schools' functioning on different dimensions: develop involvement of parents on schools' governance and to enhance teaching quality too (via attracting teachers who have more certification and higher attendance of PD training).

The analysis has some limitations. Methodologically, a more exhaustive approach for benchmarking students profiles would be helpful, producing a more robust contrasting of students' profiles, based on incremental steps (e.g., deciles) of the SHAP values index $\big[\Phi^{\text{min}}_{i}$, $\Phi^{\text{q10}}_{i}$,...,$\Phi^{\text{q90}}_{i}$, $\Phi^{\text{max}}_{i}\big]$. Rather than focusing on two students for each sample, a continuum of profiles along students' distribution of SHAP values would allow a better comparison on shift of covariates underpinned by future scaffolding interventions. A future direction of research is to address intersectionality (combinations of student$/$family covariates with school covariates) which is a driving concept behind how inequality appears and persist in resource-constrained education systems (\citealp{codiroli19,delprato24}). We succinctly address this with SHAP values interactions, though further work in this direction is indispensable so that interventions are designed using joint markers of personal and contextual disadvantages from students and schools.


\section{Conclusions}
\label{section7}

In this paper, we analyse the determinants of low performers students in the LAC region at secondary level using PISA 2022 data. Combining the performance on three subjects assessed (math, reading and science), we focus on the bottom learning distribution for three brackets of performance : level 0 (no achievement), level 1 (below basic competencies) and level 2 (achieving basic competencies), where around 83\% of the region students' performance falls. We rely on two working samples: the sub-groups of bottom performers (level 0 versus level 1, $N$ = 16,236) and low performers (level 1 versus level 2, $N$ = 9,484) to compare the determinants of only reaching either level 0 or level 1, correspondingly. Methodologically, we use the SHAP method (\citealp{chen23,lundberg17}) to pin down the main determinants of learning poverty in the LAC region. We obtain global and local versions of estimates, both at the regional and country levels.

Leading findings can be summarised as follows. First\textendash from the global analysis\textendash variables with the highest impacts (regardless of the direction of associations) are, to a certain degree, similar for bottom and low performers: primary repetition, sense of belonging to school and well-being support from teachers are key drivers of falling into the lowest achievement group, whereas student’s paid work, the intensity of homework and mother's educational level are relevant for the level 1 versus level 2 comparison (located higher in the learning distribution), with household wealth and educational support and digital devices being higher ranked than in the level 0-level 1 comparison. Second, the rate of school contextual poverty and STR (pedagogic support) are also more relevant for the reaching level 2 performance, whilst the prevalence of students with a second language being more prominent as a driver for the level 0 performance group. Third, local estimations looking for students placed at the extremes of SHAP index distribution, indicate that a student with the highest probability of being a level 0 achiever attends a school with bad school climate, speak a minority language and had repeated, has no digital devices at home, comes from a poor family and works for payment half of the week. The school profile of this student shows disadvantages such as weak ICT infrastructure and teaching quality (only a third of teachers being certified), whereas a level 1 achiever attends a private with more autonomy and a low STR for pedagogic staff, more qualified teachers and with larger connectivity (over 90\% of PCs connected to internet). Similarly findings are obtained for the level 1 and level 2 comparison where \textendash as well as a level 2 student being better off financially overall, and having more educational inputs and support\textendash, he$/$she is not in the labour market having more feasible time for doing homework (twice as much as a level 1 student), and attends a school with considerable higher rates of teachers' certification and PD attendance (80\%-95\%).

Regarding countries' estimates, we find quite homogeneous patterns as far as global average contribution of top ranked factors is concerned, with repetition at primary, household wealth, and educational ICT inputs (digital devices) being in top ten ranked covariates in at least 8 out of the 10 total countries. Other important covariates across countries are school safety, student work, books at home and mother’s education. There is, however, some between-country heterogeneity when it comes to local estimates and defining characteristics of a lowest achieving student. A level 0 (no achiever) student shares, across countries, features such as being a repeater with very few digital devices, comes from a low SES family and attends a school with poor school climate, but in some instances with some differences and specific countries' drivers. In Chile, the lowest performer also skipped school and the education of his$/$her mother is below secondary while in Panama mothers' education for this level 0 student profile is even worse (uncompleted primary); in Mexico, a new feature is that a level 0 student does paid-work for nearly the whole week (and missed school for 3 months); in Peru and Paraguay school features turn out to be prominent (very low ICT infrastructure, low autonomy and just 16\% of PD teachers attendance in the case of Paraguay, and the lowest performing student for Peru is Indigenous). A level 1 student profile is, in the majority countries, from a family with an above average level of wealth, has a fair number of digital devices, is not engaged in paid work and, on the whole, attends schools with a large degree of autonomy on their functioning decisions.


\newpage
\printendnotes


\clearpage
\setlength\bibsep{0pt}
\bibliographystyle{elsarticle-harv}
\bibliography{references}

\begin{thebibliography}{110}
\expandafter\ifx\csname natexlab\endcsname\relax\def\natexlab#1{#1}\fi
\providecommand{\url}[1]{\texttt{#1}}
\providecommand{\href}[2]{#2}
\providecommand{\path}[1]{#1}
\providecommand{\DOIprefix}{doi:}
\providecommand{\ArXivprefix}{arXiv:}
\providecommand{\URLprefix}{URL: }
\providecommand{\Pubmedprefix}{pmid:}
\providecommand{\doi}[1]{\href{http://dx.doi.org/#1}{\path{#1}}}
\providecommand{\Pubmed}[1]{\href{pmid:#1}{\path{#1}}}
\providecommand{\bibinfo}[2]{#2}
\ifx\xfnm\relax \def\xfnm[#1]{\unskip,\space#1}\fi
\bibitem[{Agasisti et~al.(2023)Agasisti, Antequera and Delprato}]{agasisti23}
\bibinfo{author}{Agasisti, T.}, \bibinfo{author}{Antequera, G.},
  \bibinfo{author}{Delprato, M.}, \bibinfo{year}{2023}.
\newblock \bibinfo{title}{Technological resources, ict use and schools
  efficiency in latin america--insights from oecd pisa 2018}.
\newblock \bibinfo{journal}{International Journal of Educational Development}
  \bibinfo{volume}{99}, \bibinfo{pages}{102757}.
\bibitem[{Ahrens et~al.(2023)Ahrens, Hansen and Schaffer}]{ahrens23}
\bibinfo{author}{Ahrens, A.}, \bibinfo{author}{Hansen, C.B.},
  \bibinfo{author}{Schaffer, M.E.}, \bibinfo{year}{2023}.
\newblock \bibinfo{title}{pystacked: Stacking generalization and machine
  learning in stata}.
\newblock \bibinfo{journal}{The Stata Journal} \bibinfo{volume}{23},
  \bibinfo{pages}{909--931}.
\bibitem[{Ainscow et~al.(2025)Ainscow, Calder{\'o}n-Almendros, Duk and
  Viola}]{ainscow25}
\bibinfo{author}{Ainscow, M.}, \bibinfo{author}{Calder{\'o}n-Almendros, I.},
  \bibinfo{author}{Duk, C.}, \bibinfo{author}{Viola, M.}, \bibinfo{year}{2025}.
\newblock \bibinfo{title}{Using professional development to promote inclusive
  education in latin america: possibilities and challenges}.
\newblock \bibinfo{journal}{Professional Development in Education}
  \bibinfo{volume}{51}, \bibinfo{pages}{149--166}.
\bibitem[{Alam and Mohanty(2022)}]{alam22}
\bibinfo{author}{Alam, A.}, \bibinfo{author}{Mohanty, A.},
  \bibinfo{year}{2022}.
\newblock \bibinfo{title}{Predicting students’ performance employing
  educational data mining techniques, machine learning, and learning
  analytics}, in: \bibinfo{booktitle}{International Conference on
  Communication, Networks and Computing}, \bibinfo{organization}{Springer}. pp.
  \bibinfo{pages}{166--177}.
\bibitem[{Albreiki et~al.(2022)Albreiki, Habuza and Zaki}]{albreiki22}
\bibinfo{author}{Albreiki, B.}, \bibinfo{author}{Habuza, T.},
  \bibinfo{author}{Zaki, N.}, \bibinfo{year}{2022}.
\newblock \bibinfo{title}{Framework for automatically suggesting remedial
  actions to help students at risk based on explainable ml and rule-based
  models}.
\newblock \bibinfo{journal}{International Journal of Educational Technology in
  Higher Education} \bibinfo{volume}{19}, \bibinfo{pages}{49}.
\bibitem[{Ali et~al.(2023)Ali, Hamraz, Gul, Khan, Aldahmani and Khan}]{ali23}
\bibinfo{author}{Ali, A.}, \bibinfo{author}{Hamraz, M.}, \bibinfo{author}{Gul,
  N.}, \bibinfo{author}{Khan, D.M.}, \bibinfo{author}{Aldahmani, S.},
  \bibinfo{author}{Khan, Z.}, \bibinfo{year}{2023}.
\newblock \bibinfo{title}{A k nearest neighbour ensemble via extended
  neighbourhood rule and feature subsets}.
\newblock \bibinfo{journal}{Pattern Recognition} \bibinfo{volume}{142},
  \bibinfo{pages}{109641}.
\bibitem[{Ali et~al.(2024)Ali, Khan, Khan and Aldahmani}]{ali24}
\bibinfo{author}{Ali, A.}, \bibinfo{author}{Khan, Z.}, \bibinfo{author}{Khan,
  D.M.}, \bibinfo{author}{Aldahmani, S.}, \bibinfo{year}{2024}.
\newblock \bibinfo{title}{An optimal random projection k nearest neighbors
  ensemble via extended neighborhood rule for binary classification}.
\newblock \bibinfo{journal}{IEEE Access} \bibinfo{volume}{12},
  \bibinfo{pages}{61401--61409}.
\bibitem[{Ali et~al.(2025)Ali, Muse, Abdi, Ali, Muse and Cumar}]{ali25}
\bibinfo{author}{Ali, J.A.}, \bibinfo{author}{Muse, A.H.},
  \bibinfo{author}{Abdi, M.K.}, \bibinfo{author}{Ali, T.A.},
  \bibinfo{author}{Muse, Y.H.}, \bibinfo{author}{Cumar, M.A.},
  \bibinfo{year}{2025}.
\newblock \bibinfo{title}{Machine learning-driven analysis of academic
  performance determinants: Geographic, socio-demographic, and subject-specific
  influences in somaliland's 2022--2023 national primary examinations}.
\newblock \bibinfo{journal}{International Journal of Educational Research Open}
  \bibinfo{volume}{8}, \bibinfo{pages}{100426}.
\bibitem[{Alves and Candido(2020)}]{alves20}
\bibinfo{author}{Alves, F.A.}, \bibinfo{author}{Candido, O.},
  \bibinfo{year}{2020}.
\newblock \bibinfo{title}{School effect and student performance: a latin
  american assessment from pisa}.
\newblock \bibinfo{journal}{Econom{\'\i}a} \bibinfo{volume}{43},
  \bibinfo{pages}{79--99}.
\bibitem[{Angrist et~al.(2021)Angrist, Djankov, Goldberg and
  Patrinos}]{angrist21}
\bibinfo{author}{Angrist, N.}, \bibinfo{author}{Djankov, S.},
  \bibinfo{author}{Goldberg, P.K.}, \bibinfo{author}{Patrinos, H.A.},
  \bibinfo{year}{2021}.
\newblock \bibinfo{title}{Measuring human capital using global learning data}.
\newblock \bibinfo{journal}{Nature} \bibinfo{volume}{592},
  \bibinfo{pages}{403--408}.
\bibitem[{Arias~Ortiz et~al.(2023)Arias~Ortiz, Bos, Giambruno and
  Zoido}]{arias23}
\bibinfo{author}{Arias~Ortiz, E.}, \bibinfo{author}{Bos, M.},
  \bibinfo{author}{Giambruno, C.}, \bibinfo{author}{Zoido, P.},
  \bibinfo{year}{2023}.
\newblock \bibinfo{title}{Latin america and the caribbean in pisa 2022: How
  many students are low performers?}
\bibitem[{Arias~Ortiz et~al.(2024a)Arias~Ortiz, Bos, Chen~Peraza, Giambruno,
  Levin, Oubi{\~n}a, Pineda and Zoido}]{arias24a}
\bibinfo{author}{Arias~Ortiz, E.}, \bibinfo{author}{Bos, M.S.},
  \bibinfo{author}{Chen~Peraza, J.}, \bibinfo{author}{Giambruno, C.},
  \bibinfo{author}{Levin, V.}, \bibinfo{author}{Oubi{\~n}a, V.},
  \bibinfo{author}{Pineda, J.A.}, \bibinfo{author}{Zoido, P.},
  \bibinfo{year}{2024}a.
\newblock \bibinfo{title}{Learning can’t wait [el aprendizaje no puede
  esperar]}.
\bibitem[{Arias~Ortiz et~al.(2024b)Arias~Ortiz, Due{\~n}as, Giambruno and
  L{\'o}pez}]{arias24b}
\bibinfo{author}{Arias~Ortiz, E.}, \bibinfo{author}{Due{\~n}as, X.},
  \bibinfo{author}{Giambruno, C.}, \bibinfo{author}{L{\'o}pez, {\'A}.},
  \bibinfo{year}{2024}b.
\newblock \bibinfo{title}{The State of Education in Latin America and the
  Caribbean: Learning Assessments}.
\newblock \bibinfo{type}{Technical Report}. Inter-American Development Bank.
\bibitem[{Asad et~al.(2023)Asad, Altaf, Ahmad, Mahmoud, Huda and
  Iqbal}]{asad23}
\bibinfo{author}{Asad, R.}, \bibinfo{author}{Altaf, S.},
  \bibinfo{author}{Ahmad, S.}, \bibinfo{author}{Mahmoud, H.},
  \bibinfo{author}{Huda, S.}, \bibinfo{author}{Iqbal, S.},
  \bibinfo{year}{2023}.
\newblock \bibinfo{title}{Machine learning-based hybrid ensemble model
  achieving precision education for online education amid the lockdown period
  of covid-19 pandemic in pakistan}.
\newblock \bibinfo{journal}{Sustainability} \bibinfo{volume}{15},
  \bibinfo{pages}{5431}.
\bibitem[{Asadullah et~al.(2023)Asadullah, Bouhlila, Chan, Draxler, Ha,
  Heyneman, Luschei, Semela and Yemini}]{asadullah23}
\bibinfo{author}{Asadullah, M.N.}, \bibinfo{author}{Bouhlila, D.S.},
  \bibinfo{author}{Chan, S.J.}, \bibinfo{author}{Draxler, A.},
  \bibinfo{author}{Ha, W.}, \bibinfo{author}{Heyneman, S.P.},
  \bibinfo{author}{Luschei, T.F.}, \bibinfo{author}{Semela, T.},
  \bibinfo{author}{Yemini, M.}, \bibinfo{year}{2023}.
\newblock \bibinfo{title}{A year of missed opportunity: Post-covid learning
  loss--a renewed call to action}.
\newblock \bibinfo{journal}{International Journal of Educational Development}
  \bibinfo{volume}{99}, \bibinfo{pages}{102770}.
\bibitem[{Athey and Imbens(2019)}]{athey19}
\bibinfo{author}{Athey, S.}, \bibinfo{author}{Imbens, G.W.},
  \bibinfo{year}{2019}.
\newblock \bibinfo{title}{Machine learning methods that economists should know
  about}.
\newblock \bibinfo{journal}{Annual Review of Economics} \bibinfo{volume}{11},
  \bibinfo{pages}{685--725}.
\bibitem[{Azevedo et~al.(2021)Azevedo, Hasan, Goldemberg, Geven and
  Iqbal}]{azevedo21}
\bibinfo{author}{Azevedo, J.P.}, \bibinfo{author}{Hasan, A.},
  \bibinfo{author}{Goldemberg, D.}, \bibinfo{author}{Geven, K.},
  \bibinfo{author}{Iqbal, S.A.}, \bibinfo{year}{2021}.
\newblock \bibinfo{title}{Simulating the potential impacts of covid-19 school
  closures on schooling and learning outcomes: A set of global estimates}.
\newblock \bibinfo{journal}{The World Bank Research Observer}
  \bibinfo{volume}{36}, \bibinfo{pages}{1--40}.
\bibitem[{Berkowitz et~al.(2017)Berkowitz, Moore, Astor and
  Benbenishty}]{berkowitz17}
\bibinfo{author}{Berkowitz, R.}, \bibinfo{author}{Moore, H.},
  \bibinfo{author}{Astor, R.A.}, \bibinfo{author}{Benbenishty, R.},
  \bibinfo{year}{2017}.
\newblock \bibinfo{title}{A research synthesis of the associations between
  socioeconomic background, inequality, school climate, and academic
  achievement}.
\newblock \bibinfo{journal}{Review of educational research}
  \bibinfo{volume}{87}, \bibinfo{pages}{425--469}.
\bibitem[{Berlanga et~al.(2020)Berlanga, Morduchowicz, Scasso and
  Vera}]{berlanga20}
\bibinfo{author}{Berlanga, C.}, \bibinfo{author}{Morduchowicz, A.},
  \bibinfo{author}{Scasso, M.}, \bibinfo{author}{Vera, A.},
  \bibinfo{year}{2020}.
\newblock \bibinfo{title}{Reabrir las escuelas en am{\'e}rica latina y el
  caribe: Claves, desaf{\'\i}os y dilemas para planificar el retorno seguro a
  las clases presenciales}.
\newblock \bibinfo{journal}{Banco Interamericano de Desarrollo}
  \bibinfo{volume}{10}, \bibinfo{pages}{0002906}.
\bibitem[{Betth{\"a}user et~al.(2023)Betth{\"a}user, Bach-Mortensen and
  Engzell}]{betthauser23}
\bibinfo{author}{Betth{\"a}user, B.A.}, \bibinfo{author}{Bach-Mortensen, A.M.},
  \bibinfo{author}{Engzell, P.}, \bibinfo{year}{2023}.
\newblock \bibinfo{title}{A systematic review and meta-analysis of the evidence
  on learning during the covid-19 pandemic}.
\newblock \bibinfo{journal}{Nature human behaviour} \bibinfo{volume}{7},
  \bibinfo{pages}{375--385}.
\bibitem[{Borchers and da~Cunha(2025)}]{borchers25}
\bibinfo{author}{Borchers, J.}, \bibinfo{author}{da~Cunha, M.S.},
  \bibinfo{year}{2025}.
\newblock \bibinfo{title}{School performance and inequality of opportunities in
  latin america}.
\newblock \bibinfo{journal}{Studies in Educational Evaluation}
  \bibinfo{volume}{86}, \bibinfo{pages}{101497}.
\bibitem[{Bracco et~al.(2025)Bracco, Ciaschi, Gasparini, Marchionni and
  Neidh{\"o}fer}]{bracco25}
\bibinfo{author}{Bracco, J.}, \bibinfo{author}{Ciaschi, M.},
  \bibinfo{author}{Gasparini, L.}, \bibinfo{author}{Marchionni, M.},
  \bibinfo{author}{Neidh{\"o}fer, G.}, \bibinfo{year}{2025}.
\newblock \bibinfo{title}{The impact of covid-19 on education in latin america:
  Long-run implications for poverty and inequality}.
\newblock \bibinfo{journal}{Review of Income and Wealth} \bibinfo{volume}{71},
  \bibinfo{pages}{e12687}.
\bibitem[{Brand et~al.(2023)Brand, Zhou and Xie}]{brand23}
\bibinfo{author}{Brand, J.E.}, \bibinfo{author}{Zhou, X.},
  \bibinfo{author}{Xie, Y.}, \bibinfo{year}{2023}.
\newblock \bibinfo{title}{Recent developments in causal inference and machine
  learning}.
\newblock \bibinfo{journal}{Annual Review of Sociology} \bibinfo{volume}{49},
  \bibinfo{pages}{81--110}.
\bibitem[{Breton and Canavire-Bacarreza(2018)}]{breton18}
\bibinfo{author}{Breton, T.R.}, \bibinfo{author}{Canavire-Bacarreza, G.},
  \bibinfo{year}{2018}.
\newblock \bibinfo{title}{Low test scores in latin america: poor schools, poor
  families or something else?}
\newblock \bibinfo{journal}{Compare: A Journal of Comparative and International
  Education} \bibinfo{volume}{48}, \bibinfo{pages}{733--748}.
\bibitem[{Brunori et~al.(2024)Brunori, Ferreira and Neidh{\"o}fer}]{brunori24}
\bibinfo{author}{Brunori, P.}, \bibinfo{author}{Ferreira, F.H.},
  \bibinfo{author}{Neidh{\"o}fer, G.}, \bibinfo{year}{2024}.
\newblock \bibinfo{title}{Inequality of opportunity and intergenerational
  persistence in Latin America}.
\newblock \bibinfo{type}{Technical Report} \bibinfo{number}{17202}. IZA
  Discussion Papers.
\bibitem[{Bu and Chen(2023)}]{bu23}
\bibinfo{author}{Bu, Y.}, \bibinfo{author}{Chen, F.}, \bibinfo{year}{2023}.
\newblock \bibinfo{title}{What key contextual factors contribute to students’
  reading literacy among top-performing countries and economies? statistical
  and machine learning analyses}.
\newblock \bibinfo{journal}{International Journal of Educational Research}
  \bibinfo{volume}{122}, \bibinfo{pages}{102267}.
\bibitem[{CEPAL(2020)}]{cepal20}
\bibinfo{author}{CEPAL}, \bibinfo{year}{2020}.
\newblock \bibinfo{title}{Universalizar el acceso a las tecnologías digitales
  para enfrentar los efectos del covid-19}.
\bibitem[{Ch{\'a}vez et~al.(2021)Ch{\'a}vez, Cebotari, Jos{\'e}~Ben{\'\i}tez,
  Richardson, Chii and Zapata}]{chavez21}
\bibinfo{author}{Ch{\'a}vez, C.}, \bibinfo{author}{Cebotari, V.},
  \bibinfo{author}{Jos{\'e}~Ben{\'\i}tez, M.}, \bibinfo{author}{Richardson,
  D.}, \bibinfo{author}{Chii, F.H.}, \bibinfo{author}{Zapata, J.},
  \bibinfo{year}{2021}.
\newblock \bibinfo{title}{School-related violence in latin america and the
  caribbean: Building an evidence base for stronger schools}.
\bibitem[{Chen et~al.(2023)Chen, Covert, Lundberg and Lee}]{chen23}
\bibinfo{author}{Chen, H.}, \bibinfo{author}{Covert, I.C.},
  \bibinfo{author}{Lundberg, S.M.}, \bibinfo{author}{Lee, S.I.},
  \bibinfo{year}{2023}.
\newblock \bibinfo{title}{Algorithms to estimate shapley value feature
  attributions}.
\newblock \bibinfo{journal}{Nature Machine Intelligence} \bibinfo{volume}{5},
  \bibinfo{pages}{590--601}.
\bibitem[{Chen et~al.(2020)Chen, Janizek, Lundberg and Lee}]{chen20}
\bibinfo{author}{Chen, H.}, \bibinfo{author}{Janizek, J.D.},
  \bibinfo{author}{Lundberg, S.}, \bibinfo{author}{Lee, S.I.},
  \bibinfo{year}{2020}.
\newblock \bibinfo{title}{True to the model or true to the data?}
\newblock \bibinfo{journal}{arXiv preprint arXiv:2006.16234} .
\bibitem[{Chen et~al.(2022)Chen, Li, Kim, Plumb and Talwalkar}]{chen22}
\bibinfo{author}{Chen, V.}, \bibinfo{author}{Li, J.}, \bibinfo{author}{Kim,
  J.S.}, \bibinfo{author}{Plumb, G.}, \bibinfo{author}{Talwalkar, A.},
  \bibinfo{year}{2022}.
\newblock \bibinfo{title}{Interpretable machine learning: Moving from mythos to
  diagnostics}.
\newblock \bibinfo{journal}{Communications of the ACM} \bibinfo{volume}{65},
  \bibinfo{pages}{43--50}.
\bibitem[{Chytas et~al.(2023)Chytas, Tsolakidis, Triperina and
  Skourlas}]{chytas23}
\bibinfo{author}{Chytas, K.}, \bibinfo{author}{Tsolakidis, A.},
  \bibinfo{author}{Triperina, E.}, \bibinfo{author}{Skourlas, C.},
  \bibinfo{year}{2023}.
\newblock \bibinfo{title}{Educational data mining in the academic setting:
  employing the data produced by blended learning to ameliorate the learning
  process}.
\newblock \bibinfo{journal}{Data Technologies and Applications}
  \bibinfo{volume}{57}, \bibinfo{pages}{366--384}.
\bibitem[{Codiroli~Mcmaster and Cook(2019)}]{codiroli19}
\bibinfo{author}{Codiroli~Mcmaster, N.}, \bibinfo{author}{Cook, R.},
  \bibinfo{year}{2019}.
\newblock \bibinfo{title}{The contribution of intersectionality to quantitative
  research into educational inequalities}.
\newblock \bibinfo{journal}{Review of Education} \bibinfo{volume}{7},
  \bibinfo{pages}{271--292}.
\bibitem[{Cortina(2017)}]{cortina17}
\bibinfo{author}{Cortina, R.}, \bibinfo{year}{2017}.
\newblock \bibinfo{title}{How to improve quality education for indigenous
  children in latin america}.
\newblock \bibinfo{journal}{Indigenous education policy, equity, and
  intercultural understanding in Latin America} , \bibinfo{pages}{3--25}.
\bibitem[{Covert et~al.(2021)Covert, Lundberg and Lee}]{covert21}
\bibinfo{author}{Covert, I.}, \bibinfo{author}{Lundberg, S.},
  \bibinfo{author}{Lee, S.I.}, \bibinfo{year}{2021}.
\newblock \bibinfo{title}{Explaining by removing: A unified framework for model
  explanation}.
\newblock \bibinfo{journal}{Journal of Machine Learning Research}
  \bibinfo{volume}{22}, \bibinfo{pages}{1--90}.
\bibitem[{Cristia et~al.(2017)Cristia, Ibarrar{\'a}n, Cueto, Santiago and
  Sever{\'\i}n}]{cristia17}
\bibinfo{author}{Cristia, J.}, \bibinfo{author}{Ibarrar{\'a}n, P.},
  \bibinfo{author}{Cueto, S.}, \bibinfo{author}{Santiago, A.},
  \bibinfo{author}{Sever{\'\i}n, E.}, \bibinfo{year}{2017}.
\newblock \bibinfo{title}{Technology and child development: Evidence from the
  one laptop per child program}.
\newblock \bibinfo{journal}{American Economic Journal: Applied Economics}
  \bibinfo{volume}{9}, \bibinfo{pages}{295--320}.
\bibitem[{Delprato(2019)}]{delprato19}
\bibinfo{author}{Delprato, M.}, \bibinfo{year}{2019}.
\newblock \bibinfo{title}{Parental education expectations and achievement for
  indigenous students in latin america: Evidence from terce learning survey}.
\newblock \bibinfo{journal}{International Journal of Educational Development}
  \bibinfo{volume}{65}, \bibinfo{pages}{10--25}.
\bibitem[{Delprato(2021)}]{delprato21}
\bibinfo{author}{Delprato, M.}, \bibinfo{year}{2021}.
\newblock \bibinfo{title}{Indigenous learning gaps and home language
  instruction: New evidence from pisa-d}.
\newblock \bibinfo{journal}{International Journal of Educational Research}
  \bibinfo{volume}{109}, \bibinfo{pages}{101800}.
\bibitem[{Delprato(2024)}]{delprato24}
\bibinfo{author}{Delprato, M.}, \bibinfo{year}{2024}.
\newblock \bibinfo{title}{A conceptual framework to assess missing data for sdg
  4}, in: \bibinfo{booktitle}{Achieving Equitable Education}.
  \bibinfo{publisher}{Edward Elgar Publishing}, pp. \bibinfo{pages}{14--31}.
\bibitem[{Delprato(2025)}]{delprato25b}
\bibinfo{author}{Delprato, M.}, \bibinfo{year}{2025}.
\newblock \bibinfo{title}{Pedagogic engagement in argentina for students
  attending primary and secondary levels--an empirical analysis of main
  barriers during the covid-19 pandemic}.
\newblock \bibinfo{journal}{OSF Preprints}
  \DOIprefix\doi{https://osf.io/preprints/osf/37hfb_v1}.
\bibitem[{Delprato et~al.(2017)Delprato, Akyeampong and Dunne}]{delprato17}
\bibinfo{author}{Delprato, M.}, \bibinfo{author}{Akyeampong, K.},
  \bibinfo{author}{Dunne, M.}, \bibinfo{year}{2017}.
\newblock \bibinfo{title}{The impact of bullying on students’ learning in
  latin america: A matching approach for 15 countries}.
\newblock \bibinfo{journal}{International journal of educational development}
  \bibinfo{volume}{52}, \bibinfo{pages}{37--57}.
\bibitem[{Delprato and Antequera(2025)}]{delprato25}
\bibinfo{author}{Delprato, M.}, \bibinfo{author}{Antequera, G.},
  \bibinfo{year}{2025}.
\newblock \bibinfo{title}{School efficiency in latin america before and after
  the covid-19 pandemic: New evidence from pisa 2018 and 2022}.
\newblock \bibinfo{journal}{International Journal of Educational Research}
  \bibinfo{volume}{129}, \bibinfo{pages}{102493}.
\bibitem[{Delprato et~al.(2022)Delprato, Frola and Antequera}]{delprato22}
\bibinfo{author}{Delprato, M.}, \bibinfo{author}{Frola, A.},
  \bibinfo{author}{Antequera, G.}, \bibinfo{year}{2022}.
\newblock \bibinfo{title}{Indigenous and non-indigenous proficiency gaps for
  out-of-school and in-school populations: A machine learning approach}.
\newblock \bibinfo{journal}{International Journal of Educational Development}
  \bibinfo{volume}{93}, \bibinfo{pages}{102631}.
\bibitem[{Delprato et~al.(2015)Delprato, K{\"o}seleci and
  Antequera}]{delprato15}
\bibinfo{author}{Delprato, M.}, \bibinfo{author}{K{\"o}seleci, N.},
  \bibinfo{author}{Antequera, G.}, \bibinfo{year}{2015}.
\newblock \bibinfo{title}{Educaci{\'o}n para todos en am{\'e}rica latina:
  Evoluci{\'o}n del impacto de la desigualdad escolar en los resultados
  educativos}.
\newblock \bibinfo{journal}{Revista latinoamericana de educaci{\'o}n comparada}
  \bibinfo{volume}{6}, \bibinfo{pages}{45--75}.
\bibitem[{Dhal and Azad(2022)}]{dhal22}
\bibinfo{author}{Dhal, P.}, \bibinfo{author}{Azad, C.}, \bibinfo{year}{2022}.
\newblock \bibinfo{title}{A comprehensive survey on feature selection in the
  various fields of machine learning}.
\newblock \bibinfo{journal}{Applied Intelligence} \bibinfo{volume}{52},
  \bibinfo{pages}{4543--4581}.
\bibitem[{D{\'\i}az et~al.(2022)D{\'\i}az, Dodel and Menese}]{diaz22}
\bibinfo{author}{D{\'\i}az, C.}, \bibinfo{author}{Dodel, M.},
  \bibinfo{author}{Menese, P.}, \bibinfo{year}{2022}.
\newblock \bibinfo{title}{Can one laptop per child reduce digital inequalities?
  ict household access patterns under plan ceibal}.
\newblock \bibinfo{journal}{Telecommunications Policy} \bibinfo{volume}{46},
  \bibinfo{pages}{102406}.
\bibitem[{Dol and Jawandhiya(2023)}]{dol23}
\bibinfo{author}{Dol, S.M.}, \bibinfo{author}{Jawandhiya, P.M.},
  \bibinfo{year}{2023}.
\newblock \bibinfo{title}{Classification technique and its combination with
  clustering and association rule mining in educational data mining—a
  survey}.
\newblock \bibinfo{journal}{Engineering applications of artificial
  intelligence} \bibinfo{volume}{122}, \bibinfo{pages}{106071}.
\bibitem[{Draeger(2021)}]{draeger21}
\bibinfo{author}{Draeger, E.}, \bibinfo{year}{2021}.
\newblock \bibinfo{title}{Do conditional cash transfers increase schooling
  among adolescents? evidence from brazil}.
\newblock \bibinfo{journal}{International Economics and Economic Policy}
  \bibinfo{volume}{18}, \bibinfo{pages}{743--766}.
\bibitem[{Elbasi et~al.(2025)Elbasi, Nadeem, Alzoubi, Topcu and
  Varghese}]{elbasi25}
\bibinfo{author}{Elbasi, E.}, \bibinfo{author}{Nadeem, M.},
  \bibinfo{author}{Alzoubi, Y.I.}, \bibinfo{author}{Topcu, A.E.},
  \bibinfo{author}{Varghese, G.}, \bibinfo{year}{2025}.
\newblock \bibinfo{title}{Machine learning in education: Innovations, impacts,
  and ethical considerations}.
\newblock \bibinfo{journal}{IEEE Access} .
\bibitem[{Ersozlu et~al.(2024)Ersozlu, Taheri and Koch}]{ersozlu24}
\bibinfo{author}{Ersozlu, Z.}, \bibinfo{author}{Taheri, S.},
  \bibinfo{author}{Koch, I.}, \bibinfo{year}{2024}.
\newblock \bibinfo{title}{A review of machine learning methods used for
  educational data}.
\newblock \bibinfo{journal}{Education and Information Technologies} ,
  \bibinfo{pages}{1--21}.
\bibitem[{Fern{\'a}ndez et~al.(2024)Fern{\'a}ndez, Pag{\'e}s, Szekely and
  Acevedo}]{fernandez24}
\bibinfo{author}{Fern{\'a}ndez, R.}, \bibinfo{author}{Pag{\'e}s, C.},
  \bibinfo{author}{Szekely, M.}, \bibinfo{author}{Acevedo, I.},
  \bibinfo{year}{2024}.
\newblock \bibinfo{title}{Education inequalities in Latin America and the
  Caribbean}.
\newblock \bibinfo{type}{Technical Report}. National Bureau of Economic
  Research.
\bibitem[{Gaentzsch(2020)}]{gaentzsch20}
\bibinfo{author}{Gaentzsch, A.}, \bibinfo{year}{2020}.
\newblock \bibinfo{title}{Do conditional cash transfers (ccts) raise
  educational attainment? an impact evaluation of juntos in peru}.
\newblock \bibinfo{journal}{Development Policy Review} \bibinfo{volume}{38},
  \bibinfo{pages}{747--765}.
\bibitem[{G{\'o}mez and Andr{\'e}s Uz{\'\i}n~P(2022)}]{gomez22}
\bibinfo{author}{G{\'o}mez, G.M.}, \bibinfo{author}{Andr{\'e}s Uz{\'\i}n~P,
  G.}, \bibinfo{year}{2022}.
\newblock \bibinfo{title}{Effects of covid-19 on education and schools’
  reopening in latin america}.
\newblock \bibinfo{journal}{COVID-19 and international development} ,
  \bibinfo{pages}{119--135}.
\bibitem[{G{\'o}mez-Talal et~al.(2024)G{\'o}mez-Talal, Bote-Curiel and
  Rojo-{\'A}lvarez}]{gomez24}
\bibinfo{author}{G{\'o}mez-Talal, I.}, \bibinfo{author}{Bote-Curiel, L.},
  \bibinfo{author}{Rojo-{\'A}lvarez, J.L.}, \bibinfo{year}{2024}.
\newblock \bibinfo{title}{Understanding the disparities in mathematics
  performance: An interpretability-based examination}.
\newblock \bibinfo{journal}{Engineering Applications of Artificial
  Intelligence} \bibinfo{volume}{133}, \bibinfo{pages}{108109}.
\bibitem[{Gonzalez(2021)}]{gonzalez21}
\bibinfo{author}{Gonzalez, O.}, \bibinfo{year}{2021}.
\newblock \bibinfo{title}{Psychometric and machine learning approaches for
  diagnostic assessment and tests of individual classification.}
\newblock \bibinfo{journal}{Psychological Methods} \bibinfo{volume}{26},
  \bibinfo{pages}{236}.
\bibitem[{Grimmer et~al.(2021)Grimmer, Roberts and Stewart}]{grimmer21}
\bibinfo{author}{Grimmer, J.}, \bibinfo{author}{Roberts, M.E.},
  \bibinfo{author}{Stewart, B.M.}, \bibinfo{year}{2021}.
\newblock \bibinfo{title}{Machine learning for social science: An agnostic
  approach}.
\newblock \bibinfo{journal}{Annual Review of Political Science}
  \bibinfo{volume}{24}, \bibinfo{pages}{395--419}.
\bibitem[{Haimovich et~al.(2021)Haimovich, Vazquez and Adelman}]{haimovich21}
\bibinfo{author}{Haimovich, F.}, \bibinfo{author}{Vazquez, E.},
  \bibinfo{author}{Adelman, M.}, \bibinfo{year}{2021}.
\newblock \bibinfo{title}{Scalable early warning systems for school dropout
  prevention: Evidence from a 4.000-school randomized controlled trial}.
\bibitem[{Hanushek and Woessmann(2023)}]{hanushek23}
\bibinfo{author}{Hanushek, E.A.}, \bibinfo{author}{Woessmann, L.},
  \bibinfo{year}{2023}.
\newblock \bibinfo{title}{The knowledge capital of nations: Education and the
  economics of growth}.
\newblock \bibinfo{publisher}{MIT press}.
\bibitem[{Hastie et~al.(2009)Hastie, Tibshirani and Friedman}]{hastie09}
\bibinfo{author}{Hastie, T.}, \bibinfo{author}{Tibshirani, R.},
  \bibinfo{author}{Friedman, J.H.}, \bibinfo{year}{2009}.
\newblock \bibinfo{title}{The elements of statistical learning: data mining,
  inference, and prediction}. volume~\bibinfo{volume}{2}.
\newblock \bibinfo{publisher}{Springer}.
\bibitem[{Hastie et~al.(2015)Hastie, Tibshirani and Wainwright}]{hastie15}
\bibinfo{author}{Hastie, T.}, \bibinfo{author}{Tibshirani, R.},
  \bibinfo{author}{Wainwright, M.}, \bibinfo{year}{2015}.
\newblock \bibinfo{title}{Statistical learning with sparsity}.
\newblock \bibinfo{journal}{Monographs on statistics and applied probability}
  \bibinfo{volume}{143}, \bibinfo{pages}{8}.
\bibitem[{Hellas et~al.(2018)Hellas, Ihantola, Petersen, Ajanovski, Gutica,
  Hynninen, Knutas, Leinonen, Messom and Liao}]{hellas18}
\bibinfo{author}{Hellas, A.}, \bibinfo{author}{Ihantola, P.},
  \bibinfo{author}{Petersen, A.}, \bibinfo{author}{Ajanovski, V.V.},
  \bibinfo{author}{Gutica, M.}, \bibinfo{author}{Hynninen, T.},
  \bibinfo{author}{Knutas, A.}, \bibinfo{author}{Leinonen, J.},
  \bibinfo{author}{Messom, C.}, \bibinfo{author}{Liao, S.N.},
  \bibinfo{year}{2018}.
\newblock \bibinfo{title}{Predicting academic performance: a systematic
  literature review}, in: \bibinfo{booktitle}{Proceedings companion of the 23rd
  annual ACM conference on innovation and technology in computer science
  education}, pp. \bibinfo{pages}{175--199}.
\bibitem[{Hilbert et~al.(2021)Hilbert, Coors, Kraus, Bischl, Lindl, Frei, Wild,
  Krauss, Goretzko and Stachl}]{hilbert21}
\bibinfo{author}{Hilbert, S.}, \bibinfo{author}{Coors, S.},
  \bibinfo{author}{Kraus, E.}, \bibinfo{author}{Bischl, B.},
  \bibinfo{author}{Lindl, A.}, \bibinfo{author}{Frei, M.},
  \bibinfo{author}{Wild, J.}, \bibinfo{author}{Krauss, S.},
  \bibinfo{author}{Goretzko, D.}, \bibinfo{author}{Stachl, C.},
  \bibinfo{year}{2021}.
\newblock \bibinfo{title}{Machine learning for the educational sciences}.
\newblock \bibinfo{journal}{Review of Education} \bibinfo{volume}{9},
  \bibinfo{pages}{e3310}.
\bibitem[{Holzinger et~al.(2020)Holzinger, Saranti, Molnar, Biecek and
  Samek}]{holzinger20}
\bibinfo{author}{Holzinger, A.}, \bibinfo{author}{Saranti, A.},
  \bibinfo{author}{Molnar, C.}, \bibinfo{author}{Biecek, P.},
  \bibinfo{author}{Samek, W.}, \bibinfo{year}{2020}.
\newblock \bibinfo{title}{Explainable ai methods-a brief overview}, in:
  \bibinfo{booktitle}{International workshop on extending explainable AI beyond
  deep models and classifiers}, \bibinfo{organization}{Springer}. pp.
  \bibinfo{pages}{13--38}.
\bibitem[{Huang et~al.(2024)Huang, Zhou, Chen and Wu}]{huang24}
\bibinfo{author}{Huang, Y.}, \bibinfo{author}{Zhou, Y.}, \bibinfo{author}{Chen,
  J.}, \bibinfo{author}{Wu, D.}, \bibinfo{year}{2024}.
\newblock \bibinfo{title}{Applying machine learning and shap method to identify
  key influences on middle-school students’ mathematics literacy
  performance}.
\newblock \bibinfo{journal}{Journal of Intelligence} \bibinfo{volume}{12},
  \bibinfo{pages}{93}.
\bibitem[{Jayachandran et~al.(2023)Jayachandran, Biradavolu and
  Cooper}]{jayachandran23}
\bibinfo{author}{Jayachandran, S.}, \bibinfo{author}{Biradavolu, M.},
  \bibinfo{author}{Cooper, J.}, \bibinfo{year}{2023}.
\newblock \bibinfo{title}{Using machine learning and qualitative interviews to
  design a five-question survey module for women’s agency}.
\newblock \bibinfo{journal}{World Development} \bibinfo{volume}{161},
  \bibinfo{pages}{106076}.
\bibitem[{Karalar et~al.(2021)Karalar, Kapucu and G{\"u}r{\"u}ler}]{Karalar21}
\bibinfo{author}{Karalar, H.}, \bibinfo{author}{Kapucu, C.},
  \bibinfo{author}{G{\"u}r{\"u}ler, H.}, \bibinfo{year}{2021}.
\newblock \bibinfo{title}{Predicting students at risk of academic failure using
  ensemble model during pandemic in a distance learning system}.
\newblock \bibinfo{journal}{International Journal of Educational Technology in
  Higher Education} \bibinfo{volume}{18}, \bibinfo{pages}{63}.
\bibitem[{Khan et~al.(2024)Khan, Ali, Khan and Aldahmani}]{khan24}
\bibinfo{author}{Khan, Z.}, \bibinfo{author}{Ali, A.}, \bibinfo{author}{Khan,
  D.M.}, \bibinfo{author}{Aldahmani, S.}, \bibinfo{year}{2024}.
\newblock \bibinfo{title}{Regularized ensemble learning for prediction and risk
  factors assessment of students at risk in the post-covid era}.
\newblock \bibinfo{journal}{Scientific reports} \bibinfo{volume}{14},
  \bibinfo{pages}{16200}.
\bibitem[{Kr{\"u}ger(2019)}]{kruger19}
\bibinfo{author}{Kr{\"u}ger, N.S.}, \bibinfo{year}{2019}.
\newblock \bibinfo{title}{La segregaci{\'o}n por nivel socioecon{\'o}mico como
  dimensi{\'o}n de la exclusi{\'o}n educativa: 15 a{\~n}os de evoluci{\'o}n en
  am{\'e}rica latina}.
\bibitem[{Van~der Laan et~al.(2007)Van~der Laan, Polley and
  Hubbard}]{vanderlaan07}
\bibinfo{author}{Van~der Laan, M.J.}, \bibinfo{author}{Polley, E.C.},
  \bibinfo{author}{Hubbard, A.E.}, \bibinfo{year}{2007}.
\newblock \bibinfo{title}{Super learner}.
\newblock \bibinfo{journal}{Statistical applications in genetics and molecular
  biology} \bibinfo{volume}{6}.
\bibitem[{Lara and Saracostti(2019)}]{lara19}
\bibinfo{author}{Lara, L.}, \bibinfo{author}{Saracostti, M.},
  \bibinfo{year}{2019}.
\newblock \bibinfo{title}{Effect of parental involvement on children’s
  academic achievement in chile}.
\newblock \bibinfo{journal}{Frontiers in psychology} \bibinfo{volume}{10},
  \bibinfo{pages}{1464}.
\bibitem[{Linardatos et~al.(2020)Linardatos, Papastefanopoulos and
  Kotsiantis}]{linardatos20}
\bibinfo{author}{Linardatos, P.}, \bibinfo{author}{Papastefanopoulos, V.},
  \bibinfo{author}{Kotsiantis, S.}, \bibinfo{year}{2020}.
\newblock \bibinfo{title}{Explainable ai: A review of machine learning
  interpretability methods}.
\newblock \bibinfo{journal}{Entropy} \bibinfo{volume}{23}, \bibinfo{pages}{18}.
\bibitem[{Lisboa et~al.(2023)Lisboa, Saralajew, Vellido, Fern{\'a}ndez-Domenech
  and Villmann}]{lisboa23}
\bibinfo{author}{Lisboa, P.J.}, \bibinfo{author}{Saralajew, S.},
  \bibinfo{author}{Vellido, A.}, \bibinfo{author}{Fern{\'a}ndez-Domenech, R.},
  \bibinfo{author}{Villmann, T.}, \bibinfo{year}{2023}.
\newblock \bibinfo{title}{The coming of age of interpretable and explainable
  machine learning models}.
\newblock \bibinfo{journal}{Neurocomputing} \bibinfo{volume}{535},
  \bibinfo{pages}{25--39}.
\bibitem[{Liu et~al.(2022)Liu, Chen and Liu}]{liu22}
\bibinfo{author}{Liu, H.}, \bibinfo{author}{Chen, X.}, \bibinfo{author}{Liu,
  X.}, \bibinfo{year}{2022}.
\newblock \bibinfo{title}{Factors influencing secondary school students’
  reading literacy: An analysis based on xgboost and shap methods}.
\newblock \bibinfo{journal}{Frontiers in Psychology} \bibinfo{volume}{13},
  \bibinfo{pages}{948612}.
\bibitem[{L{\'o}pez(2021)}]{lopez21}
\bibinfo{author}{L{\'o}pez, L.E.}, \bibinfo{year}{2021}.
\newblock \bibinfo{title}{What is educaci{\'o}n intercultural biling{\"u}e in
  latin america nowadays: results and challenges}.
\newblock \bibinfo{journal}{Journal of Multilingual and Multicultural
  Development} \bibinfo{volume}{42}, \bibinfo{pages}{955--968}.
\bibitem[{Lundberg and Lee(2017)}]{lundberg17}
\bibinfo{author}{Lundberg, S.M.}, \bibinfo{author}{Lee, S.I.},
  \bibinfo{year}{2017}.
\newblock \bibinfo{title}{A unified approach to interpreting model
  predictions}, in: \bibinfo{booktitle}{Proceedings of the 31st International
  Conference on Neural Information Processing Systems},
  \bibinfo{publisher}{Curran Associates Inc.}, \bibinfo{address}{Red Hook, NY,
  USA}. p. \bibinfo{pages}{4768–4777}.
\bibitem[{Mar{\'\i}n et~al.(2025)Mar{\'\i}n, Huatangari, Tuesta, Caro, Guevara,
  Bardales and Santos}]{marin25}
\bibinfo{author}{Mar{\'\i}n, Y.R.}, \bibinfo{author}{Huatangari, L.Q.},
  \bibinfo{author}{Tuesta, J.N.A.}, \bibinfo{author}{Caro, O.C.},
  \bibinfo{author}{Guevara, J.L.M.}, \bibinfo{author}{Bardales, E.S.},
  \bibinfo{author}{Santos, R.C.}, \bibinfo{year}{2025}.
\newblock \bibinfo{title}{Analysis of factors affecting the academic
  performance of university students using machine learning}.
\newblock \bibinfo{journal}{Scientific Reports} .
\bibitem[{McGinn and Schiefelbein(2023)}]{mcginn23}
\bibinfo{author}{McGinn, N.}, \bibinfo{author}{Schiefelbein, E.},
  \bibinfo{year}{2023}.
\newblock \bibinfo{title}{The political economy of teacher training in latin
  america: A review of the research literature}.
\newblock \bibinfo{journal}{The Palgrave Handbook of Teacher Education
  Research} , \bibinfo{pages}{1379--1401}.
\bibitem[{Melo et~al.(2017)Melo, Machado and Miranda}]{melo17}
\bibinfo{author}{Melo, G.d.}, \bibinfo{author}{Machado, A.},
  \bibinfo{author}{Miranda, A.}, \bibinfo{year}{2017}.
\newblock \bibinfo{title}{El impacto en el aprendizaje del programa una laptop
  por ni{\~n}o. la evidencia de uruguay}.
\newblock \bibinfo{journal}{El trimestre econ{\'o}mico} \bibinfo{volume}{84},
  \bibinfo{pages}{383--409}.
\bibitem[{Mereles and Canese(2022)}]{mereles22}
\bibinfo{author}{Mereles, J.I.}, \bibinfo{author}{Canese, V.},
  \bibinfo{year}{2022}.
\newblock \bibinfo{title}{Dificultades docentes durante la educaci{\'o}n remota
  en paraguay: Teaching difficulties during remote education in paraguay}.
\newblock \bibinfo{journal}{Revista cient{\'\i}fica en ciencias sociales-ISSN:
  2708-0412} \bibinfo{volume}{4}, \bibinfo{pages}{8--22}.
\bibitem[{Miranda et~al.(2025)Miranda, Aravena, de~Oliveira and
  Pineda-B{\'a}ez}]{miranda25}
\bibinfo{author}{Miranda, R.M.}, \bibinfo{author}{Aravena, F.},
  \bibinfo{author}{de~Oliveira, A.C.P.}, \bibinfo{author}{Pineda-B{\'a}ez, C.},
  \bibinfo{year}{2025}.
\newblock \bibinfo{title}{Reimagining professional development for school
  leaders in brazil, chile, and colombia: an examination of current approaches
  and future directions}.
\newblock \bibinfo{journal}{Professional Development in Education}
  \bibinfo{volume}{51}, \bibinfo{pages}{54--68}.
\bibitem[{Molnar(2020)}]{molnar20}
\bibinfo{author}{Molnar, C.}, \bibinfo{year}{2020}.
\newblock \bibinfo{title}{Interpretable machine learning}.
\newblock \bibinfo{publisher}{Lulu. com}.
\bibitem[{Munguia(2023)}]{munguia23}
\bibinfo{author}{Munguia, N.}, \bibinfo{year}{2023}.
\newblock \bibinfo{title}{Covid-19 and its influence on sustainable development
  goal 4: Latin america and caribbean region}, in: \bibinfo{booktitle}{SDGs in
  the Americas and Caribbean Region}. \bibinfo{publisher}{Springer}, pp.
  \bibinfo{pages}{1--17}.
\bibitem[{Murillo et~al.(2023)Murillo, Mart{\'\i}nez-Garrido and
  Gra{\~n}a}]{murillo23}
\bibinfo{author}{Murillo, J.}, \bibinfo{author}{Mart{\'\i}nez-Garrido, C.},
  \bibinfo{author}{Gra{\~n}a, R.}, \bibinfo{year}{2023}.
\newblock \bibinfo{title}{Segregaci{\'o}n escolar por nivel socioecon{\'o}mico
  en educaci{\'o}n primaria en am{\'e}rica latina y el caribe}.
\bibitem[{Naidu et~al.(2023)Naidu, Zuva and Sibanda}]{naidu23}
\bibinfo{author}{Naidu, G.}, \bibinfo{author}{Zuva, T.},
  \bibinfo{author}{Sibanda, E.M.}, \bibinfo{year}{2023}.
\newblock \bibinfo{title}{A review of evaluation metrics in machine learning
  algorithms}.
\bibitem[{Neidh{\"o}fer et~al.(2021)Neidh{\"o}fer, Lustig and
  Tommasi}]{neidhofer21}
\bibinfo{author}{Neidh{\"o}fer, G.}, \bibinfo{author}{Lustig, N.},
  \bibinfo{author}{Tommasi, M.}, \bibinfo{year}{2021}.
\newblock \bibinfo{title}{Intergenerational transmission of lockdown
  consequences: prognosis of the longer-run persistence of covid-19 in latin
  america}.
\newblock \bibinfo{journal}{The Journal of Economic Inequality}
  \bibinfo{volume}{19}, \bibinfo{pages}{571--598}.
\bibitem[{Neidh{\"o}fer et~al.(2018)Neidh{\"o}fer, Serrano and
  Gasparini}]{neidhofer18}
\bibinfo{author}{Neidh{\"o}fer, G.}, \bibinfo{author}{Serrano, J.},
  \bibinfo{author}{Gasparini, L.}, \bibinfo{year}{2018}.
\newblock \bibinfo{title}{Educational inequality and intergenerational mobility
  in latin america: A new database}.
\newblock \bibinfo{journal}{Journal of development economics}
  \bibinfo{volume}{134}, \bibinfo{pages}{329--349}.
\bibitem[{OECD(2023a)}]{oecd23a}
\bibinfo{author}{OECD}, \bibinfo{year}{2023}a.
\newblock \bibinfo{title}{Pisa 2022 assessment and analytical framework}.
\newblock \DOIprefix\doi{https://doi.org/10.1787/dfe0bf9c-en}.
\bibitem[{OECD(2023b)}]{oecd23b}
\bibinfo{author}{OECD}, \bibinfo{year}{2023}b.
\newblock \bibinfo{title}{Pisa 2022 results (volume i): The state of learning
  and equity in education}.
\newblock \DOIprefix\doi{https://doi.org/10.1787/53f23881-en}.
\bibitem[{Olsen et~al.(2024)Olsen, Glad, Jullum and Aas}]{olsen24}
\bibinfo{author}{Olsen, L.H.B.}, \bibinfo{author}{Glad, I.K.},
  \bibinfo{author}{Jullum, M.}, \bibinfo{author}{Aas, K.},
  \bibinfo{year}{2024}.
\newblock \bibinfo{title}{A comparative study of methods for estimating
  model-agnostic shapley value explanations}.
\newblock \bibinfo{journal}{Data Mining and Knowledge Discovery} ,
  \bibinfo{pages}{1--48}.
\bibitem[{{\"O}z et~al.(2024){\"O}z, Bulut, Cellat and Y{\"u}rekli}]{oz24}
\bibinfo{author}{{\"O}z, E.}, \bibinfo{author}{Bulut, O.},
  \bibinfo{author}{Cellat, Z.F.}, \bibinfo{author}{Y{\"u}rekli, H.},
  \bibinfo{year}{2024}.
\newblock \bibinfo{title}{Stacking: An ensemble learning approach to predict
  student performance in pisa 2022}.
\newblock \bibinfo{journal}{Education and Information Technologies} ,
  \bibinfo{pages}{1--27}.
\bibitem[{Pedregosa et~al.(2011)Pedregosa, Varoquaux, Gramfort, Michel,
  Thirion, Grisel, Blondel, Prettenhofer, Weiss, Dubourg et~al.}]{pedregosa11}
\bibinfo{author}{Pedregosa, F.}, \bibinfo{author}{Varoquaux, G.},
  \bibinfo{author}{Gramfort, A.}, \bibinfo{author}{Michel, V.},
  \bibinfo{author}{Thirion, B.}, \bibinfo{author}{Grisel, O.},
  \bibinfo{author}{Blondel, M.}, \bibinfo{author}{Prettenhofer, P.},
  \bibinfo{author}{Weiss, R.}, \bibinfo{author}{Dubourg, V.}, et~al.,
  \bibinfo{year}{2011}.
\newblock \bibinfo{title}{Scikit-learn: Machine learning in python}.
\newblock \bibinfo{journal}{the Journal of machine Learning research}
  \bibinfo{volume}{12}, \bibinfo{pages}{2825--2830}.
\bibitem[{Pritchett(2025)}]{pritchett25}
\bibinfo{author}{Pritchett, L.}, \bibinfo{year}{2025}.
\newblock \bibinfo{title}{Addressing the learning crisis: an emergent
  consensus}.
\bibitem[{Qiao and Hu(2020)}]{qiao20}
\bibinfo{author}{Qiao, C.}, \bibinfo{author}{Hu, X.}, \bibinfo{year}{2020}.
\newblock \bibinfo{title}{A joint neural network model for combining
  heterogeneous user data sources: An example of at-risk student prediction}.
\newblock \bibinfo{journal}{Journal of the Association for Information Science
  and Technology} \bibinfo{volume}{71}, \bibinfo{pages}{1192--1204}.
\bibitem[{Randall and Anderson(2016)}]{randall16}
\bibinfo{author}{Randall, L.}, \bibinfo{author}{Anderson, J.B.},
  \bibinfo{year}{2016}.
\newblock \bibinfo{title}{Schooling for success: Preventing repetition and
  dropout in Latin American primary schools}.
\newblock \bibinfo{publisher}{Routledge}.
\bibitem[{Raschka and Mirjalili(2019)}]{raschka19}
\bibinfo{author}{Raschka, S.}, \bibinfo{author}{Mirjalili, V.},
  \bibinfo{year}{2019}.
\newblock \bibinfo{title}{Python machine learning: Machine learning and deep
  learning with Python, scikit-learn, and TensorFlow 2}.
\newblock \bibinfo{publisher}{Packt publishing ltd}.
\bibitem[{Rodr{\'\i}guez et~al.(2023)Rodr{\'\i}guez, Villanueva, Dombrovskaia
  and Valenzuela}]{rodriguez23}
\bibinfo{author}{Rodr{\'\i}guez, P.}, \bibinfo{author}{Villanueva, A.},
  \bibinfo{author}{Dombrovskaia, L.}, \bibinfo{author}{Valenzuela, J.P.},
  \bibinfo{year}{2023}.
\newblock \bibinfo{title}{A methodology to design, develop, and evaluate
  machine learning models for predicting dropout in school systems: the case of
  chile}.
\newblock \bibinfo{journal}{Education and Information Technologies}
  \bibinfo{volume}{28}, \bibinfo{pages}{10103--10149}.
\bibitem[{Ruding(2019)}]{rudin19}
\bibinfo{author}{Ruding, C.}, \bibinfo{year}{2019}.
\newblock \bibinfo{title}{Stop explaining black box machine learning models for
  high stakes decisions and use interpretable models instead}.
\newblock \bibinfo{journal}{Nat Mach Intell} \bibinfo{volume}{1},
  \bibinfo{pages}{206--215}.
\bibitem[{Sarker(2021)}]{sarker21}
\bibinfo{author}{Sarker, I.H.}, \bibinfo{year}{2021}.
\newblock \bibinfo{title}{Machine learning: Algorithms, real-world applications
  and research directions}.
\newblock \bibinfo{journal}{SN computer science} \bibinfo{volume}{2},
  \bibinfo{pages}{160}.
\bibitem[{Shapley(1953)}]{shapley53}
\bibinfo{author}{Shapley, L.S.}, \bibinfo{year}{1953}.
\newblock \bibinfo{title}{A value for n-person games}.
\newblock \bibinfo{journal}{Contribution to the Theory of Games}
  \bibinfo{volume}{2}.
\bibitem[{Shi et~al.(2024)Shi, Caskurlu, Zhang and Na}]{shi24}
\bibinfo{author}{Shi, H.}, \bibinfo{author}{Caskurlu, S.},
  \bibinfo{author}{Zhang, N.}, \bibinfo{author}{Na, H.}, \bibinfo{year}{2024}.
\newblock \bibinfo{title}{To what extent has machine learning achieved in
  predicting online at-risk students? evidence from quantitative
  meta-analysis}.
\newblock \bibinfo{journal}{Journal of Research on Technology in Education} ,
  \bibinfo{pages}{1--20}.
\bibitem[{Shi et~al.(2025)Shi, Zhang, Caskurlu and Na}]{shi25}
\bibinfo{author}{Shi, H.}, \bibinfo{author}{Zhang, N.},
  \bibinfo{author}{Caskurlu, S.}, \bibinfo{author}{Na, H.},
  \bibinfo{year}{2025}.
\newblock \bibinfo{title}{Applications of machine learning for at-risk student
  prediction in online education: A 10-year systematic review of literature}.
\newblock \bibinfo{journal}{Journal of Computer Assisted Learning}
  \bibinfo{volume}{41}, \bibinfo{pages}{e70058}.
\bibitem[{{\v{S}}trumbelj and Kononenko(2014)}]{vstrumbelj14}
\bibinfo{author}{{\v{S}}trumbelj, E.}, \bibinfo{author}{Kononenko, I.},
  \bibinfo{year}{2014}.
\newblock \bibinfo{title}{Explaining prediction models and individual
  predictions with feature contributions}.
\newblock \bibinfo{journal}{Knowledge and information systems}
  \bibinfo{volume}{41}, \bibinfo{pages}{647--665}.
\bibitem[{Torrecilla and Hern{\'a}ndez-Castilla(2020)}]{torrecilla20}
\bibinfo{author}{Torrecilla, F.J.M.}, \bibinfo{author}{Hern{\'a}ndez-Castilla,
  R.}, \bibinfo{year}{2020}.
\newblock \bibinfo{title}{Does parental involvement matter in children’s
  performance? a latin american primary school study}.
\newblock \bibinfo{journal}{Revista de Psicodid{\'a}ctica (English ed.)}
  \bibinfo{volume}{25}, \bibinfo{pages}{13--22}.
\bibitem[{UNESCO(2024)}]{unesco24}
\bibinfo{author}{UNESCO}, \bibinfo{year}{2024}.
\newblock \bibinfo{title}{La urgencia de la recuperación educativa en América
  Latina y el Caribe}.
\newblock \bibinfo{publisher}{UNESCO}.
\bibitem[{UNICEF et~al.(2021)}]{unicef21}
\bibinfo{author}{UNICEF}, et~al., \bibinfo{year}{2021}.
\newblock \bibinfo{title}{Los aprendizajes fundamentales en am{\'e}rica latina
  y el caribe. evaluaci{\'o}n de logros de los estudiantes. estudio regional
  comparativo y explicativo (erce 2019).}
\bibitem[{Varela et~al.(2024)Varela, Berger, Chaux, Miranda, Lisboa and
  Orellana}]{varela24}
\bibinfo{author}{Varela, J.J.}, \bibinfo{author}{Berger, C.},
  \bibinfo{author}{Chaux, E.}, \bibinfo{author}{Miranda, R.},
  \bibinfo{author}{Lisboa, C.}, \bibinfo{author}{Orellana, C.T.L.},
  \bibinfo{year}{2024}.
\newblock \bibinfo{title}{School violence, bullying, and cyberbullying in south
  america: risks, protective factors, and interventions}, in:
  \bibinfo{booktitle}{Handbook of School Violence, Bullying and Safety}.
  \bibinfo{publisher}{Edward Elgar Publishing}, pp. \bibinfo{pages}{256--268}.
\bibitem[{Vinas-Forcade et~al.(2021)Vinas-Forcade, Mels, Van~Houtte, Valcke and
  Derluyn}]{vinas21}
\bibinfo{author}{Vinas-Forcade, J.}, \bibinfo{author}{Mels, C.},
  \bibinfo{author}{Van~Houtte, M.}, \bibinfo{author}{Valcke, M.},
  \bibinfo{author}{Derluyn, I.}, \bibinfo{year}{2021}.
\newblock \bibinfo{title}{Can failure be prevented? using longitudinal data to
  identify at-risk students upon entering secondary school}.
\newblock \bibinfo{journal}{British Educational Research Journal}
  \bibinfo{volume}{47}, \bibinfo{pages}{205--225}.
\bibitem[{Wang et~al.(2023)Wang, Perry, Malpique and Ide}]{wang23}
\bibinfo{author}{Wang, X.S.}, \bibinfo{author}{Perry, L.B.},
  \bibinfo{author}{Malpique, A.}, \bibinfo{author}{Ide, T.},
  \bibinfo{year}{2023}.
\newblock \bibinfo{title}{Factors predicting mathematics achievement in pisa: a
  systematic review}.
\newblock \bibinfo{journal}{Large-Scale Assessments in Education}
  \bibinfo{volume}{11}, \bibinfo{pages}{24}.
\bibitem[{Wolpert(1992)}]{wolpert92}
\bibinfo{author}{Wolpert, D.H.}, \bibinfo{year}{1992}.
\newblock \bibinfo{title}{Stacked generalization}.
\newblock \bibinfo{journal}{Neural networks} \bibinfo{volume}{5},
  \bibinfo{pages}{241--259}.
\bibitem[{Zhu et~al.(2025)Zhu, You, Hong and Fang}]{zhu25}
\bibinfo{author}{Zhu, L.}, \bibinfo{author}{You, H.}, \bibinfo{author}{Hong,
  M.}, \bibinfo{author}{Fang, Z.}, \bibinfo{year}{2025}.
\newblock \bibinfo{title}{Predictive insights into us students’ mathematics
  performance on pisa 2022 using ensemble tree-based machine learning models}.
\newblock \bibinfo{journal}{International Journal of Educational Research}
  \bibinfo{volume}{130}, \bibinfo{pages}{102537}.

\end{thebibliography}


\newpage

\clearpage

\clearpage

\clearpage

\clearpage

\clearpage


\newpage
\clearpage

\begin{table}[htbp]
\begin{threeparttable}
\caption{Levels of achievement and rates of learning poverty}
\begin{tabular}{lllllllll}
\hline
Levels & \multicolumn{2}{c}{Math} &  & \multicolumn{2}{c}{Reading} &  & \multicolumn{2}{c}{Science} \\
 \cmidrule{2-3}\cmidrule{5-6}\cmidrule{8-9}
 & Rate & Interval &  & Rate & Interval &  & Rate & Interval \\
 \hline
 & (1) & (2) &  & (3) & (4) &  & (5) & (6) \\
\textit{Panel A—levels of achievement} &  & &  & & &  & &  \\
\textbf{No competency achieved} & 39.05 & $<$ 356 &  & 21.64 & $<$ 335 &  & 19.56 &  $<$ 335 \\
\textbf{Level 1} & 31.99 & [356, 421) &  & 29.13 & [335, 408) &  & 33.55 & [335, 410) \\
\textbf{Level 2} & 18.42 & [421, 483) &  & 26.68 & [408, 481) &  & 27.75 & [410, 485) \\
Level 3 & 7.92 & (483, 545] &  & 16.14 & [481, 553) &  & 14.21 &  [485, 559) \\
Level 4 & 2.3 & [545, 610) &  & 5.7 & [553, 626) &  & 4.35 &  [559, 634) \\
Level 5 & 0.31 & [610, 670) &  & 0.7 & [626, 699) &  & 0.57 &  [634, 708) \\
Level 6 & 0.01 & $\ge$ 670 &  & 0.03 & $\ge$ 699 &  & 0.02 & $\ge$ 708 \\
 &  & &  & & &  & &  \\
\textit{Panel B—levels of achievement combined}  &  & &  & & &  & &  \\
& Level 1 & Level 0  & & Level 2 & Level 1 &  & & \\
 \cmidrule{2-3}\cmidrule{5-6}
Number of observations & 5,627 & 10,609 &  & 3,857 & 5,627 &  &  \\
Rate of learning poverty (\%) & \multicolumn{2}{c}{65.3} &  &   \multicolumn{2}{c}{59.3} &  &  &  \\
\hline
\end{tabular}
\label{table1}
\begin{tablenotes}[para,flushleft]
\medskip
\footnotesize{
(1) The focus is on the students whose achievement falls into the levels highlighted in bold. (2) For details on competency and achievement levels, see: \cite{oecd23a,oecd23b}.}
\end{tablenotes}
\end{threeparttable}
\end{table}

\clearpage
\begin{table}[htbp]
\begin{threeparttable}
  \centering
\caption{Differences by level of achievement for students and family explanatory variables}
\begin{tabular}{llllllll}
\hline
 & \multicolumn{3}{c}{Bottom performers}  & &  \multicolumn{3}{c}{Low performers}   \\
  \cmidrule{2-4}\cmidrule{6-8}
 & Level 1 & Level 0 & Difference &    & Level 2 & Level 1 & Difference   \\
 & (1) & (2) & (3) &    & (4) & (5) & (6)   \\
\textit{Panel  A—Students} &  & & & & & &  \\
Gender – male & 0.55 & 0.53 & -0.02* &  & 0.60 & 0.55 & -0.05*** \\
Satisfaction with life & 7.23 & 7.20 & -0.04 &  & 7.03 & 7.23 & 0.20*** \\
Homework – all subjects & 3.32 & 3.03 & -0.29*** &  & 3.59 & 3.32 & -0.28*** \\
Digital devices at home-number & 7.83 & 6.12 & -1.72*** &  & 8.57 & 7.83 & -0.73*** \\
Books – number & 1.46 & 1.11 & -0.35*** &  & 1.91 & 1.46 & -0.45*** \\
Immigrant & 0.04 & 0.09 & 0.04*** &  & 0.03 & 0.04 & 0.01** \\
Speak other language at home & 0.05 & 0.15 & 0.10*** &  & 0.03 & 0.05 & 0.03*** \\
Late entry – primary & 0.19 & 0.26 & 0.07*** &  & 0.14 & 0.19 & 0.05*** \\
Repetition – primary & 0.09 & 0.33 & 0.24*** &  & 0.04 & 0.09 & 0.05*** \\
Repetition – lower secondary & 0.09 & 0.22 & 0.13*** &  & 0.04 & 0.09 & 0.05*** \\
Missed school (3 months in a row) & 0.06 & 0.14 & 0.08*** &  & 0.03 & 0.06 & 0.03*** \\
Skipped school (1 day last 2 weeks) & 0.29 & 0.36 & 0.08*** &  & 0.23 & 0.29 & 0.06*** \\
Teacher support for well being (index, low) & 0.49 & 0.58 & 0.10*** &  & 0.48 & 0.49 & 0.01   \\
Sense of belonging to school (index, low) & 0.50 & 0.48 & -0.02*** &  & 0.50 & 0.50 & 0.00   \\
School climate – bad (index, low) & 0.49 & 0.43 & -0.06*** &  & 0.52 & 0.49 & -0.03*** \\
Safety in school and surroundings area (index, low) & 0.48 & 0.57 & 0.09*** &  & 0.49 & 0.48 & -0.01   \\
Work – unpaid, days & 3.07 & 2.95 & -0.12*** &  & 3.00 & 3.07 & 0.07* \\
Work – paid, days & 1.45 & 1.66 & 0.22*** &  & 1.02 & 1.44 & 0.42*** \\
 &  &  &  &  &  &  &      \\
\textit{Panel  B—Family}  &  & & & & & &  \\
Mother education & 5.25 & 5.20 & -0.06 &   & 5.88 & 5.25 & -0.63*** \\
Father education & 5.10 & 5.15 & 0.05 &   & 5.74 & 5.10 & -0.64*** \\
SES (index, low) & 0.57 & 0.69 & 0.11*** &  & 0.41 & 0.57 & 0.16*** \\
Siblings number & 2.16 & 2.37 & 0.21*** &  & 1.93 & 2.16 & 0.23*** \\
Family support for education (index, low) & 0.43 & 0.42 & -0.01 &   & 0.46 & 0.43 & -0.02** \\
 &  &  &  &  &  &  &      \\
N & 5,627 & 10,609 &  &    & 3,857 & 5,627 &   ` \\
\hline
\end{tabular}
\label{table2}
\begin{tablenotes}[para,flushleft]
\medskip
\footnotesize{
(1) Significance levels: * p $\leq$ 0.10, ** p $\leq$ 0.05, ***p $\leq$ 0.01.}
\end{tablenotes}
\end{threeparttable}
\end{table}

\clearpage
\begin{table}[htbp]
\begin{threeparttable}
  \centering
\caption{Differences by level of achievement for school explanatory variables}
\begin{tabular}{llllllll}
\hline
 & \multicolumn{3}{c}{Bottom performers}  & & \multicolumn{3}{c}{Low performers}   \\
  \cmidrule{2-4}\cmidrule{6-8}
 & Level 1 & Level 0 & Difference &    & Level 2 & Level 1 & Difference   \\
 & (1) & (2) & (3) &    & (4) & (5) & (6)   \\
\textit{Panel  C—Schools} &  & & & & & &  \\
Urban  & 0.68 & 0.56 & -0.13*** &  & 0.79 & 0.68 & -0.11*** \\
Private & 0.18 & 0.06 & -0.12*** &  & 0.33 & 0.18 & -0.15*** \\
Town size & 2.98 & 2.62 & -0.36*** &  & 3.25 & 2.98 & -0.27*** \\
Competition & 0.81 & 0.74 & -0.07*** &  & 0.85 & 0.81 & -0.05*** \\
Funding government (\%) & 73.92 & 83.35 & 9.44*** &  & 64.80 & 73.92 & 9.12*** \\
Funding fees (\%) & 19.14 & 9.01 & -10.13*** &  & 29.19 & 19.14 & -10.05*** \\
Students – 2nd language (\%) & 0.07 & 0.12 & 0.05*** &  & 0.05 & 0.07 & 0.02*** \\
Students – learning needs  (\%) & 0.04 & 0.05 & 0.00*** &  & 0.04 & 0.04 & 0.00   \\
Students – disadvantaged (\%) & 0.41 & 0.44 & 0.03*** &  & 0.32 & 0.41 & 0.09*** \\
School size (enrolment) & 785.95 & 577.31 & -208.63*** &  & 923.23 & 785.95 & -137.28*** \\
STR & 20.03 & 19.09 & -0.94*** &  & 19.56 & 20.03 & 0.47   \\
Teachers attended PD programme (last 3 months) (\%) & 0.41 & 0.41 & 0.00 &    & 0.41 & 0.41 & 0.00   \\
STR non-teaching staff, pedagogic support & 250.10 & 218.38 & -31.73*** &  & 259.03 & 250.10 & -8.92  \\
Teachers – certified (\%) & 0.59 & 0.56 & -0.02*** &  & 0.59 & 0.59 & 0.00   \\
Teachers – degree or higher (\%) & 0.81 & 0.85 & 0.04*** &  & 0.76 & 0.81 & 0.05*** \\
ICT – infrastructure per student (index, low) & 0.53 & 0.43 & -0.11*** &  & 0.54 & 0.53 & -0.01   \\
PC connected web-teacher ratio & 0.30 & 0.25 & -0.05*** &  & 0.33 & 0.30 & -0.02* \\
Admission criteria – none & 0.21 & 0.27 & 0.06*** &  & 0.18 & 0.21 & 0.03*** \\
Transfer policies – none & 0.52 & 0.57 & 0.05*** &  & 0.52 & 0.52 & 0.00   \\
Autonomy (index, low) & 0.51 & 0.56 & 0.05*** &  & 0.47 & 0.51 & 0.05*** \\
Social and emotional skills policy & 0.77 & 0.76 & -0.02** &  & 0.78 & 0.77 & -0.01   \\
Teacher monitoring – methods  & 2.40 & 2.35 & -0.05** &  & 2.45 & 2.40 & -0.06** \\
Testing students learning, all methods & 0.93 & 0.80 & -0.13*** &  & 0.96 & 0.93 & -0.03*** \\
Staff students support for inclusivity (index, low) & 0.53 & 0.49 & -0.04*** &  & 0.55 & 0.53 & -0.02** \\
Parents involvement fostered by school (index, low) & 0.49 & 0.43 & -0.06*** &  & 0.52 & 0.49 & -0.02** \\
School policy - ability grouping & 0.38 & 0.32 & -0.06*** &  & 0.37 & 0.38 & 0.01 \\
 &  &  &  &  &  &  &      \\
N & 5,627 & 10,609 &   &  & 3,857 & 5,627 &   ` \\
\hline
\end{tabular}
\label{table3}
\begin{tablenotes}[para,flushleft]
\medskip
\footnotesize{
Notes: (1) Significance levels: * p $\leq$ 0.10, ** p $\leq$ 0.05, ***p $\leq$ 0.01.
}
\end{tablenotes}
\end{threeparttable}
\end{table}

\clearpage
\begin{table}[htbp]
\begin{threeparttable}
  \centering
\caption{Evaluation metrics of selected ML models and stacking weights}
\begin{tabular}{lllllllll}
\hline
 & \multicolumn{6}{c}{Evaluation metrics} &  Stacking weights \\
  \cmidrule{2-7}
 & ACC & RC & F1 & PR & SP & AUC &  \\
 & (1) & (2) & (3) & (4) & (5) & (6) & (7) \\
\textit{1) Panel A—level 0 versus level 1} &  &  &  &  &  &  &  \\
\textit{Interaction 1} &  &  &  &  &  &  &  \\
Logit & 0.7305 & 0.8609 & 0.7591 & 0.4846 & 0.8068 & 0.7825 & 0.0000 \\
Lasso (with cross-validated penalty) & 0.7524 & 0.8632 & 0.7811 & 0.5435 & 0.8201 & 0.8144 & 0.1125 \\
Gradient boosting  & 0.7759 & 0.8557 & 0.8116 & 0.6251 & 0.8331 & 0.9293 & \textbf{0.5907} \\
Random forest  & 0.7707 & 0.8637 & 0.8010 & 0.5953 & 0.8312 & 0.9325 & 0.1999 \\
Neural net & 0.7564 & 0.8218 & 0.8086 & 0.6330 & 0.8151 & 0.8824 & 0.0969 \\
 &  &  &  &  &  &  &  \\
Interaction 2 &  &  &  &  &  &  &  \\
Stacked models & \textbf{0.7795} & \textbf{0.8730} & 0.8058 & 0.6032 & \textbf{0.8381} & 0.9309 & n/a \\
Lasso (with cross-validated penalty) & 0.7546 & 0.8702 & 0.7798 & 0.5364 & 0.8225 & 0.8135 & 0.1361 \\
Gradient boosting  & 0.7755 & 0.8553 & \textbf{0.8115} & \textbf{0.6251} & 0.8328 & 0.9292 & \textbf{0.6066} \\
Random forest  & 0.7686 & 0.8646 & 0.7981 & 0.5874 & 0.8300 & \textbf{0.9392} & 0.1704 \\
Neural net & 0.7394 & 0.8613 & 0.7680 & 0.5092 & 0.8120 & 0.8806 & 0.0869 \\
 &  &  &  &  &  &  &  \\
\textit{2) Panel B—level 1 versus level 2} &  &  &  &  &  &  &  \\
\textit{Interaction  1} &  &  &  &  &  &  &  \\
Logit & 0.6383 & 0.7868 & 0.6630 & 0.4246 & 0.7196 & 0.7021 & 0.0000 \\
Lasso (with cross-validated penalty) & 0.6486 & 0.7850 & 0.6735 & 0.4525 & 0.7250 & 0.7233 & \textbf{0.4287} \\
Gradient boosting & 0.6689 & 0.7586 & 0.7034 & 0.5399 & 0.7300 & 0.9057 & 0.3172 \\
Random forest  & 0.6570 & 0.7930 & 0.6792 & 0.4613 & 0.7317 & 0.9373 & 0.2541 \\
Neural net & 0.6388 & 0.8361 & 0.6509 & 0.3549 & 0.7320 & 0.7400 & 0.0000 \\
 &  &  &  &  &  &  &  \\
\textit{Interaction  2} &  &  &  &  &  &  &  \\
Stacked models & \textbf{0.6731} & \textbf{0.7938} & 0.6952 & 0.4994 & \textbf{0.7413} & 0.8827 & n/a \\
Lasso (with cross-validated penalty) & 0.6492 & 0.7877 & 0.6732 & 0.4499 & 0.7259 & 0.7233 & \textbf{0.4257} \\
Gradient boosting & 0.6694 & 0.7595 & \textbf{0.7037} & \textbf{0.5399} & 0.7305 & 0.9058 & 0.3529 \\
Random forest  & 0.6544 & 0.7868 & 0.6786 & 0.4639 & 0.7287 & \textbf{0.9375} & 0.2214 \\
\hline
\end{tabular}
\label{table4}
\begin{tablenotes}[para,flushleft]
\medskip
\footnotesize{
Notes: (1) For details on evaluation metrics, see \ref{appendixA}. (2) Additional stacked models comparison can be found in the Appendix C (Table \ref{tableC3}).
}
\end{tablenotes}
\end{threeparttable}
\end{table}

\clearpage
\begin{table}[htbp]
\begin{threeparttable}
\caption{Covariates ranked by mean SHAP (absolute) values for the two working samples}
\begin{tabular}{lllllllll}
\hline
\multicolumn{4}{c}{Bottom performers (level 0 vs  level 1)} &  &  \multicolumn{4}{c}{Low performers (level 1 vs level 2)}  \\
  \cmidrule{1-4} \cmidrule{6-9}
\underline{Student$/$family} & (1) & \underline{School}  & (2) &  & \underline{Student$/$family} & (3) & \underline{School} & (4) \\
 &  &  &  &  &  &  &  &  \\
repeprim\_yes & 0.464 & enrolment & 0.269 &  & escs & 0.345 & enrolment & 0.226 \\
devices & 0.393 & teststud\_yes & 0.213 &  & famsuppedu & 0.333 & disadvantaged & 0.199 \\
schclimabad & 0.370 & str & 0.202 &  & workpaid & 0.299 & private & 0.181 \\
satislife & 0.331 & ictinfra & 0.199 &  & sex\_f & 0.298 & strpedag & 0.166 \\
famsuppedu & 0.320 & pcwebtr & 0.167 &  & books & 0.270 & ictinfra & 0.162 \\
escs & 0.311 & autonomy & 0.150 &  & satislife & 0.255 & townsize & 0.146 \\
tchsuppwb & 0.245 & disadvantaged & 0.145 &  & devices & 0.230 & fundgovt & 0.136 \\
belongsch & 0.241 & suppinclusity & 0.135 &  & homework & 0.213 & pcwebtr & 0.129 \\
sex\_f & 0.227 & strpedag & 0.135 &  & motheredu & 0.174 & autonomy & 0.123 \\
books & 0.224 & nolangtest & 0.133 &  & schclimabad & 0.174 & str & 0.119 \\
schsafelack & 0.214 & tchcertified & 0.131 &  & belongsch & 0.167 & parentinvol & 0.114 \\
workpaid & 0.208 & private & 0.121 &  & tchsuppwb & 0.162 & suppinclusity & 0.106 \\
otherlang\_yes & 0.186 & townsize & 0.118 &  & siblings & 0.139 & tchpdatt & 0.079 \\
fatheredu & 0.183 & learnneeds & 0.093 &  & schsafelack & 0.138 & tchcertified & 0.079 \\
repelowsec\_yes & 0.181 & parentinvol & 0.089 &  & lateprim\_yes & 0.116 & tchdegree & 0.074 \\
homework & 0.177 & fundfees & 0.089 &  & repeprim\_yes & 0.090 & learnneeds & 0.065 \\
workunpaid & 0.171 & tchmonitor & 0.089 &  & workunpaid & 0.088 & competition & 0.049 \\
lateprim\_yes & 0.166 & tchpdatt & 0.083 &  & repelowsec\_yes & 0.081 & tchmonitor & 0.044 \\
motheredu & 0.162 & fundgovt & 0.073 &  & skippsch\_yes & 0.067 & nolangtest & 0.043 \\
siblings & 0.157 & tchdegree & 0.067 &  & fatheredu & 0.063 & fundfees & 0.043 \\
missclass\_yes & 0.111 & transfernone\_yes & 0.042 &  & missclass\_yes & 0.050 & urban & 0.042 \\
immigrant\_yes & 0.047 & abilitygroup\_yes & 0.026 &  & otherlang\_yes & 0.030 & transfernone\_yes & 0.039 \\
skippsch\_yes & 0.036 & polsocemskill\_yes & 0.022 &  & immigrant\_yes & 0.016 & abilitygroup\_yes & 0.034 \\
 &  & competition & 0.020 &  &  &  & teststud\_yes & 0.025 \\
 &  & polsocemskill\_yes & 0.013 &  &  &  & polsocemskill\_yes & 0.014 \\
 &  & urban & 0.010 &  &  &  & admissionnone\_yes & 0.007 \\
 &  & admissionnone\_yes & 0.008 &  &  &  &  &  \\
 \hline
\end{tabular}
\label{table5}
\begin{tablenotes}[para,flushleft]
\medskip
\footnotesize{
Notes: (1) The table shows SHAP average absolute values, i.e., $I_{j} = \big( (1/n) \times \sum_{i=1}^{n}|\phi^{(i)}_{j}| \big)$) (2). Estimates are obtained by running two models with the different student$/$family and school covariates. (3) Details on covariates can be found in Appendix B (Tables \ref{tableB1} and \ref{tableB2}).
}
\end{tablenotes}
\end{threeparttable}
\end{table}

 \setlength{\tabcolsep}{1.8pt}
\begin{sidewaystable}
\begin{threeparttable}
\scriptsize
  \centering
\caption{Model—level 0 versus level 1 (low performing group). Covariates individual SHAP values for students with the lowest and highest average SHAP values contributions. First country set: Argentina, Brazil, Chile, Colombia and Dominican Rep.}
\begin{tabular}{llllll|llllll|llllll}
\hline
\multicolumn{6}{c}{Argentina} & \multicolumn{6}{c}{Brazil} & \multicolumn{6}{c}{Chile} \\
\hline
\multicolumn{3}{c}{Low} & \multicolumn{3}{c}{High} & \multicolumn{3}{c}{Low} & \multicolumn{3}{c}{High} & \multicolumn{3}{c}{Low} & \multicolumn{3}{c}{High} \\
\hline
value & variable & SV &  value & variable & SV & value & variable & SV & value & variable & SV & value & variable & SV & value & variable & SV \\
 (1) & (2)  & (3)  & (4)  & (5)  & (6)  & (7)  & (8)  & (9)  & (10)  & (11)  & (12)  & (13)  & (14)  & (15)  & (16)  & (17) & (18) \\
value & variable & SV &  value & variable & SV & value & variable & SV & value & variable & SV & value & variable & SV & value & variable & SV \\
0.343 & escs & -1.837 & 1 & repeprim\_yes & 2.304 & 4 & homework & -1.805 & 1 & disadvantaged & 3.716 & 0.303 & schsafelack & -1.594 & 1 & repeprim\_yes & 5.034 \\
4 & books & -1.132 & 0 & devices & 2.142 & -0.568 & schclimabad & -1.787 & 10 & satislife & 1.920 & 0.69 & disadvantaged & -1.515 & 1 & skippsch\_yes & 4.812 \\
1 & private & -0.914 & 84 & enrolment & 1.528 & 5 & satislife & -1.561 & -1.922 & escs & 1.303 & 1 & siblings & -1.486 & 308 & enrolment & 2.675 \\
0.12 & disadvantaged & -0.772 & -0.668 & schclimabad & 1.519 & 1 & motheredu & -0.942 & 0 & books & 1.155 & 0 & repeprim\_yes & -1.119 & 3 & motheredu & 2.300 \\
11 & devices & -0.594 & -1.901 & parentinvol & 1.348 & 5 & workpaid & -0.920 & 10 & fatheredu & 1.109 & -0.461 & schclimabad & -0.888 & -1.308 & schsafelack & 1.144 \\
2.32 & schsafelack & -0.586 & 0.92 & disadvantaged & 1.275 & 0 & tchpdatt & -0.697 & 1.819 & schsafelack & 1.038 & 3 & books & -0.690 & 0.9 & disadvantaged & 1.141 \\
0.119 & pcwebtr & -0.549 & -1.211 & schsafelack & 0.744 & 2 & books & -0.528 & 3 & workpaid & 0.678 & -0.808 & ictinfra & -0.671 & -0.96 & parentinvol & 1.046 \\
5 & workunpaid & -0.535 & 0.082 & tchcertified & 0.696 & 0.213 & schsafelack & -0.527 & 5 & devices & 0.394 & 10 & devices & -0.520 & 5 & satislife & 0.805 \\
-0.069 & ictinfra & -0.462 & 0.02 & tchpdatt & 0.690 & 1 & siblings & -0.460 & 3 & homework & 0.328 & 1 & homework & -0.444 & 0.11 & ictinfra & 0.717 \\
-0.796 & suppinclusity & -0.425 & -0.462 & escs & 0.376 & 0 & repeprim\_yes & -0.419 & 8 & motheredu & 0.307 & 7 & fatheredu & -0.442 & 0.498 & schclimabad & 0.714 \\
9 & satislife & -0.376 & 0 & books & 0.365 & 0 & sex\_f & -0.415 & 0.3 & autonomy & 0.305 & 15.432 & str & -0.428 & 1 & workunpaid & 0.681 \\
-0.135 & parentinvol & -0.319 & 0 & private & 0.357 & 0.515 & autonomy & -0.402 & 0.3 & tchpdatt & 0.290 & 0 & sex\_f & -0.337 & -1.355 & suppinclusity & 0.636 \\
1 & abilitygroup\_yes & -0.255 & 5 & workunpaid & 0.287 & 1 & fatheredu & -0.280 & 22.667 & str & 0.214 & 0 & skippsch\_yes & -0.327 & -2.827 & escs & 0.546 \\
1 & urban & -0.217 & 0 & satislife & 0.275 & 1 & abilitygroup\_yes & -0.255 & 0.09 & tchcertified & 0.149 & 7 & motheredu & -0.325 & 0.05 & tchpdatt & 0.505 \\
10 & motheredu & -0.135 & 3 & siblings & 0.217 & 0 & lateprim\_yes & -0.248 & 1 & skippsch\_yes & 0.132 & 1 & sex\_m & -0.307 & -0.506 & autonomy & 0.411 \\
  &  & &  & & &  &  &  &  &  & &  &  &  &  &   \\
  \hline
\multicolumn{6}{c}{Colombia} & \multicolumn{6}{c}{Dominican Rep.} \\
\hline
\multicolumn{3}{c}{Low} & \multicolumn{3}{c}{High} & \multicolumn{3}{c}{Low} & \multicolumn{3}{c}{High} \\
\hline
value & variable & SV &  value & variable & SV & value & variable & SV & value & variable & SV & & &\\
(1) & (2)  & (3)  & (4)  & (5)  & (6)  & (7)  & (8)  & (9)  & (10)  & (11)  & (12)  &   &   &   &   &  &  \\
3 & books & -2.231 & 6.581 & schclimabad & 4.357 & 0 & workpaid & -2.403 & -0.66 & schclimabad & 3.045 &  &  &  &  &  &  \\
1.171 & escs & -2.136 & 1 & repeprim\_yes & 2.840 & -0.529 & schclimabad & -1.944 & 1 & repeprim\_yes & 2.300 &  &  &  &  &  &  \\
0 & workpaid & -1.433 & 0 & devices & 1.667 & 1.396 & suppinclusity & -1.659 & -2.83 & escs & 2.046 &  &  &  &  &  &  \\
1.415 & autonomy & -1.411 & 10 & motheredu & 1.571 & 0.421 & schsafelack & -1.426 & 10 & satislife & 1.046 &  &  &  &  &  &  \\
0 & disadvantaged & -1.257 & -1.338 & escs & 1.462 & 9 & satislife & -1.363 & 0.818 & suppinclusity & 0.963 &  &  &  &  &  &  \\
1 & siblings & -1.052 & 1 & otherlang\_yes & 1.420 & 0.269 & escs & -1.215 & -1.223 & autonomy & 0.948 &  &  &  &  &  &  \\
17.389 & str & -0.941 & 0 & books & 0.991 & 0.847 & autonomy & -0.894 & 0 & devices & 0.947 &  &  &  &  &  &  \\
-0.506 & schclimabad & -0.796 & 1 & immigrant\_yes & 0.735 & 4 & homework & -0.885 & 0.077 & parentinvol & 0.671 &  &  &  &  &  &  \\
5 & homework & -0.632 & 0 & sex\_m & 0.589 & 5 & motheredu & -0.860 & 24.43 & str & 0.640 &  &  &  &  &  &  \\
0.25 & pcwebtr & -0.630 & 1 & sex\_f & 0.588 & 7 & devices & -0.602 & 965 & enrolment & 0.579 &  &  &  &  &  &  \\
0.669 & suppinclusity & -0.596 & 4 & workpaid & 0.423 & 2 & workunpaid & -0.568 & 1 & workpaid & 0.559 &  &  &  &  &  &  \\
0 & repeprim\_yes & -0.577 & -1.193 & parentinvol & 0.373 & 0.141 & ictinfra & -0.523 & 0.81 & pcwebtr & 0.509 &  &  &  &  &  &  \\
13 & devices & -0.533 & 0.824 & autonomy & 0.368 & -0.2 & parentinvol & -0.501 & 3 & homework & 0.441 &  &  &  &  &  &  \\
-0.949 & parentinvol & -0.463 & 3 & siblings & 0.323 & 0 & repeprim\_yes & -0.405 & 0 & books & 0.339 &  &  &  &  &  &  \\
0.1 & tchpdatt & -0.317 & 2650 & enrolment & 0.321 & 1 & polsocemskill\_yes & -0.265 & 1 & transfernone\_yes & 0.315 &  &  &  &  &  &   \\
\hline
\end{tabular}
\label{table6}
\begin{tablenotes}[para,flushleft]
\medskip
\scriptsize{
Notes: (1) Index-high: student with the highest chance of reaching level 0 and whose sum of SHAP values in the country $c$ sample is: $\Phi^{\text{max},c}_{i,c}$. (2) Index-low: student with the highest chance of reaching level 1, and whose sum of SHAP values in the country $c$ sample is: $\Phi^{\text{min},c}_{i,c}$. (3) Details on covariates can be found in Appendix B (Tables \ref{tableB1} and \ref{tableB2}).
}
\end{tablenotes}
\end{threeparttable}
\end{sidewaystable}

 \setlength{\tabcolsep}{1.8pt}
\begin{sidewaystable}
\begin{threeparttable}
\scriptsize
  \centering
\caption{Model—level 0 versus level 1 (low performing group). Covariates individual SHAP values for students with the lowest and highest average SHAP values contributions. Second country set: Mexico, Panama, Peru, Paraguay and Uruguay}
\begin{tabular}{llllll|llllll|llllll}
\hline
\multicolumn{6}{c}{Mexico} & \multicolumn{6}{c}{Panama} & \multicolumn{6}{c}{Peru} \\
\hline
\multicolumn{3}{c}{Low} & \multicolumn{3}{c}{High} & \multicolumn{3}{c}{Low} & \multicolumn{3}{c}{High} & \multicolumn{3}{c}{Low} & \multicolumn{3}{c}{High} \\
\hline
value & variable & SV &  value & variable & SV & value & variable & SV & value & variable & SV & value & variable & SV & value & variable & SV \\
 (1) & (2)  & (3)  & (4)  & (5)  & (6)  & (7)  & (8)  & (9)  & (10)  & (11)  & (12)  & (13)  & (14)  & (15)  & (16)  & (17) & (18) \\
-1.983 & escs & -1.690 & 1 & repeprim\_yes & 4.294 & -2.281 & autonomy & -5.764 & 0.95 & disadvantaged & 4.018 & 1.574 & autonomy & -1.653 & 1 & otherlang\_yes & 2.311 \\
0.819 & schsafelack & -1.431 & 10 & motheredu & 3.154 & 0.319 & escs & -3.789 & 0 & pcwebtr & 2.885 & -0.586 & escs & -0.991 & 1 & repeprim\_yes & 2.035 \\
0 & workpaid & -1.321 & 1 & missclass\_yes & 1.279 & 0.028 & pcwebtr & -2.901 & 1 & homework & 2.416 & 0 & workpaid & -0.931 & 3.906 & schclimabad & 1.486 \\
-1.812 & parentinvol & -0.933 & 5 & workpaid & 1.202 & 9 & devices & -1.883 & -2.425 & escs & 2.293 & 8 & devices & -0.858 & -3.273 & ictinfra & 1.351 \\
4 & homework & -0.895 & 0 & books & 1.182 & 0 & tchcertified & -1.790 & 1 & motheredu & 1.615 & 32 & str & -0.823 & 2 & homework & 0.776 \\
5 & workunpaid & -0.564 & -0.892 & autonomy & 1.045 & 8 & motheredu & -1.323 & 1.496 & schclimabad & 1.381 & 0.95 & pcwebtr & -0.784 & 1 & missclass\_yes & 0.688 \\
-0.172 & schclimabad & -0.405 & -0.298 & schsafelack & 0.808 & -0.384 & schclimabad & -1.247 & 0 & devices & 1.043 & 1.396 & suppinclusity & -0.724 & 7 & motheredu & 0.591 \\
0 & repeprim\_yes & -0.339 & 1.265 & schclimabad & 0.785 & 0 & repeprim\_yes & -1.032 & 1 & repeprim\_yes & 0.851 & 4 & workunpaid & -0.497 & 1 & workunpaid & 0.439 \\
2 & siblings & -0.312 & 6 & devices & 0.735 & 12.512 & str & -0.879 & 1 & missclass\_yes & 0.759 & 1 & siblings & -0.483 & 5 & workpaid & 0.351 \\
1102 & enrolment & -0.255 & -1.096 & escs & 0.683 & 0.68 & disadvantaged & -0.699 & -0.223 & ictinfra & 0.519 & 1 & tchpdatt & -0.477 & 17.385 & str & 0.326 \\
8 & satislife & -0.175 & 0 & polsocemskill\_yes & 0.522 & 3 & fatheredu & -0.642 & 3 & siblings & 0.514 & 1 & private & -0.430 & 113 & enrolment & 0.311 \\
0 & missclass\_yes & -0.171 & -0.03 & ictinfra & 0.479 & 5 & workpaid & -0.637 & 1 & otherlang\_yes & 0.374 & 0 & repeprim\_yes & -0.397 & 0.462 & tchdegree & 0.288 \\
1 & tchpdatt & -0.160 & -0.454 & suppinclusity & 0.265 & -0.974 & schsafelack & -0.512 & 0.034 & suppinclusity & 0.281 & 0 & lateprim\_yes & -0.314 & 10 & satislife & 0.277 \\
7 & devices & -0.159 & 0 & sex\_m & 0.201 & 0 & lateprim\_yes & -0.432 & 1 & workpaid & 0.252 & -0.487 & schclimabad & -0.247 & 0 & pcwebtr & 0.252 \\
24.49 & str & -0.157 & 1 & sex\_f & 0.191 & 0 & immigrant\_yes & -0.246 & 1 & workunpaid & 0.250 & -0.362 & ictinfra & -0.272 & -0.992 & escs & 0.229 \\
  &  & &  & & &  &  &  &  &  & &  &  &  &  &   \\
  \hline
\multicolumn{6}{c}{Paraguay} & \multicolumn{6}{c}{Uruguay} \\
\hline
\multicolumn{3}{c}{Low} & \multicolumn{3}{c}{High} & \multicolumn{3}{c}{Low} & \multicolumn{3}{c}{High} \\
\hline
value & variable & SV &  value & variable & SV & value & variable & SV & value & variable & SV & & &\\
(1) & (2)  & (3)  & (4)  & (5)  & (6)  & (7)  & (8)  & (9)  & (10)  & (11)  & (12)  &   &   &   &   &  &  \\
1.028 & autonomy & -6.389 & 3 & devices & 4.142 & 0 & repeprim\_yes & -2.530 & 1.248 & str & 4.465 &  &  &  &  &  &  \\
1 & private & -5.746 & 0.231 & escs & 3.072 & 0.485 & escs & -2.375 & 10 & fatheredu & 3.285 &  &  &  &  &  &  \\
904 & enrolment & -5.032 & 1 & workunpaid & 2.961 & 0.714 & tchcertified & -1.987 & 0 & devices & 2.518 &  &  &  &  &  &  \\
11 & devices & -4.791 & -1.636 & autonomy & 2.022 & 0.75 & tchpdatt & -0.872 & 1 & repeprim\_yes & 2.473 &  &  &  &  &  &  \\
0.655 & escs & -1.884 & 0.16 & tchpdatt & 1.407 & 3 & books & -0.783 & 1 & satislife & 1.901 &  &  &  &  &  &  \\
1 & urban & -1.786 & 1 & homework & 1.286 & 0 & repelowsec\_yes & -0.634 & 156 & enrolment & 1.164 &  &  &  &  &  &  \\
7 & satislife & -1.629 & 140 & enrolment & 1.165 & 10 & devices & -0.572 & 0.3 & tchcertified & 1.053 &  &  &  &  &  &  \\
-0.42 & schclimabad & -1.455 & 0 & tchdegree & 1.058 & -1.286 & schsafelack & -0.559 & 0.032 & pcwebtr & 0.765 &  &  &  &  &  &  \\
9 & fatheredu & -1.265 & -0.58 & schclimabad & 0.981 & 3 & siblings & -0.478 & -0.704 & schclimabad & 0.459 &  &  &  &  &  &  \\
5 & homework & -1.023 & 1 & otherlang\_yes & 0.835 & 5 & workunpaid & -0.439 & 0.75 & disadvantaged & 0.398 &  &  &  &  &  &  \\
1 & books & -0.987 & 56 & str & 0.789 & 8 & motheredu & -0.393 & 0.629 & schsafelack & 0.376 &  &  &  &  &  &  \\
0 & otherlang\_yes & -0.934 & 1 & sex\_f & 0.718 & 0.095 & pcwebtr & -0.391 & 3 & motheredu & 0.339 &  &  &  &  &  &  \\
0 & disadvantaged & -0.836 & 0 & sex\_m & 0.617 & 10.476 & str & -0.390 & 0.274 & parentinvol & 0.319 &  &  &  &  &  &  \\
0 & tchdegree & -0.770 & -0.102 & ictinfra & 0.609 & 2 & fatheredu & -0.380 & -0.75 & escs & 0.255 &  &  &  &  &  &  \\
0 & repeprim\_yes & -0.720 & 0 & private & 0.457 & 0.01 & tchdegree & -0.338 & 1 & siblings & 0.192 &  &  &  &  &  &  \\
\hline
\end{tabular}
\label{table7}
\begin{tablenotes}[para,flushleft]
\medskip
\scriptsize{
Notes: (1) Index-high: student with the highest chance of reaching level 0 and whose sum of SHAP values in the country $c$ sample is: $\Phi^{\text{max},c}_{i,c}$. (2) Index-low: student with the highest chance of reaching level 1, and whose sum of SHAP values in the country $c$ sample is: $\Phi^{\text{min},c}_{i,c}$. (3) Details on covariates can be found in Appendix B (Tables \ref{tableB1} and \ref{tableB2}).
}
\end{tablenotes}
\end{threeparttable}
\end{sidewaystable}


\clearpage
\appendix


\newpage
\renewcommand{\thesubsection}{\Alph{subsection}}
\section{Evaluation metrics}\label{appendixA}

Here I describe the different metrics to assess the five ML models used with a distinct hyperparameters configurations. The Accuracy (ACC) metric equation is as follows:

\begin{equation}
  ACC = \frac{\text{TP + TN}}{\text{TP + TN + FP + FN}}
\end{equation}

where TP (TN) is the true (negative) positive rate, and FP (FN) is the false positive (negative) rate. The second metric, recall (RC), measures the capacity of the model to efficiently identify all positive cases, and it is defined as:

\begin{equation}
  RC = \frac{\text{TP}}{\text{TP + FN}}
\end{equation}

The precision metric (PR) quantifies, among all instances of classified positive responses, those accurately identified as positive responses:

\begin{equation}
  PR = \frac{\text{TP}}{\text{TP + FP}}
\end{equation}

The complement of the PR metric is the specificity metric (SP) which measures the percentage of correctly classified negative samples (true negatives) with respect to all samples classified as negative, and it is given by:

\begin{equation}
  SP = \frac{\text{TN}}{\text{TN + FP}}
\end{equation}

The F1 score metric combines the precision and recall scores of a model by using the harmonic mean of PR and RC, yielding a balance between the model accuracy and
its completeness. F1 is obtained as:

\begin{equation}
  ACC = \frac{2 \times (\text{PR} \times \text{RC})}{\text{PR + RC}}
\end{equation}

Finally, the Area Under the Curve (AUC) metric is defined as the integral of the ROC curve, allowing a assessment of the classification
performance for varying discrimination thresholds, with a AUC equals to one as perfect classification and a AUC of 0.5 equals to a random classification. The AUC is calculated as follows:

\begin{equation}
AUC = \int_{0}^{1} ROC(t) dt
\end{equation}


\setcounter{table}{0}
\renewcommand{\thetable}{B\arabic{table}}

\setlength{\tabcolsep}{1.1pt}
\begin{sidewaystable}
\scriptsize
  \centering
\caption{Students and family variables$\textendash$description}
\begin{tabular}{llllll}
\hline
Variable & Name & Question & Original variable & Variable type & Variable label \\
\hline
 & (1) & (2) & (3) & (4) & (5) \\
Gender – female & sex\_f & Q3 & st004d01t & Binary & 1=female, 0=male \\
Satisfaction with life & satislife & Q42 & st016q01na  & Discrete & 1=lowest,…,10=highest \\
Homework – all subjects (time spent) & homework & Q44 & st296q04ja & Discrete & 1 (< 30min), 2 (30min-1hour), 3 (1- 2 hours),  \\
 &  &  &  &  & 4 (2-3 hours), 5 (3-4 hours), 6 (> 4 hours)   \\
Digital devices at home-number & devices & Q8 & st254q02ja to st254q06ja & Discrete & 1 to 20 \\
Books – number & books & Q9 & st255q01ja  & Discrete & 0 (none), 1 (1 to 10), 2 (11 to 25), 3 (26 to100),  \\
 &  &  &  &  & 4 (101 to 200), 5 (201 to 500), 6 (> 500)   \\
Immigrant & immigrant\_yes & Q20 & st021q01ta  & Binary & 1=immigrant, 0=no immigrant \\
Speak other language at home & otherlang\_yes & Q21 & st022q01ta & Binary & 1=speak other language at home, 0=otherwise \\
Late entry – primary & lateprim\_yes & Q24 & st126q01ta  & Binary & 1=late entrant in primary, 0=otherwise \\
Repetition – primary & repeprim\_yes & Q25 & st127q01ta & Binary & 1=repeated primary, 0=otherwise \\
Repetition – lower secondary & repelowsec\_yes & Q25 & st127q02ta & Binary & 1=repeated lower secondary, 0=otherwise \\
Missed school (3 months in a row) & missclass\_yes & Q26 & st260q02ja & Binary & 1=missed school 3 months in a row (lower secondary), 0=otherwise \\
Skipped school (1 day last 2 weeks) & skippsch\_yes & Q28 & st062q01ta & Binary & 1=skipped school (1 whole day last 2 weeks), 0=otherwise \\
Teacher support for well being (index) & tchsuppwb & Q29 & st267q01ja to st267q08ja & Continuous & Intervals: low (-4.634, 0.053) and high (0.054, -14.024) \\
Sense of belonging to school (index) & belongsch & Q30 & St034q01ta to st034q06ta & Continuous & Intervals: low (-5.110, -0.039) and high (-0.038, 4.326) \\
School climate – bad (index) & schclimabad & Q31 & st038q03na to st038q11ja & Continuous & Intervals: low (-0.753, -0.296) and high (-0.297, 9.336) \\
Safety in school and surroundings (index) & schsafelack & Q32 & st265q01ja to st265q04ja & Continuous & Intervals: low (-1.257, 0.357) and high (0.358, 4.077) \\
Work – unpaid, days & workunpaid & Q34 & st294q03ja & Discrete & 0 (0 day), 1 (1 day), 2 (2 days), 3 (3 days), 4 (4 days), 5 ( $ge$ 5 days) \\
Work – paid, days & workpaid & Q35 & st294q04ja & Discrete & As above \\
Mother education & motheredu & Q11 & st005q01ja & Discrete & 1 ($<$ ISCED1), 2 (ISCED level 1), 3 (ISCED level 2), 4 (ISCED level 3.3),  \\
 &  &  &  &  & 5 (ISCED level 3.4), 6 (ISCED level 4),…,8 (ISCED level 8)  \\
Father education & fatheredu & Q13 & st007q01ja & Discrete & As above \\
Siblings number & siblings & Q10 & st230q01ja  & Discrete & 1 (one), 2 (two), 3 (3 or more) \\
SES (index) & ses & derived & na & Continuous & Intervals: low (-6.654, -0.852), high (-0.851, 2.039) \\
Family support for education (index) & famsuppedu & Q55 & st300q04ja to st300q10ja & Continuous & Intervals: low (-2.728, 0.174), high (0.174, 1.185) \\
 \hline
\end{tabular}
\label{tableB1}
\begin{tablenotes}[para,flushleft]
\medskip
(1) For further details on student and family variables' definition and measurement, check PISA 2022 Student Questionnaire:
\href{https://www.oecd.org/en/data/datasets/pisa-2022-database.html#questionnaires}{link}.
\end{tablenotes}
\end{sidewaystable}

\setlength{\tabcolsep}{1.1pt}
\begin{sidewaystable}
\scriptsize
  \centering
\caption{School variables$\textendash$description}
\begin{tabular}{llllll}
\hline
Variable & Name & Question & Original variable & Variable type & Variable label \\
\hline
 & (1) & (2) & (3) & (4) & (5) \\
Urban  & urban & Q1 & sc001q01ta & Binary & 1=urban, 0=rural \\
Private & private & Q2 & sc013q01ta & Binary & 1=private, 0=public \\
Town size & townsize & Q1 & sc001q01ta & Discrete & 1 (village: < 3000), 2 (small town: 3000-15000),  \\
 &  &  &  &  & 3 (town: 15000-100000), 4 (city, large city: \$ge\$ 1million) \\
Competition & competition & Q5 & sc011q01ta & Binary & 1=competition, 0=no competition \\
Funding government  & fundgovt & Q4 & sc016q01ta & Continuous & 0-1 (rate) \\
Funding fees  & fundfees & Q4 & sc016q02ta & Continuous & 0-1 (rate) \\
Students – second language  & nolangtest & Q7 & sc211q01ja & Continuous & 0-1 (rate) \\
Students - learning needs   & learnneeds & Q8 & sc211q02ja  & Continuous & 0-1 (rate) \\
Students – disadvantaged  & disadvantaged & Q9 & sc211q03ja  & Continuous & 0-1 (rate) \\
School size (enrolment) & enrolment & Q6 & sc002q01ta to sc002q02ta & Continuous & 0 to 7000 \\
STR & srt & Q6,Q8 & sc018q01ta01, sc018q01ta02 & Continuous & 0 to720 \\
Teachers attended PD programme (last 3 months)  & tchpdatt & Q21 & sc025q01na & Continuous & 0-1 (rate) \\
STR non-teaching staff, pedagogic support & strpedag & Q6, Q10 & sc168q01ja & Continuous & 0-1 (rate) \\
Teachers – certified  & tchcertified & Q8 & Sc018q02ta01, sc018q02ta02  & Continuous & 0-1 (rate) \\
Teachers – degree or higher  & tchdegree & Q8 & sc018q08ja01 to sc018q09ja02  & Continuous & 0-1 (rate) \\
ICT – infrastructure per student (index) & ictinfra & Q15 & sc004q01ta,sc004q02ta, sc004q03ta, & Continuous & Intervals: low (-4.683, -0.060) and high (0.061, 37.96)  \\
 &  &  &  sc004q07na, sc004q08ja &  &  \\
PC connected web-teacher ratio & pcwebtr & Q15 & Sc004q01ta, sc004q03ta & Continuous & 0-1 (rate) \\
Admission criteria & admissionnone\_yes & Q11 & sc012q01ta to sc012q12ja & Binary & 1=admission (any criteria used), 0=otherwise \\
Transfer policies  & transfernone\_yes & Q12 & sc185q01wa sc185q03wa sc185q04wa  & Binary & 1=transfer (due to performance and behaviour), 0=otherwise \\
Autonomy (index) & autonomy & Q13 & sc202q01ja to sc202q12ja & Continuous & Intervals: low (-2.268, 0.068) and high (0.069, 1.794)  \\
Social and emotional skills policy & polsocemskill\_yes & Q48 & sc189q01ja & Binary & 1=policy exists, 0=otherwise \\
Teacher monitoring – methods  & tchmonitor & Q19 & sc032q01ta to sc032q04ta & Discrete & 0, 1, 2, 3 and 4 \\
Testing students learning, all methods & teststud\_yes & Q37 & sc034q01na sc034q02na sc034q03ta & Binary & 1=test, 0=otherwise \\
Staff students support for inclusivity (index) & suppinclusity & Q28 & sc173q01ja to sc173q06ja & Continuous & Intervals: low (-2.290, 0.084) and high (0.085, 1.399)  \\
Parents involvement fostered by school (index) & parentinvol & Q30 & sc192q01ja to sc192q06ja & Continuous & Intervals: low (-2.666, -0.043) and high (-0.042, 2.490)  \\
School policy  – ability grouping & abilitygroup\_yes & Q39 & sc042q01ta, sc042q02ta & Binary & 1=ability grouping, 0=otherwise \\
 \hline
\end{tabular}
\label{tableB2}
\begin{tablenotes}[para,flushleft]
\medskip
(1) For further details on school variables' definition and measurement, check PISA 2022 School Questionnaire:
\href{https://www.oecd.org/en/data/datasets/pisa-2022-database.html#questionnaires}{link}.
\end{tablenotes}
\end{sidewaystable}


\setcounter{table}{0}
\renewcommand{\thetable}{C\arabic{table}}

\begin{sidewaystable}
\footnotesize
  \centering
\caption{Outcome (performance level 0 versus level 1). Comparison of ML models – parameters search}
\begin{tabular}{llllllll}
\hline
Model & Parameters & \multicolumn{6}{c}{Evaluation metrics}  \\
  \cmidrule{3-8}
 &  & ACC & RC & F1 & PR & SP & AUC \\
 \hline
 &  & (1) & (2) & (3) & (4) & (5) & (6) \\
Logistic  & \textbf{penalty = none} & \textbf{0.7305} & \textbf{0.8609} & 0.7591 & 0.4846 & \textbf{0.8068} & \textbf{0.7825} \\
regression & penalty = l2 & 0.7299 & 0.8548 & \textbf{0.7613} & \textbf{0.4943} & 0.8053 & 0.7823 \\
\hline
 &  &  &  &  &  & &   \\
Lasso  & C = 10 & 0.7533 & 0.8641 & 0.7816 & 0.5443 & 0.8208 & 0.8143 \\
with CV penalty & C = 50 & \textbf{0.7549} & \textbf{0.8702} & 0.7801 & 0.5373 & \textbf{0.8227} & 0.8139 \\
 & \textbf{C = 100} & 0.7527 & 0.8576 & \textbf{0.7843} & \textbf{0.5549} & 0.8193 & \textbf{0.8147} \\
 & C = 200 & 0.7524 & 0.8571 & 0.7842 & \textbf{0.5549} & 0.8190 & \textbf{0.8147} \\
 \hline
 &  &  &  &  &  & &   \\
Gradient boosting & Trees = 100, learning rate = 0.01 & 0.6989 & \textbf{0.9725} & 0.6918 & 0.1826 & 0.8085 & 0.7871 \\
 & Trees = 100, learning rate = 0.10 & 0.7752 & 0.8730 & 0.8010 & 0.5909 & 0.8354 & 0.8560 \\
 & Trees = 500, learning rate = 0.01 & 0.7588 & 0.8813 & 0.7788 & 0.5277 & 0.8269 & 0.8378 \\
 & Trees = 500, learning rate = 0.10 & 0.7737 & 0.8632 & 0.8048 & 0.6049 & 0.8330 & 0.9037 \\
 & Trees = 1000, learning rate = 0.01 & 0.7752 & 0.8748 & 0.8000 & 0.5874 & \textbf{0.8357} & 0.8559 \\
 & \textbf{Trees = 1000, learning rate = 0.10} & \textbf{0.7755} & 0.8553 & \textbf{0.8115} & \textbf{0.6251} & 0.8328 & \textbf{0.9292} \\
 \hline
 &  &  &  &  &  & &   \\
Random forest  & Trees = 100, max depth = 5 & 0.7503 & 0.8474 & 0.7869 & 0.5672 & 0.8160 & 0.8074 \\
 & Trees = 100, max depth = 10 & 0.7616 & 0.8571 & 0.7943 & 0.5812 & 0.8245 & 0.9305 \\
 & Trees = 500, max depth = 5 & 0.7533 & 0.8483 & 0.7899 & 0.5742 & 0.8180 & 0.8083 \\
 & Trees = 500, max depth = 10 & 0.7673 & 0.8609 & 0.7988 & 0.5909 & 0.8287 & 0.9319 \\
 & Trees = 1000, max depth = 5 & 0.7530 & 0.8492 & 0.7890 & 0.5716 & 0.8180 & 0.8086 \\
 & \textbf{Trees = 1000, max depth = 10} & \textbf{0.7692} & \textbf{0.8627} & \textbf{0.7998} & \textbf{0.5926} & \textbf{0.8301} & \textbf{0.9330} \\
 \hline
 &  &  &  &  &  & &   \\
Neural net & 1 hidden layer (200 nodes),  activation = relu & 0.6676 & 0.6598 & 0.7966 & \textbf{0.6822} & 0.7218 & 0.7565 \\
 & 1 hidden layer (200 nodes),  activation = logistic & \textbf{0.7488} & 0.8395 & 0.7895 & 0.5777 & \textbf{0.8137} & 0.8704 \\
 & 1 hidden layer (200 nodes),  activation = tanh & 0.7296 & 0.7604 & \textbf{0.8137} & 0.6716 & 0.7861 & 0.8573 \\
 & 3 hidden layers (200, 100 and 50 nodes),  activation = relu & 0.7302 & \textbf{0.8432} & 0.7671 & 0.5171 & 0.8034 & 0.8676 \\
 & \textbf{3 hidden layers (200, 100 and 50 nodes),  activation = logistic} & 0.7479 & 0.8423 & 0.7870 & 0.5698 & 0.8137 & \textbf{0.8773} \\
 & 3 hidden layers (200, 100 and 50 nodes),  activation = tanh & 0.7403 & 0.8399 & 0.7797 & 0.5522 & 0.8087 & 0.8599 \\
\hline
\end{tabular}
\label{tableC1}
\begin{tablenotes}[para,flushleft]
\medskip
\scriptsize{
(1) Within each learner$/$model, highest metrics are shown in bold. (2) Also, the chosen most efficient configuration of each model is highlighted in bold (that is, models with the largest number of top preforming metrics) and in case of a draw, we rely on the AUC metric (column 6) to reach a final selection. (3) Some of the additional (default) parameters are as follows. Logitlasso: penalty(l1)  solver(saga); GB: subsample(1), max\_depth(3), max\_features(None); RF: max\_features(1.0), min\_samples\_split(2); NN: solver(adam), batch\_size(auto).}
\end{tablenotes}
\end{sidewaystable}

\begin{sidewaystable}
\footnotesize
  \centering
\caption{Outcome (performance level 1 versus level 2). Comparison of ML models – parameters search}
\begin{tabular}{llllllll}
\hline
Model & Parameters & \multicolumn{6}{c}{Evaluation metrics}  \\
  \cmidrule{3-8}
 &  & ACC & RC & F1 & PR & SP & AUC \\
 \hline
 &  & (1) & (2) & (3) & (4) & (5) & (6) \\
Logistic  & \textbf{penalty = none} & \textbf{0.6383} & 0.7868 & \textbf{0.6630} & \textbf{0.4246} & \textbf{0.7196} & \textbf{0.7021} \\
regression & penalty = l2 & 0.6284 & \textbf{0.7947} & 0.6517 & 0.3891 & 0.7162 & 0.6941 \\
\hline
 &  &  &  &  &  &  & \\
Lasso  & \textbf{C = 10} & 0.6486 & 0.7850 & \textbf{0.6735} & \textbf{0.4525} & 0.7250 & \textbf{0.7233} \\
with CV penalty & C = 50 & \textbf{0.6492} & 0.7877 & 0.6732 & 0.4499 & 0.7259 & \textbf{0.7233} \\
 & C = 100 & 0.6476 & \textbf{0.8106} & 0.6652 & 0.4132 & \textbf{0.7307} & 0.7198 \\
 & C = 200 & 0.6486 & 0.7938 & 0.6709 & 0.4398 & 0.7272 & 0.7231 \\
 \hline
 &  &  &  &  &  & &  \\
Gradient boosting & Trees = 100, learning rate = 0.01 & 0.6268 & \textbf{0.9507} & 0.6198 & 0.1610 & 0.7503 & 0.713 \\
 & Trees = 100, learning rate = 0.10 & 0.6622 & 0.7912 & 0.6850 & 0.4766 & 0.7343 & 0.7868 \\
 & Trees = 500, learning rate = 0.01 & 0.6512 & 0.8176 & 0.6667 & 0.4119 & \textbf{0.7345} & 0.7625 \\
 & Trees = 500, learning rate = 0.10 & 0.6684 & 0.7692 & 0.6990 & 0.5234 & 0.7324 & 0.8647 \\
 & Trees = 1000, learning rate = 0.01 & 0.6590 & 0.7956 & 0.6805 & 0.4626 & 0.7335 & 0.7863 \\
 & \textbf{Trees = 1000, learning rate = 0.10} & \textbf{0.6694} & 0.7595 & \textbf{0.7037} & \textbf{0.5399} & 0.7305 & \textbf{0.9057} \\
 \hline
 &  &  &  &  &  &  & \\
Random forest  & Trees = 100, max depth = 5 & 0.6440 & \textbf{0.8564} & 0.6506 & 0.3384 & \textbf{0.7394} & 0.742 \\
 & Trees = 100, max depth = 10 & 0.6481 & 0.7833 & 0.6735 & 0.4537 & 0.7242 & 0.9354 \\
 & Trees = 500, max depth = 5 & 0.6429 & 0.8520 & 0.6507 & 0.3422 & 0.7379 & 0.7439 \\
 & Trees = 500, max depth = 10 & 0.6533 & 0.7877 & 0.6773 & 0.4601 & 0.7283 & \textbf{0.9378} \\
 & Trees = 1000, max depth = 5 & 0.6450 & 0.8520 & 0.6525 & 0.3473 & 0.7390 & 0.7432 \\
 & \textbf{Trees = 1000, max depth = 10} & \textbf{0.6585} & 0.7947 & \textbf{0.6802} & \textbf{0.4626} & 0.7330 & 0.9377 \\
 \hline
 &  &  &  &  &  & &  \\
Neural net & 1 hidden layer (200 nodes),  activation = relu & 0.6133 & \textbf{0.9568} & 0.6098 & 0.1191 & \textbf{0.7449} & 0.6999 \\
 & 1 hidden layer (200 nodes),  activation = logistic & 0.6429 & 0.6502 & 0.7179 & 0.6324 & 0.6824 & 0.7307 \\
 & \textbf{1 hidden layer (200 nodes),  activation = tanh} & \textbf{0.6481} & 0.6485 & \textbf{0.7258} & \textbf{0.6477} & 0.6850 & 0.7502 \\
 & 3 hidden layers (200, 100 and 50 nodes),  activation = relu & 0.6263 & 0.8564 & 0.6361 & 0.2953 & 0.7300 & 0.7105 \\
 & 3 hidden layers (200, 100 and 50 nodes),  activation = logistic & 0.6502 & 0.7137 & 0.6995 & 0.5589 & 0.7065 & \textbf{0.7979} \\
 & 3 hidden layers (200, 100 and 50 nodes),  activation = tanh & 0.6388 & 0.7101 & 0.6877 & 0.5361 & 0.6987 & 0.7874 \\
\hline
\end{tabular}
\label{tableC2}
\begin{tablenotes}[para,flushleft]
\medskip
\scriptsize{
(1) Within each learner$/$model, highest metrics are shown in bold. (2) Also, the chosen most efficient configuration of each model is highlighted in bold (that is, models with the largest number of top preforming metrics) and in case of a draw, we rely on the AUC metric (column 6) to reach a final selection. (3) Some of the additional (default) parameters are as follows. Logitlasso: penalty(l1)  solver(saga); GB: subsample(1), max\_depth(3), max\_features(None); RF: max\_features(1.0), min\_samples\_split(2); NN: solver(adam), batch\_size(auto).}
\end{tablenotes}
\end{sidewaystable}

\begin{sidewaystable}
\footnotesize
  \centering
\caption{Comparison of different stacked models}
\begin{tabular}{llllllll}
\hline
 &  & ACC & RC & PR & SP & F1 & AUC \\
 \hline
 &  & (1) & (2) & (3) & (4) & (5) & (6) \\
\textit{Panel A – Model Gradient boosting} &  &  &  &  &  &  &  \\
Outcome – level 0 versus level 1 &  &  &  &  &  &  &  \\
trees & learning rates &  &  &  &  &  &  \\
100 & 0.01 & 0.7746 & 0.8688 & 0.8027 & 0.5970 & 0.8344 & 0.9191 \\
100 & 0.10 & 0.7719 & \textbf{0.8795} & 0.7938 & 0.5689 & 0.8344 & 0.8792 \\
500 & 0.01 & 0.7698 & 0.8725 & 0.7952 & 0.5759 & 0.8320 & 0.9125 \\
500 & 0.10 & \textbf{0.7795} & 0.8776 & 0.8032 & 0.5944 & \textbf{0.8388} & 0.9145 \\
1000 & 0.01 & 0.7746 & 0.8790 & 0.7970 & 0.5777 & 0.8360 & 0.8822 \\
1000 & 0.10 & 0.7783 & 0.8734 & \textbf{0.8042} & \textbf{0.5988} & 0.8374 & \textbf{0.9299} \\
 &  &  &  &  &  &  &  \\
\textit{Panel B – Lasso (with CV penalty)} &  &  &  &  &  &  &  \\
Outcome – level 1 versus level 2 &  &  &  &  &  &  &  \\
C &  &  &  &  &  &  &  \\
10 &  & 0.6741 & 0.7947 & \textbf{0.6960} & \textbf{0.5006} & 0.7421 & 0.8825 \\
50 &  & 0.6741 & 0.7947 & \textbf{0.6960} & \textbf{0.5006} & 0.7421 & 0.8827 \\
100 &  & \textbf{0.6752} & \textbf{0.8009} & 0.6950 & 0.4943 & \textbf{0.7442} & \textbf{0.8894} \\
200 &  & 0.6736 & 0.7956 & 0.6952 & 0.4981 & 0.7420 & 0.8852 \\
\hline
\end{tabular}
\label{tableC3}
\begin{tablenotes}[para,flushleft]
\medskip
\scriptsize{
(1) For assessment of different stacked models we select the model with the highest weight from interaction 2 (see Table \ref{table4}) for each outcome, and then calculate evaluation metrics for the stacked model using the parameters grid of each model. (2) Within each stacked model, highest metrics are shown in bold.}
\end{tablenotes}
\end{sidewaystable}


\setcounter{figure}{0}
\renewcommand{\thefigure}{D\arabic{figure}}

\begin{figure}
\includegraphics[clip,width=\columnwidth]{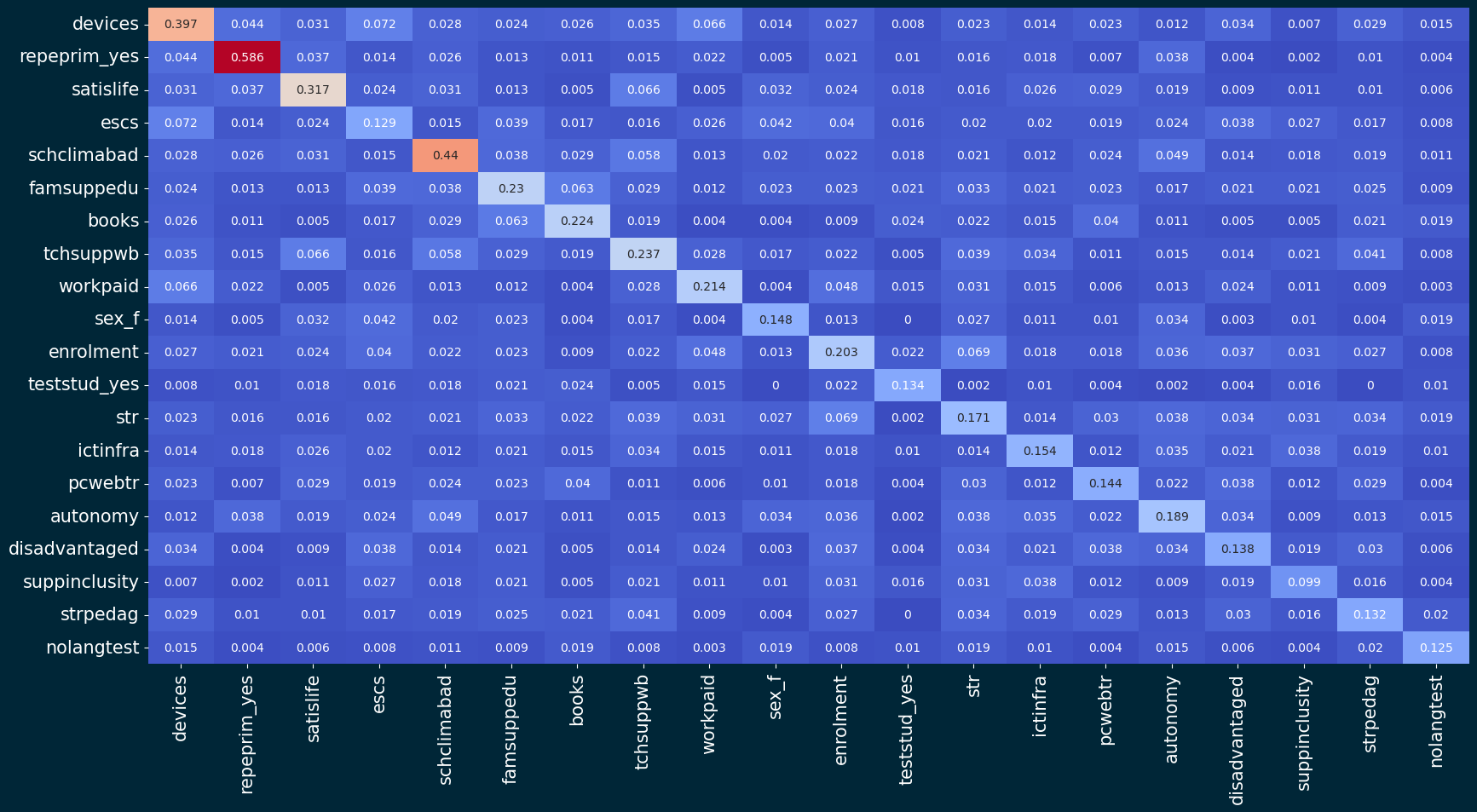}
\includegraphics[clip,width=\columnwidth]{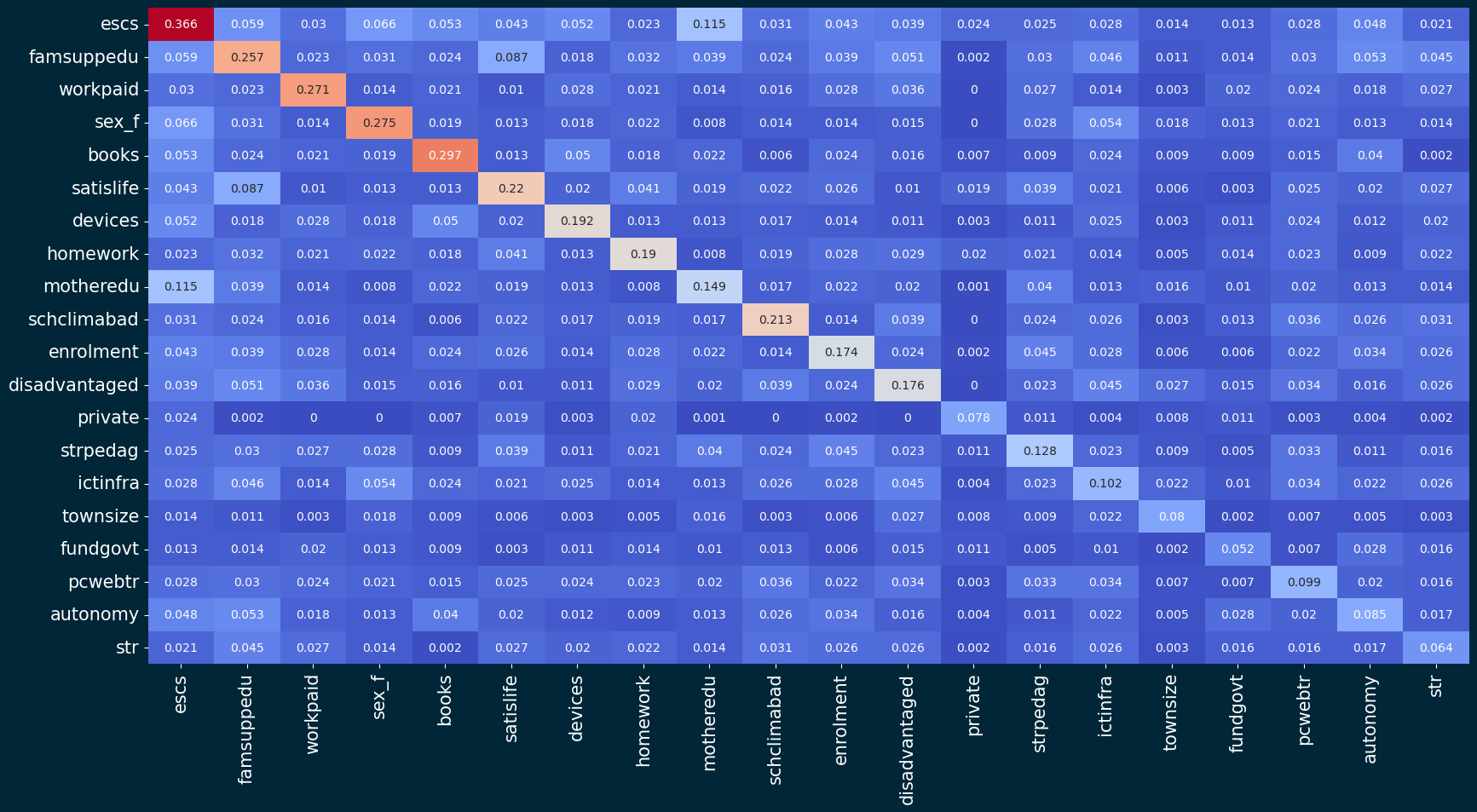}
\caption{Intersectionality$\textendash$SHAP values interactions}
\label{figureD1}
        \medskip
\begin{minipage}{0.99\textwidth}
\scriptsize{
Notes: (1) The heatmap plots of correlations across SHAP values include the top 20 covariates across students$/$family and school features. (2) Top plot (level 0 versus level 1) and bottom plot (level 1 versus level 2). 
}
\end{minipage}
\end{figure}

\begin{figure}
\begin{multicols}{2}
\centering
\includegraphics[width=0.50\textwidth]{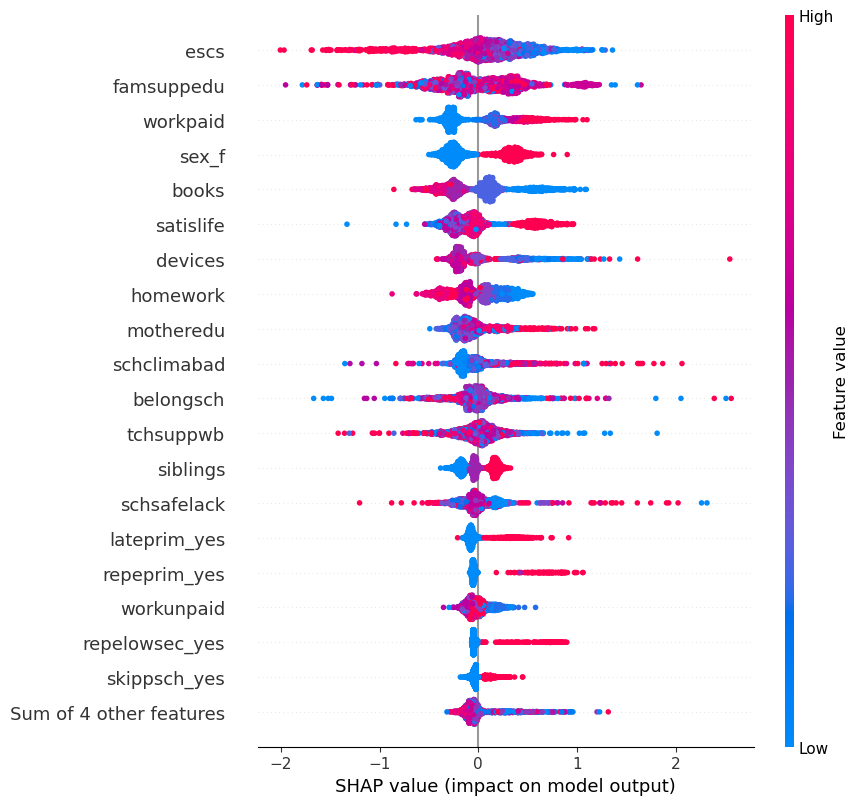}
\includegraphics[width=0.50\textwidth]{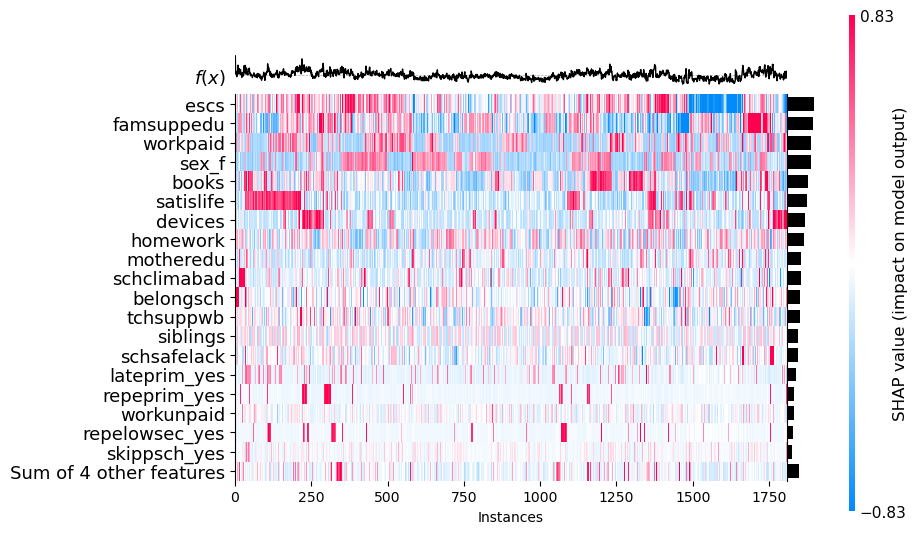}
\includegraphics[width=0.50\textwidth]{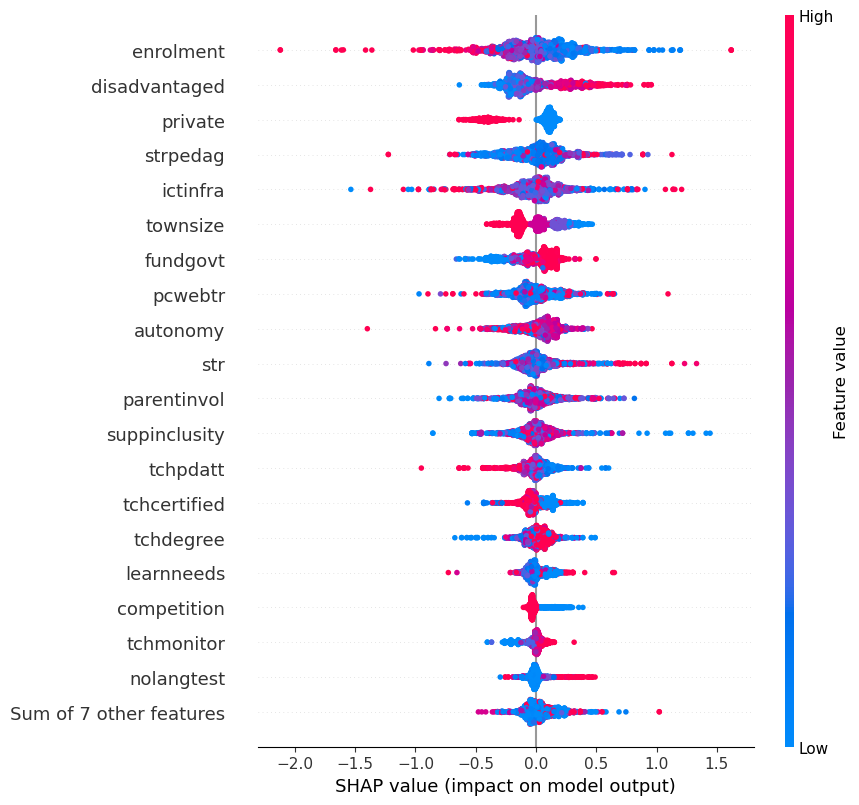}
\includegraphics[width=0.50\textwidth]{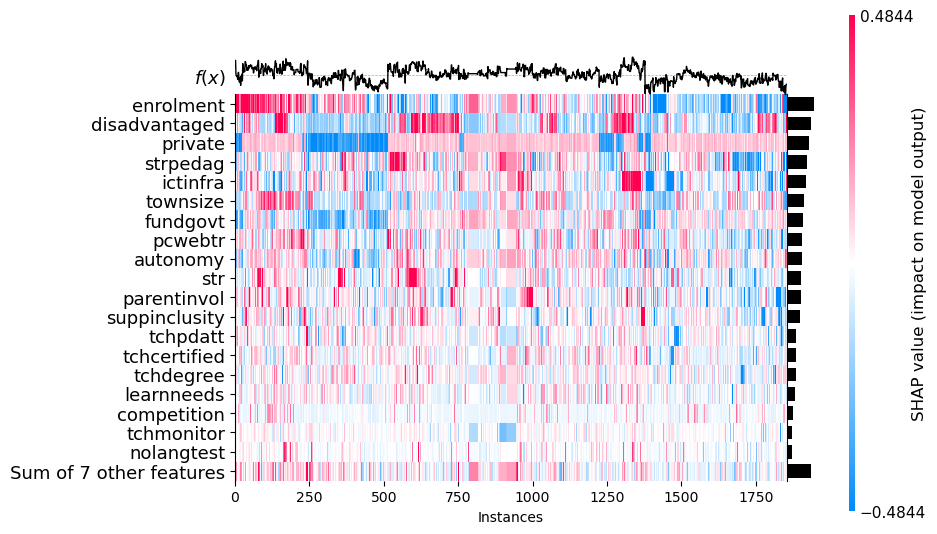}
\end{multicols}
\caption{Model: level 1 versus level 2 (low performing group). Beeswarm dot plot and heatmap of individual SHAP values for each student$/$famly and school covariate across observations}
\label{figureD2}
   \medskip
\begin{minipage}{0.99\textwidth}
\scriptsize{
Notes: (1) Beeswarm plot. Red represents that covariates is = 1 and, blue (= 0) otherwise. (2) Heatmaps. Colour intensity reflects the degree of impact on student performance, ranging from low (blue) to high (red).
}
\end{minipage}
\end{figure}


     \begin{figure*}[hp!]
        \centering
        \subfloat[Argentina]{%
            \includegraphics[width=0.26\textwidth]{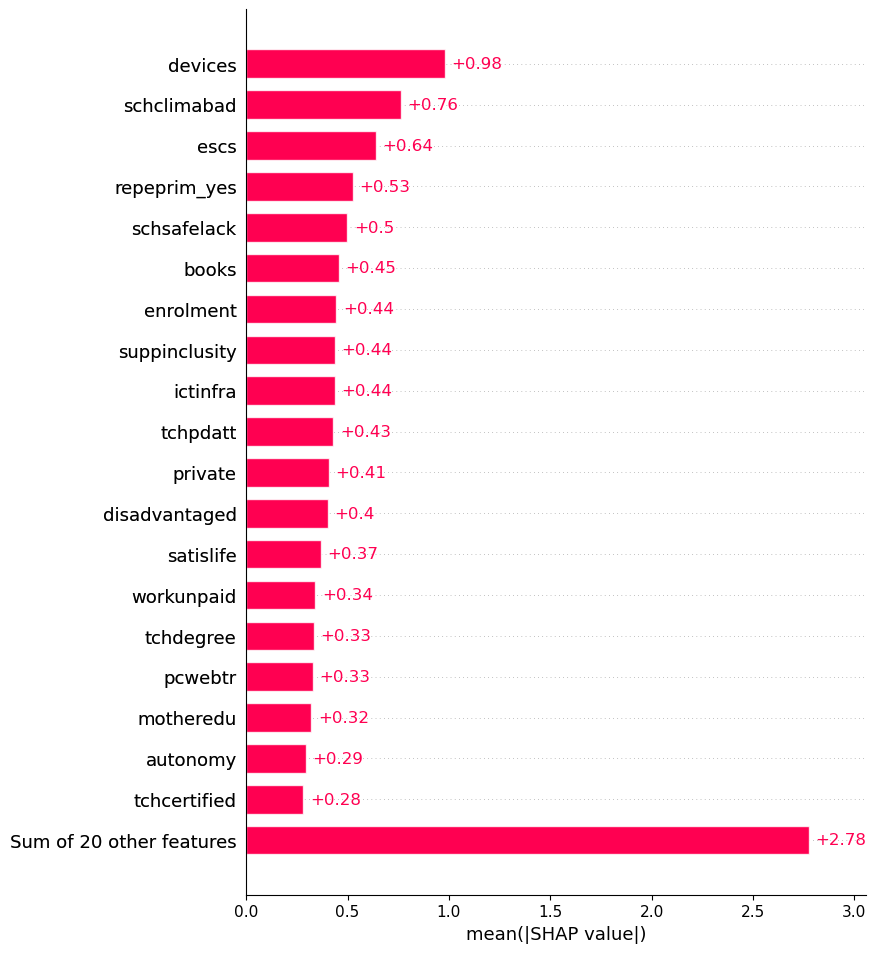}
        }
        \subfloat[Brazil]{%
            \includegraphics[width=0.26\textwidth]{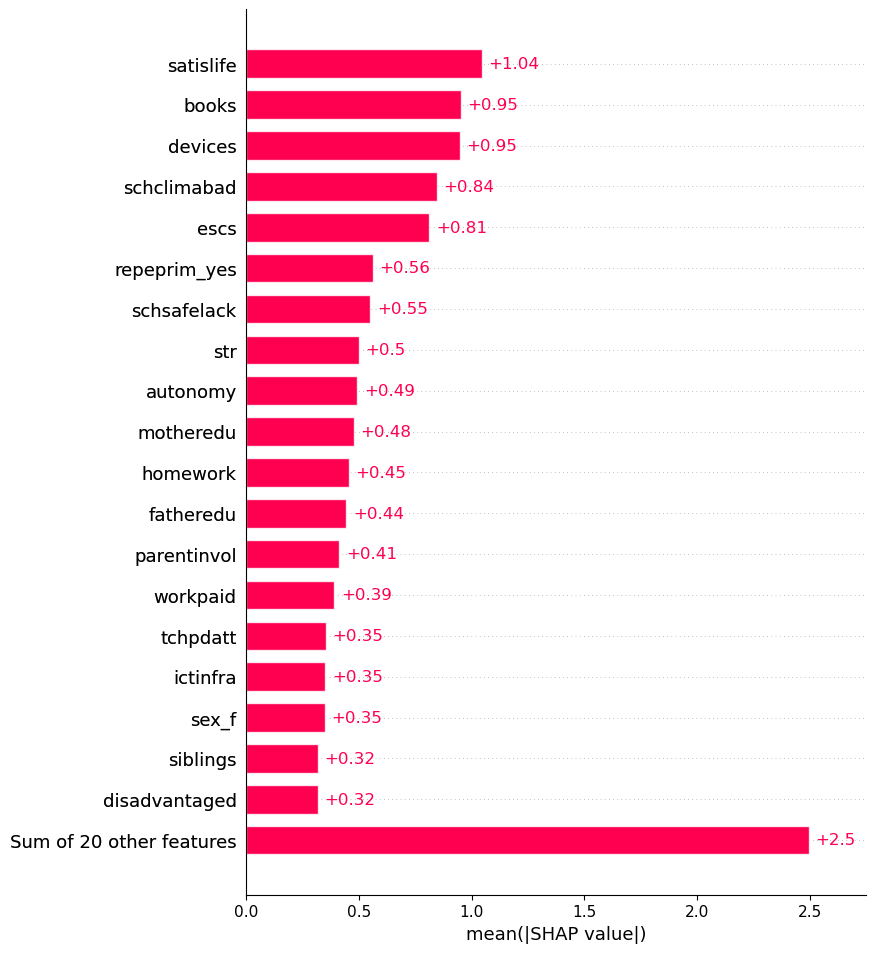}
        }
        \subfloat[Chile]{%
            \includegraphics[width=0.26\textwidth]{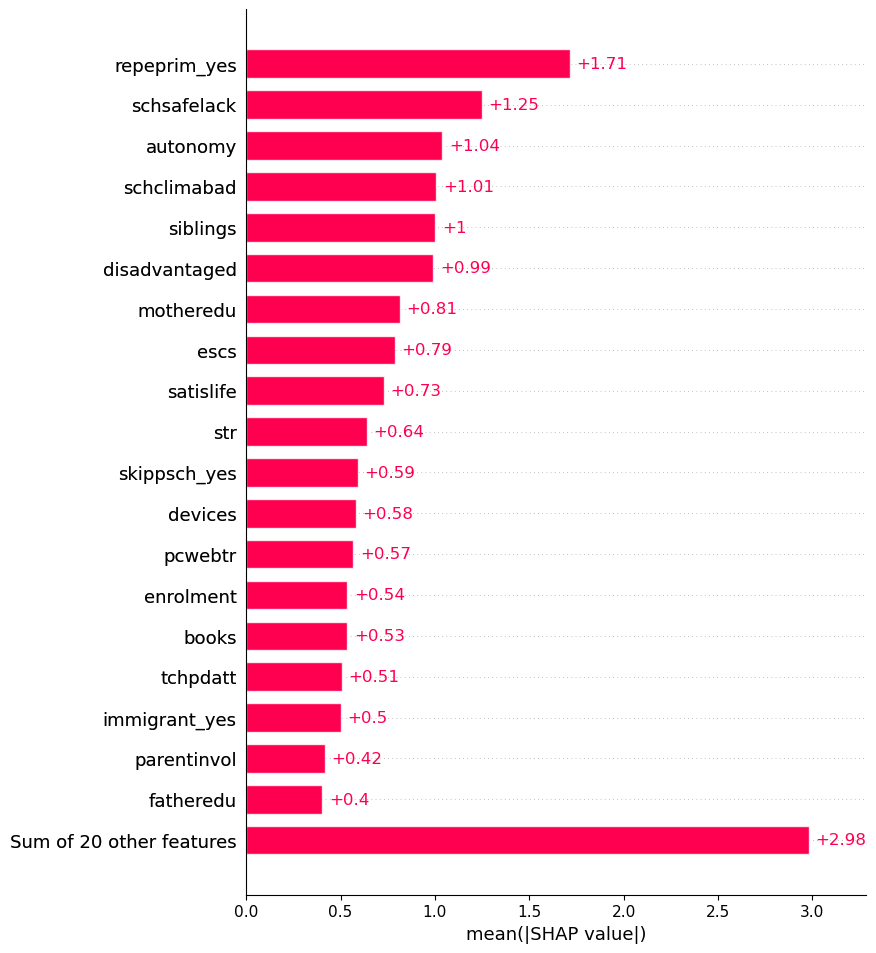}
        }
        \subfloat[Colombia]{%
            \includegraphics[width=0.26\textwidth]{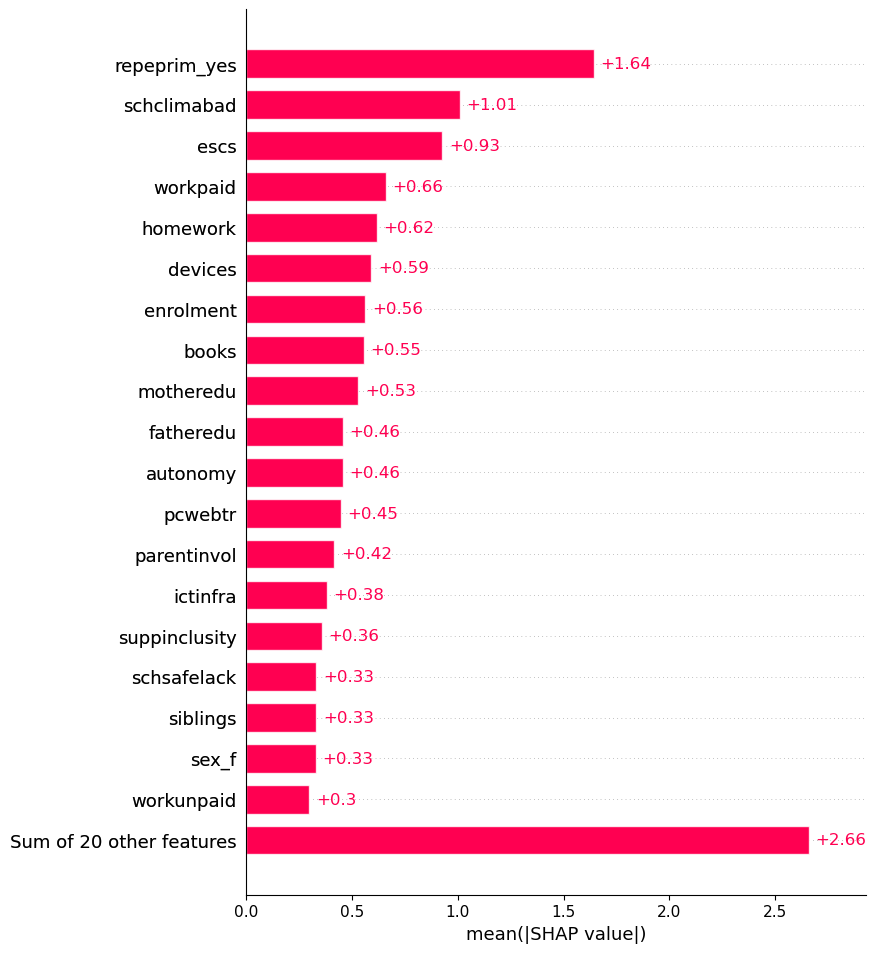}
        }\\
        \subfloat[Dominican Rep.]{%
            \includegraphics[width=0.26\textwidth]{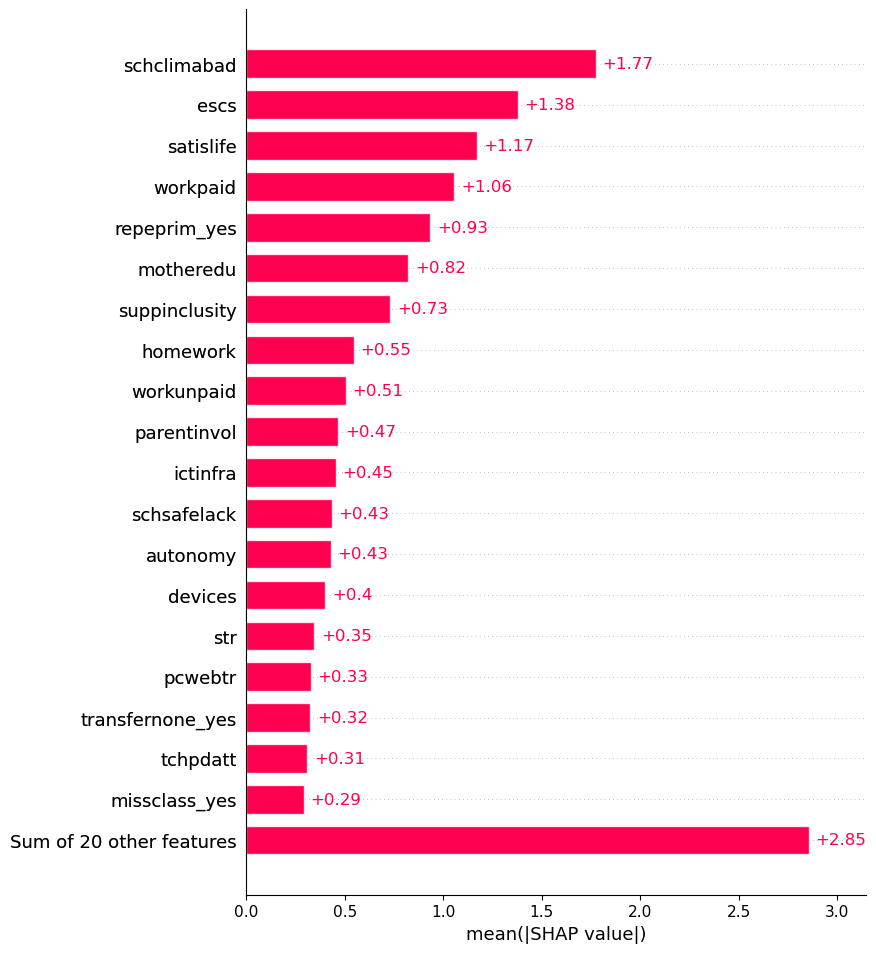}
        }
        \subfloat[Mexico]{%
            \includegraphics[width=0.26\textwidth]{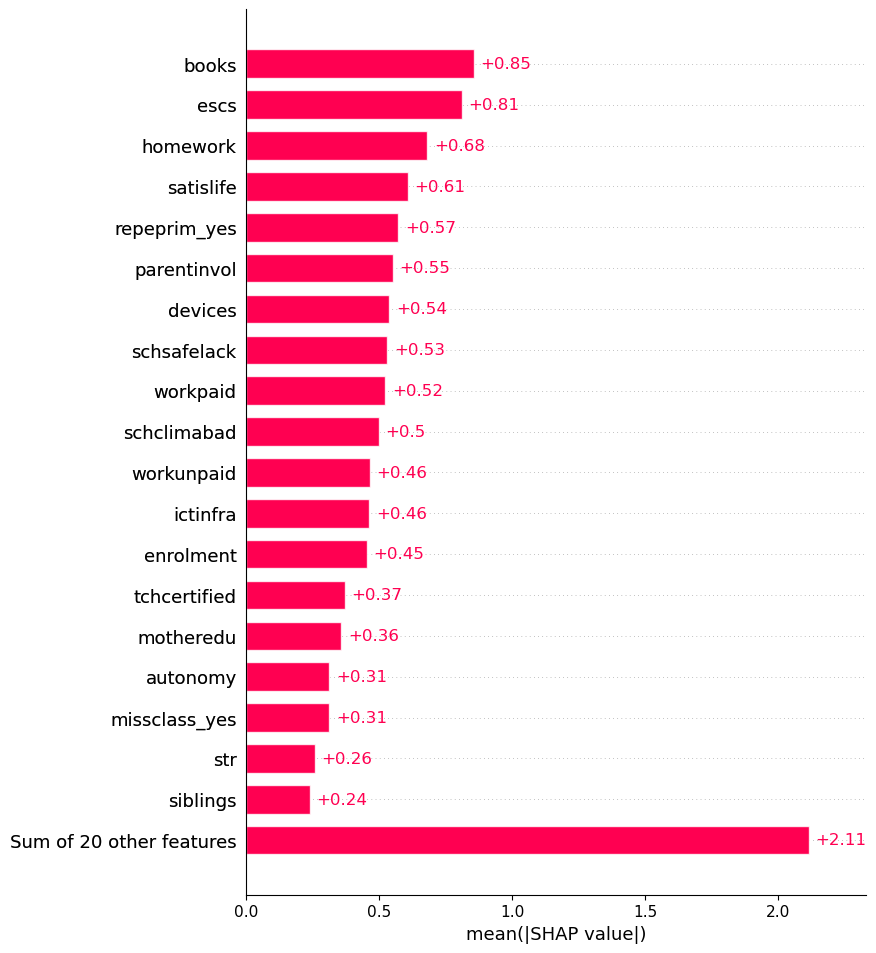}
        }
        \subfloat[Panama]{%
            \includegraphics[width=0.26\textwidth]{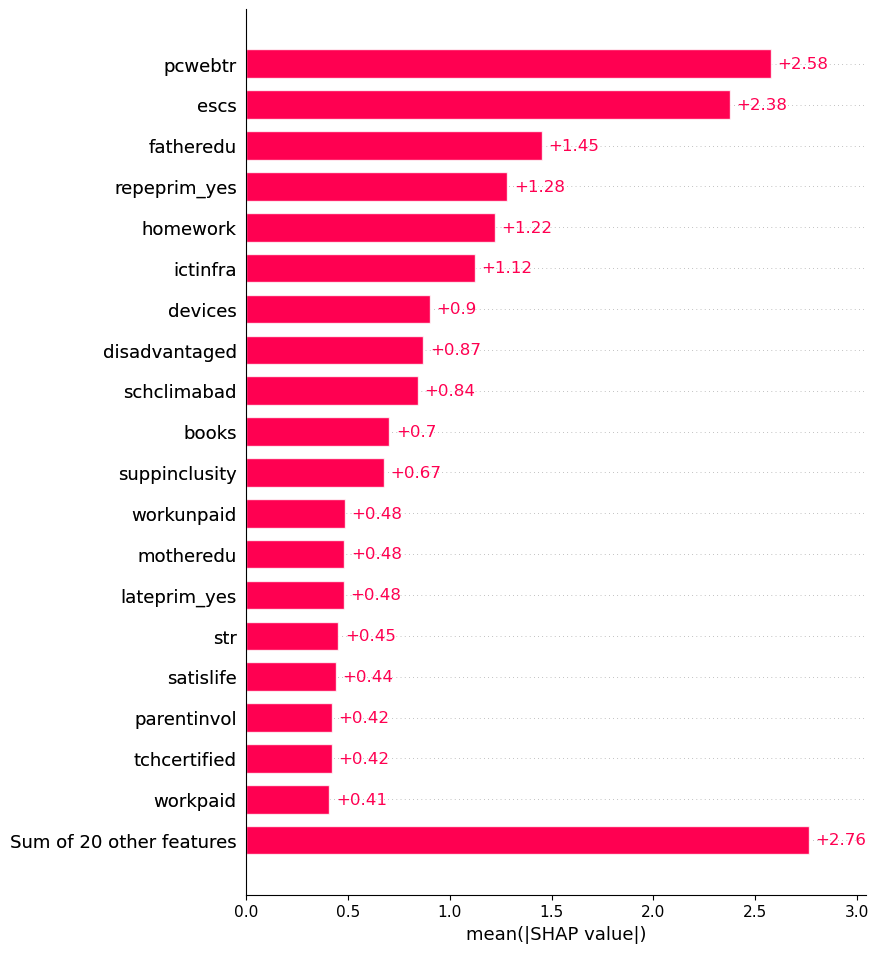}
        }
        \subfloat[Peru]{%
            \includegraphics[width=0.26\textwidth]{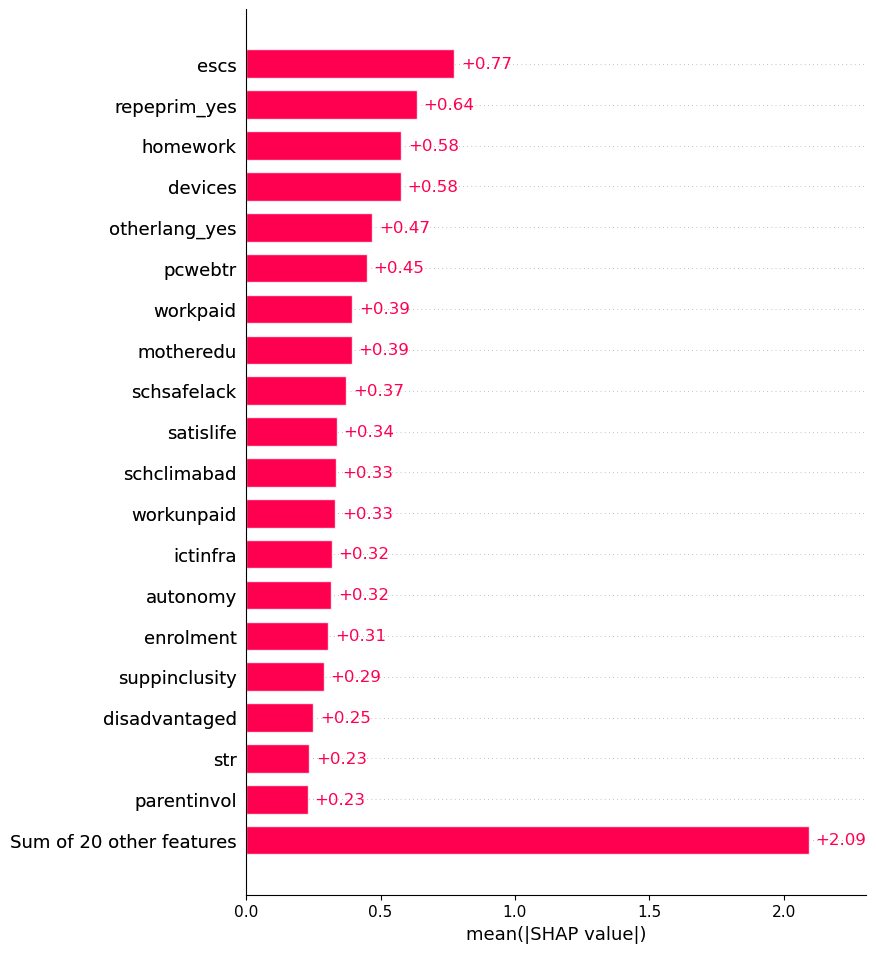}
        }\\
        \subfloat[Paraguay]{%
            \includegraphics[width=0.26\textwidth]{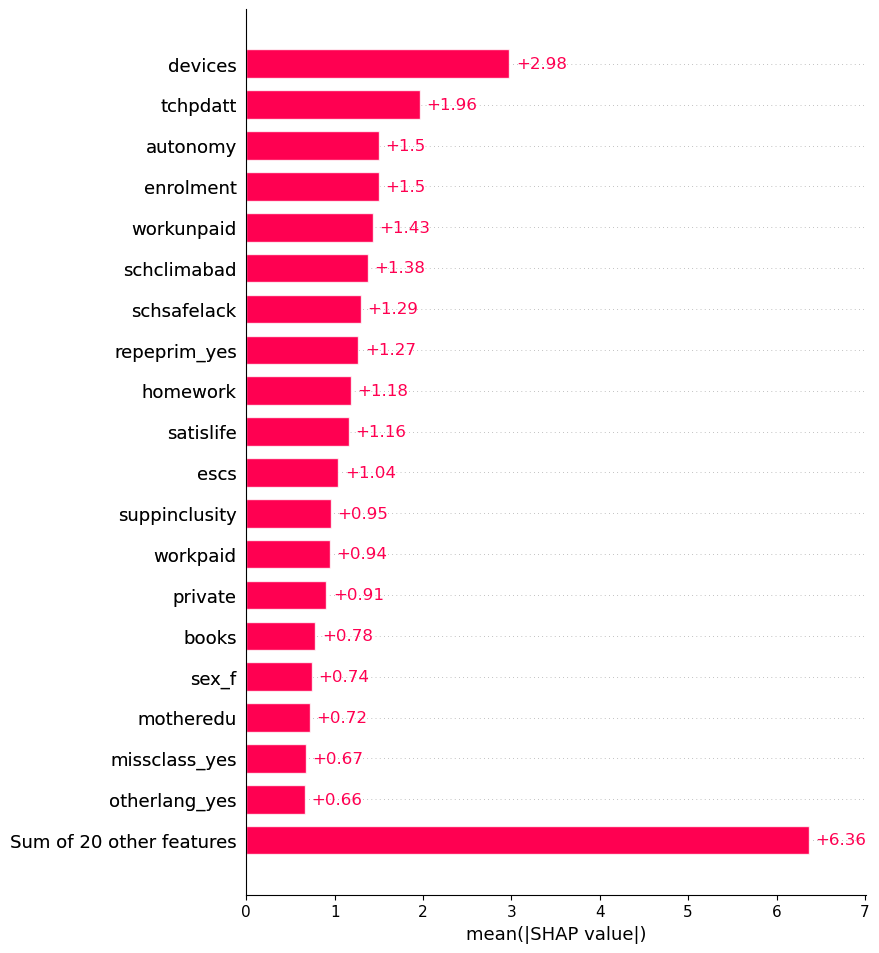}
        }
        \subfloat[Uruguay]{%
            \includegraphics[width=0.26\textwidth]{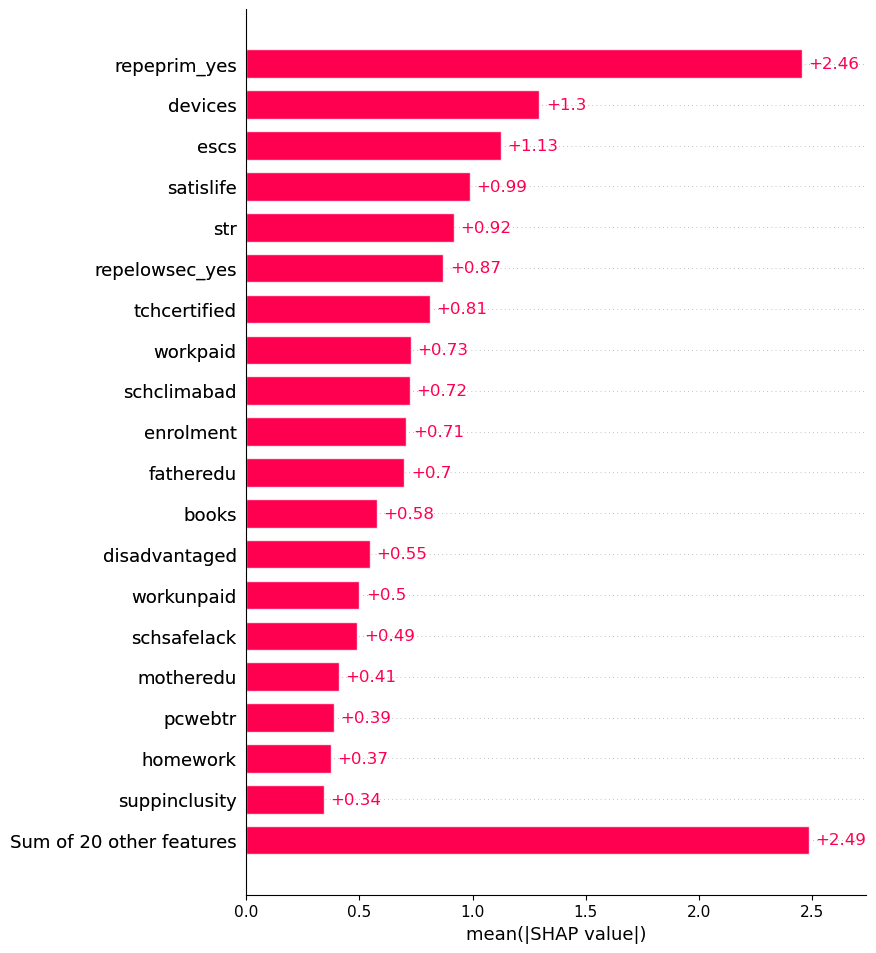}
        }
        \caption{Countries' results for model: level 0 versus level 1 (bottom performing group). Student$/$famliy and school covariates average SHAP values (absolute values) }
        \label{figureD3}
    \end{figure*}


\clearpage
\newpage
\setcounter{figure}{0}
\renewcommand{\thefigure}{E\arabic{figure}}
\setcounter{equation}{0}
\renewcommand{\theequation}{E.\arabic{equation}}

\section*{Appendix E. Analysis for the sub-sample of close down schools during the COVID-19 pandemic}

Here, we look into the pandemic main determinants on students' learning poverty chances for the sub-sample of schools which were shutdown (around 75\%). We rely on the school questionnaire (from question 58) which gathers information only for the sub-sample of schools closed. For the level 0-level 1 comparison sample, the number of observations are reduced from 16,236 to 13,489 (with 668 out of the total 2,747 school remaining opened); and, for the level1-level2 comparison sample, this reduction is from 9,484 to 6,995 observations.

\begin{figure}[ht!]
        \centering
        \subfloat[\scriptsize{School days closed}]{%
            \includegraphics[width=0.33\textwidth]{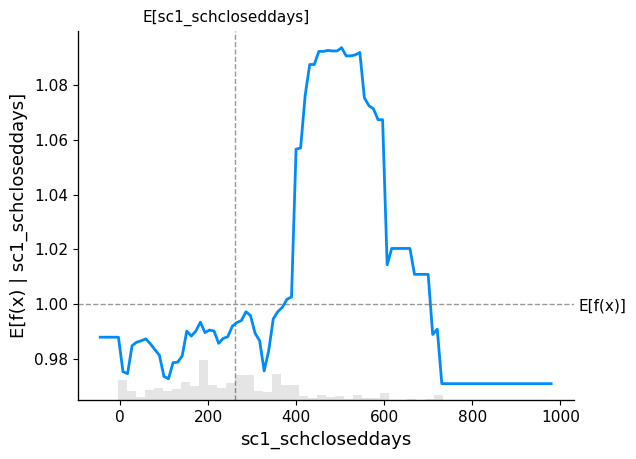}\label{figureE1a}
        }
        \subfloat[\scriptsize{Remote learning-students rate}]{%
            \includegraphics[width=0.33\textwidth]{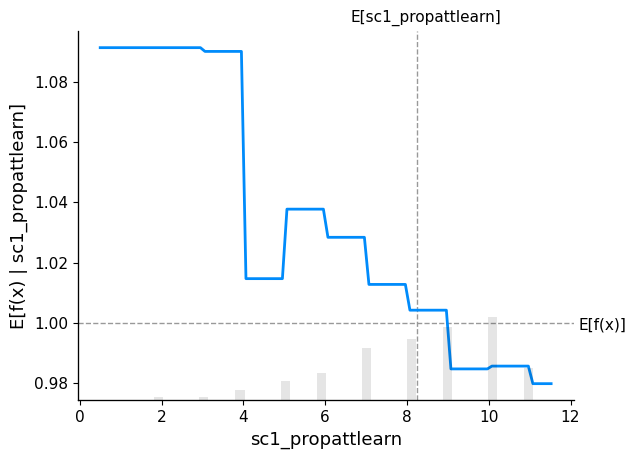}\label{figureE1b}
        }\\
        \subfloat[\scriptsize{Barrier-ICT}]{%
            \includegraphics[width=0.33\textwidth]{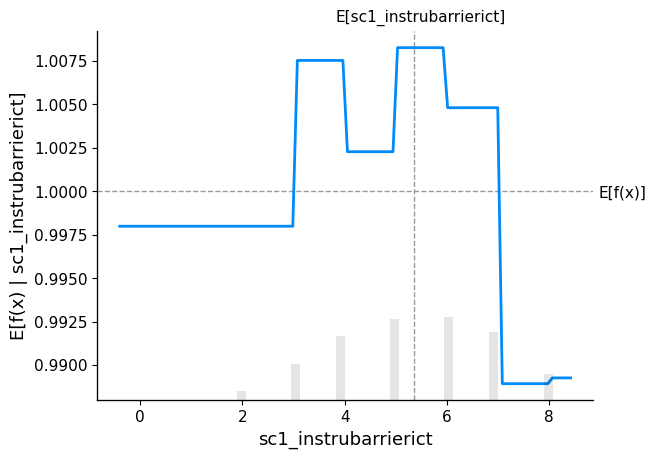}\label{figureE1c}
        }
        \subfloat[\scriptsize{Barrier-connectivity}]{%
            \includegraphics[width=0.33\textwidth]{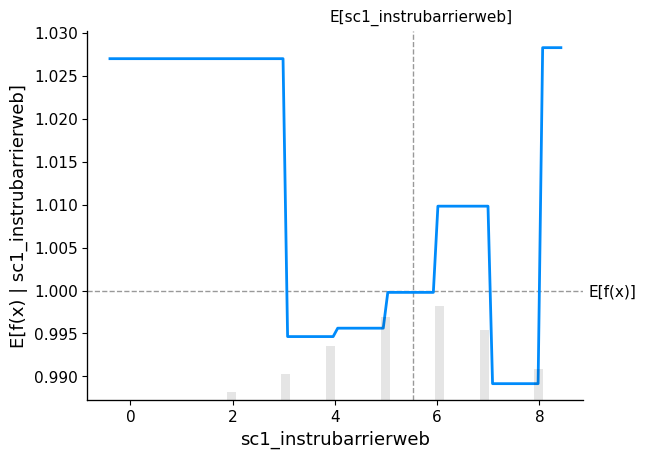}\label{figureE1d}
        }
        \subfloat[\scriptsize{Barrier-system}]{%
            \includegraphics[width=0.33\textwidth]{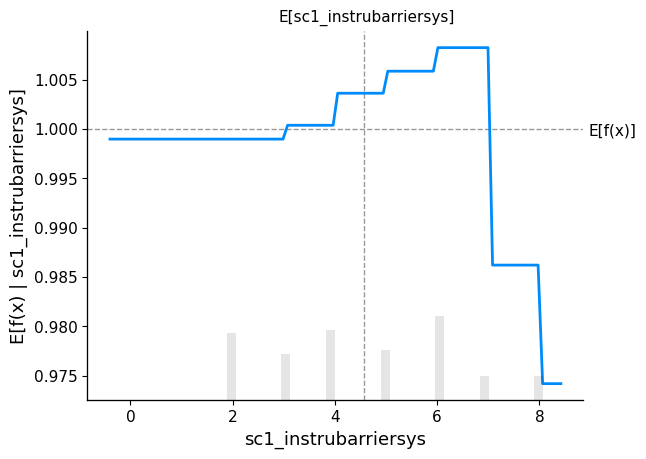}\label{figureE1e}
        }\\
        \subfloat[\scriptsize{Barrier-ICT-interaction}]{%
            \includegraphics[width=0.33\textwidth]{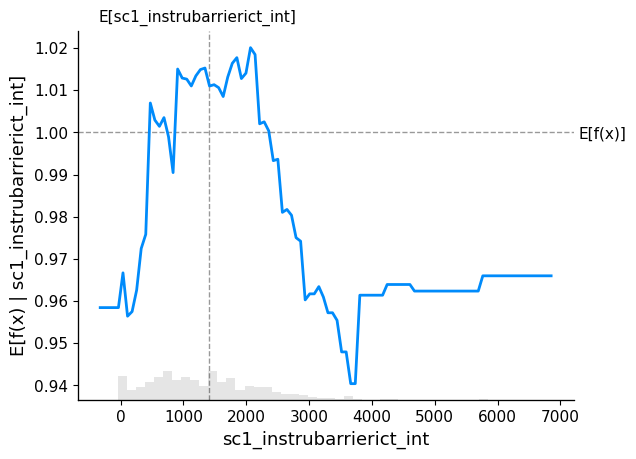}\label{figureE1f}
        }
        \subfloat[\scriptsize{Barrier-connectivity-interaction}]{%
            \includegraphics[width=0.33\textwidth]{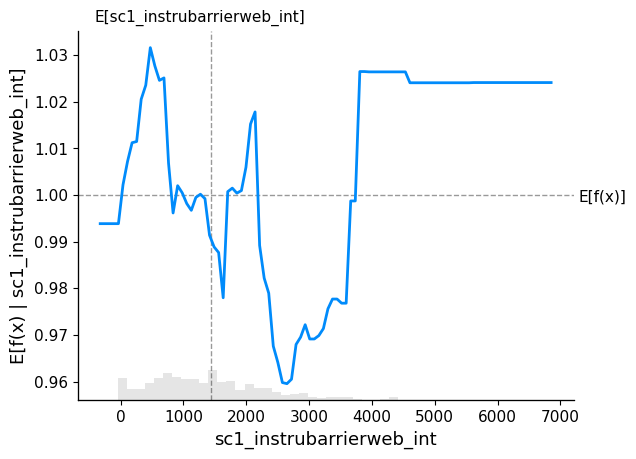}\label{figureE1g}
        }
        \subfloat[\scriptsize{Barrier-system-interaction}]{%
            \includegraphics[width=0.33\textwidth]{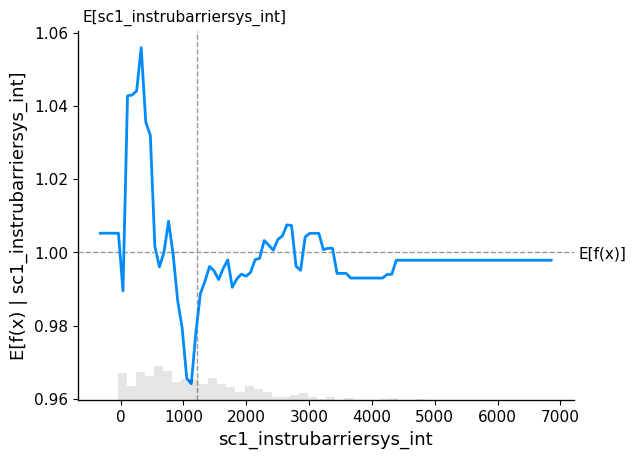}\label{figureE1h}
        }\\
        \caption{Model: level 0 versus level 1 (bottom performing group). Partial dependence plots showing predicted relative ratios (RP) for school days of closure, rate of engagement on remote learning and barriers for remote learning}
        \label{figureE1}
\end{figure}

\begin{figure}[ht!]
        \centering
        \subfloat[\scriptsize{School days closed}]{%
            \includegraphics[width=0.33\textwidth]{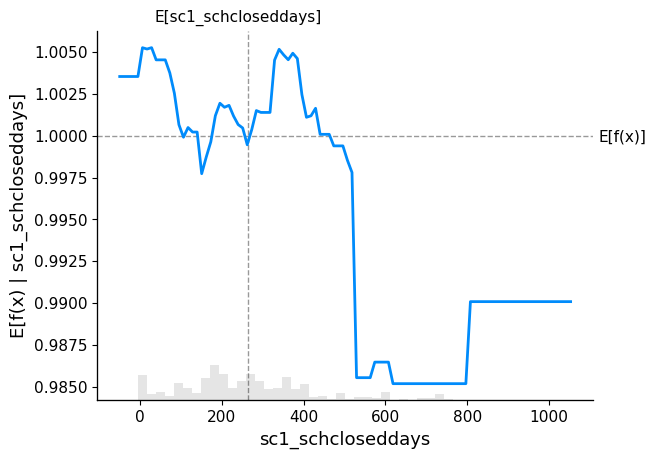}\label{figureE2a}
        }
        \subfloat[\scriptsize{Remote learning-students rate}]{%
            \includegraphics[width=0.33\textwidth]{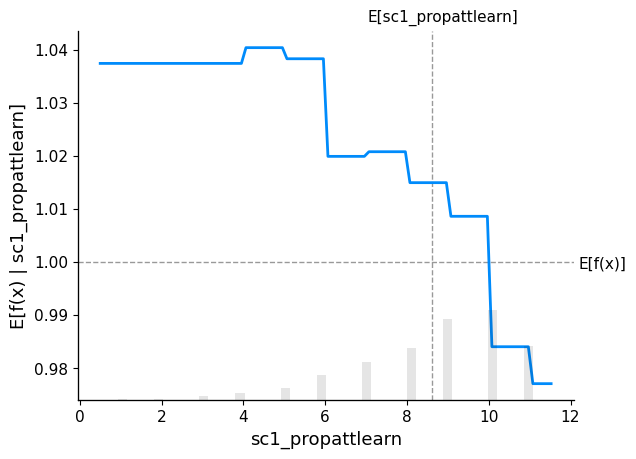}\label{figureE2b}
        }\\
        \subfloat[\scriptsize{Barrier-ICT}]{%
            \includegraphics[width=0.33\textwidth]{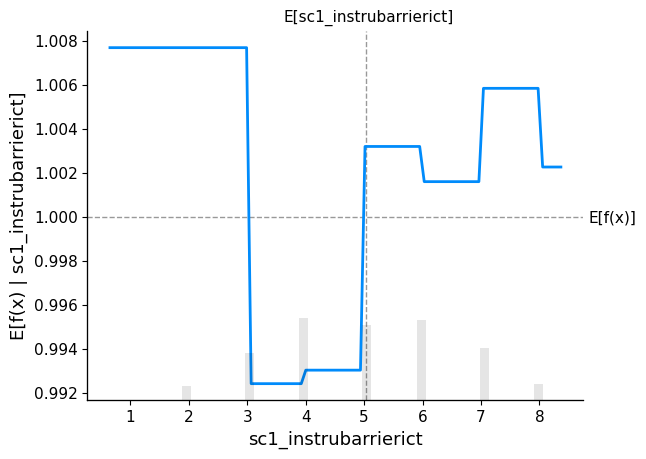}\label{figureE2c}
        }
        \subfloat[\scriptsize{Barrier-connectivity}]{%
            \includegraphics[width=0.33\textwidth]{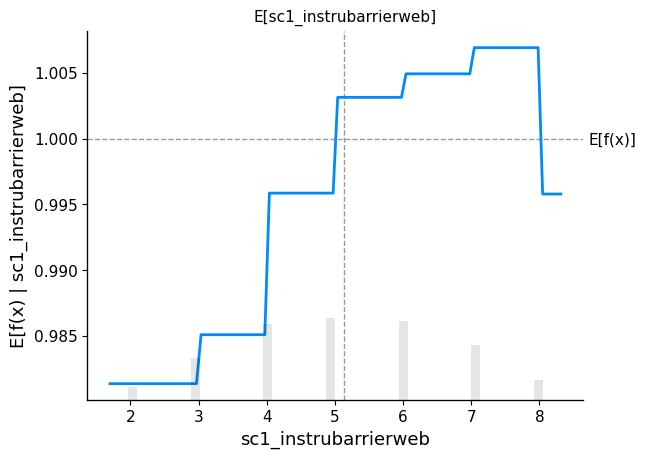}\label{figureE2d}
        }
        \subfloat[\scriptsize{Barrier-system}]{%
            \includegraphics[width=0.33\textwidth]{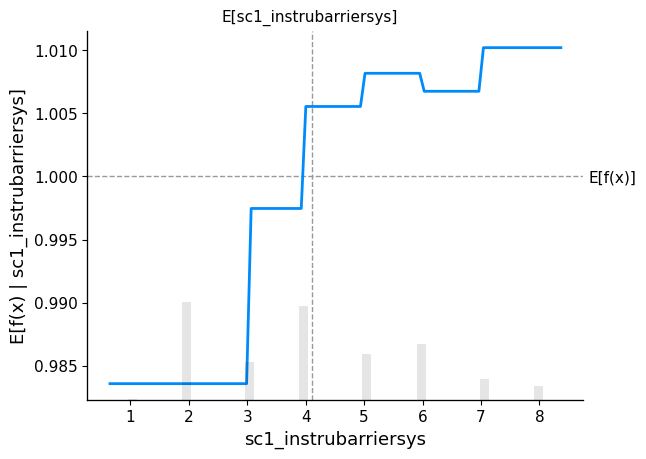}\label{figureE2e}
        }\\
        \subfloat[\scriptsize{Barrier-ICT-interaction}]{%
            \includegraphics[width=0.33\textwidth]{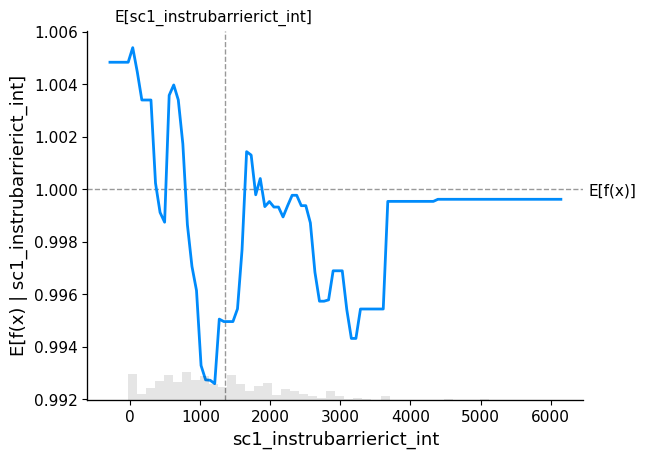}\label{figureE2f}
        }
        \subfloat[\scriptsize{Barrier-connectivity-interaction}]{%
            \includegraphics[width=0.33\textwidth]{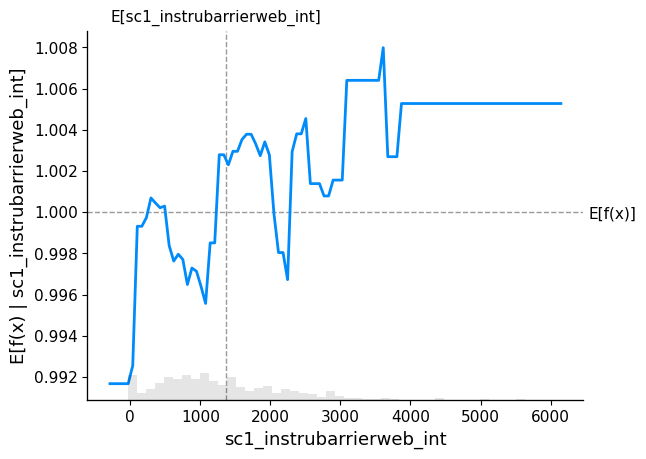}\label{figureE2g}
        }
        \subfloat[\scriptsize{Barrier-system-interaction}]{%
            \includegraphics[width=0.33\textwidth]{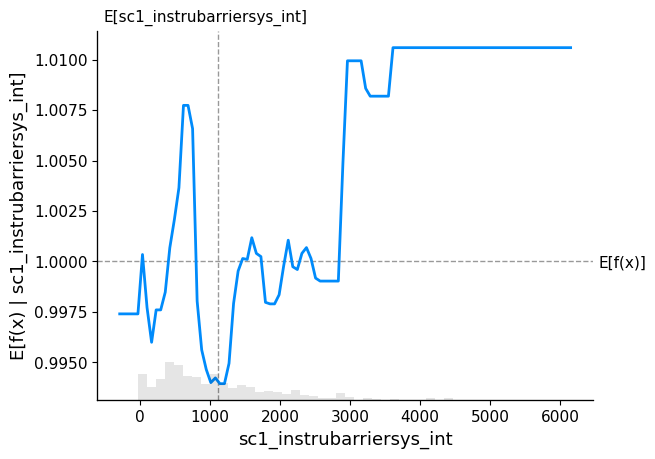}\label{figureE2h}
        }\\
        \caption{Model: level 1 versus level 2 (low performing group). Partial dependence plots showing predicted relative ratios (RP) for school days of closure, rate of engagement on remote learning and barriers for remote learning}
        \label{figureE2}
\end{figure}

We assess the associations of the following covariates: school days of closure (SDC), school prevalence of students engaged in remote learning, and three different barriers for online learning (as well as their interactions with school days of closure) through dependence plots yielding `predicted probability ratios'. Specifically, partial dependence plots show the marginal effect of a given covariate on the prediction of the ML model. We obtain the predicted relative probability (RP) of the educational outcome measuring learning poverty (level 0 =1, level 1 = 1; and level1 = 1 and level 2 = 0). The RP of a given COVID-19 pandemic related covariate is defined as:

\begin{equation}\label{RP}
  \text{RP} = \frac{f_{\mathcal{S}}(x_{\mathcal{S}})}{(1/n) \sum_{i=1}^{n}f(x^{(i)})}
\end{equation}

which is the mean value of the stacked model predicted probability when fixing the specific covariate divided by the mean value of the model's predicted probability, and $f_{\mathcal{S}}(x_{\mathcal{S}})$ is the partial function of the specific covariate. Results are displayed in Figure \ref{figureE1} (for bottom performers) and in Figure \ref{figureE2} (for low performers).

First, for the bottom performers comparison, on the one hand, Figure \ref{figureE1a} shows how longer SDC (above the average of $\approx$ 270) pushes students chances towards the lowest bracket of performance (level 0). In particular, $\widehat{\text{RP}}$ are above one for the 400-700 SDC interval, reaching its peak at 500 days with a $\widehat{\text{RP}}$ = 1.08 where students attending schools closed by 500 days have 8\% higher probability of falling into the level 0 performing group (instead of the level 1 achievement group). Besides, Figure \ref{figureE1b} highlights the relevance of schools reaching significant and larger rates of students actually engaged in remote learning. Indeed, all estimated RP are above one if this rate is below 8 (where the percentage of students attended distance learning being between 61\%-70\%).

On the other hand, as far as barriers for students online learning during the pandemic is concerned (Figures \ref{figureE1c} to \ref{figureE1e}), RP estimates show an overall tendency where chances of students only reaching level 0 performance increases the higher the extent of barriers are. But, with $\widehat{\text{RP}}$ between 1.00-1.02 at given interval of barriers covariates, these effects are not as significant as the effects of SDC and a school's rate of students involved remote learning. Furthermore, in the last three plots (i.e., Figures \ref{figureE1f} to \ref{figureE1h}) we interact each of these three barriers with SDC so that to capture whether the impact of barriers on level 0 performing likelihood is potentiated by longer periods of school closure. Estimated RP for the three variables reach maximum values of 1.02, 1.03 and 1.06, without an increasing tendency along higher values of barriers and higher values of SDC. Overall, the bottom performing analysis shown in the plots of Figure \ref{figureE1} conveys more relevance to the degree of school shutdowns and rates of remote learning rather than barriers for engagement and pedagogic continuity during the pandemic.

Second, with regards to the low performing group comparison, pandemic covariates' impact on the probability of students falling into level 1 (instead of level 2) has a distinct pattern when compared to what we found earlier for students chances falling into level 0 (shown in Figure \ref{figureE1}). To begin with, SDC impacts on level 1 achievement is much smaller in magnitude, RP below 1.005 (Figures \ref{figureE2a}), and for the variable school's rate of engagement on remote learning of students (Figure \ref{figureE2b}), there is a similar decreasing tendency showing the importance of students' being engaged in distance learning for balancing level 1 and level 2 performance chances, though this become equal at a school's higher rate for remote learning (when the covariate takes the value of 10: 81\%-90\%). In addition, barriers' estimates (Figures \ref{figureE2c}-Figure \ref{figureE2e}) for distance learning have a clear increasing tendency (specially connectivity and systemic barriers) and, when they are above their mean values, the $\widehat{\text{RP}}$ are above 1, but again these values are not so large ($\le$ 1.01). The same holds for the interacted versions of barriers (i.e., Figures \ref{figureE2f} to Figure \ref{figureE2h}).

\end{document}